%% file: paper.tex
\documentclass[aps,prd,reprint,showpacs,floatfix,longbibliography,nofootinbib,superscriptaddress]{revtex4-1}
\usepackage{tikz,graphicx,booktabs,multirow,array,microtype,mathtools,float}
\usepackage{enumerate}  
\extrafloats{200}
\usepackage[colorlinks=true,backref=false, linktocpage=true,
citecolor=blue,
urlcolor=blue,linkcolor=blue,
pdfpagemode=UseOutlines]{hyperref}
\usepackage{dcolumn}

\usepackage{anyfontsize}
\usepackage[shortlabels]{enumitem}

\usepackage[noabbrev,capitalize]{cleveref} 
\usepackage{adjustbox}
\usepackage{longtable}
\newcolumntype{L}{>{$}l<{$}}
\newcolumntype{C}{>{$}c<{$}}
\newcolumntype{R}{>{$}r<{$}}
\renewcommand{\thetable}{\arabic{table}}  

\usepackage{comment} 
\excludecomment{supplement}
\usepackage{braket}
\usepackage{amsfonts}
\AtBeginDocument{
\heavyrulewidth=.08em
\lightrulewidth=.05em
\cmidrulewidth=.03em
\belowrulesep=.65ex
\belowbottomsep=0pt
\aboverulesep=.4ex
\abovetopsep=0pt
\cmidrulesep=\doublerulesep
\cmidrulekern=.5em
\defaultaddspace=.5em
}
\def\fm {\mathop{\hbox{fm}}}
\def\MeV {\mathop{\hbox{MeV}}}
\def\GeV {\mathop{\hbox{GeV}}}

\usepackage{amsmath} 
\usepackage{dsfont}  
\usepackage{bm}      

\def\beq{\begin{equation}}
\def\eeq{\end{equation}}
\def\beqs#1\eeqs{\beq\begin{split} #1 \end{split}\eeq}

\long\def\comment#1{}

\usepackage{placeins}

\usepackage{mleftright,xparse}
\NewDocumentCommand\xDeclarePairedDelimiter{mmm}
{%
	\NewDocumentCommand#1{som}{%
		\IfNoValueTF{##2}
		{\IfBooleanTF{##1}{#2##3#3}{\mleft#2##3\mright#3}}
		{\mathopen{\csname##2\endcsname#2}##3\mathclose{\csname##2\endcsname#3}}%
	}%
}
\xDeclarePairedDelimiter{\av}{\langle}{\rangle}

\NewDocumentCommand\opbraket{sommm}{%
	\IfNoValueTF{#2}
	{\IfBooleanTF{#1}{\langle#3|#4|#5\rangle}{\mleft\langle #3 \left| #4 \right| #5 \mright\rangle}}
	{\mathopen{\csname#2\endcsname\langle}#3\mathopen{\csname#2\endcsname|} #4 \mathclose{\csname#2\endcsname|} #5\mathclose{\csname#2\endcsname\rangle}}
}

\begin{document}

\title{Higher order quantization conditions for two-body scattering with spin}

\author{Lucas\ Chandler}
\email{lchandler27@gwu.edu}
\author{Frank X.\ Lee}
\email{fxlee@gwu.edu}
\author{Andrei\ Alexandru}
\email{aalexan@gwu.edu}
\affiliation{Physics Department, The George Washington University, Washington, DC 20052, USA}

\date{Oct 23, 2025}

\begin{abstract}
We examine the L\"uscher quantization condition to high order for the scattering of a spinless particle and a spin-1/2 particle in a periodic box.
First, we derive the quantization conditions in a non-relativistic framework up to total angular momentum $J=11/2$ in both cubic and elongated geometries, and for both rest and moving frames.
Then, we introduce a method to transparently cross-check their convergence, using both quantized energy levels in the box and infinite-volume phase shifts for the same potential.  We clarify how to incorporate spin-orbit coupling into the formalism and show in detail how the quantization conditions converge order by order in the various irreducible representations.  In all, we validated 19 quantization conditions (12 in cubic box, 7 in elongated box). This is a necessary step in applying the method in precision studies of systems in finite volume  with half-integer spin, such as meson-baryon scattering.
\medskip 

\end{abstract}

\maketitle

\section{Introduction}
Scattering is an indispensable tool in our understanding of nature at deeper and deeper levels, from the original Rutherford experiment on the structure of the atom, to modern nuclear and particle physics experiments on the structure and interaction of hadrons. Such physics is governed by quantum chromodynamics (QCD), the fundemantal theory of the strong interaction. Theoretically, determination of hadronic properties at low to medium energies from the first principles of QCD remains fundamental but challenging. 
Because the interaction is strong, it requires non-perturbative methods. 

In lattice QCD, numerical simulations based on Monte Carlo evaluation of the path integral of QCD are performed by placing the system on a discrete lattice with periodic boundary conditions, which leads to quantized energy spectrum in the system. 
Continuum but finite-volume results can be obtained by increasing the number of grid points and decreasing the lattice spacing at the same time. 
The energy spectrum in the finite volume is then  connected  to the infinite volume scattering phase shifts through quantization conditions (QC). 
A number of finite-volume approaches have been proposed for two-body scattering. The most well-known is the L\"uscher method \cite{Luscher:1990ux} which establishes a direct connection between the energy levels and phase shifts.
Alternative approaches based on the finite volume include the  HALQCD method~\cite{PhysRevLett.99.022001,PhysRevD.99.014514,ISHII2012437}
which extracts the interaction potential from Bethe-Salpeter amplitudes, 
 and the integrated correlation function method (ICF)~\cite{Guo:2023ecc,Guo:2025lmd,guo2026}
 which works directly with correlation functions and bypasses the energy spectrum determination. 

The original L\"uscher method has been extended to a variety of scenarios (such as moving frames, asymmetric lattices, multi- and coupled-channels, effective field theories, etc) to extract information about hadron-hadron interactions~\cite{Rummukainen:1995vs,Kim_2005,Fu:2011xz,Leskovec:2012gb,Gockeler:2012yj,Bernard:2008ax,Li:2021mob,Doring:2012eu,Li:2012bi,Guo:2012hv,Feng:2004ua,Lee:2017igf,LIU_2006,Briceno:2012yi,Briceno:2014oea,Li:2019qvh,Doring:2011wz,Hansen_2012,Guo:2024pvt}. 
In the meson sector, such methods have been successfully applied in lattice QCD simulations~\cite{Aoki:2007rd,Feng:2010es,Lang:2011mn,Aoki:2011yj,Guo:2016zos,Dudek:2016cru,Luu:2011ep,BULAVA2016842,Gunnar2016,Morningstar_2017,Mai2019,wang2025}.
With increasing computing power and algorithm improvements, such approaches are starting to bear fruit in the baryon sector~\cite{Bulava_2023,Bulava_2024,Alexandrou_2024,Kiratidis:2015vpa,morningstar2025nucleonscatteringlatticeqcd,Leskovec_2018,m32c-wk63} (see Ref.\cite{morningstar2025} for a recent review). The main challenge is to overcome the deterioration of signal-to-noise ratio when baryon are involved. 
Remarkable progress has also been made in finite volume approaches in the three-body sector~\cite{Beane:2007es,Guo:2019ogp,Mai:2019fba,Horz:2019rrn,Romero-Lopez:2019qrt,Blanton:2019igq,Mai:2021nul,Alexandru:2020xqf,Mai:2021lwb,Yan2024,Yan:2025mdm,Feng:2026ixm,Bubna:2026iga,alharazin2026} (see Ref.~\cite{sharpe2026} for a recent review).

In the present work, we focus on the scattering of a spinless particle and a spin-1/2 particle in the L\"uscher method. The study is motivated by the desire to apply the method in the baryon sector, especially meson-baryon scattering. 
We aim to conduct a comprehensive examination of the QCs in this sector, 
complete with  cubic and elongated boxes, rest and moving frames, unequal masses, and higher partial waves (up to $\ell = 6$ or $J = 11/2$). 
Furthermore, after deriving the QCs, we validate each one numerically to high precision. This is accomplished by separately computing the two main ingredients of the QCs for a toy potential: energy spectrum in the periodic box and infinite-volume phase shifts.
It follows a similar study for systems with no spin~\cite{Lee_2022}. 

The work is presented in four steps. In Sec. \ref{section:QuantizationConditions} we outline the derivation of the QCs and their projection to the irreducible representations (irreps) of various symmetry groups. In  Sec. \ref{section:two-part energies} we detail how to solve the Schr\"odinger equation in the box to obtain the  energy spectrum for a test potential that includes a spin-orbit interaction.  In Sec. \ref{section:numerical checks} we compute infinite volume phase shifts, discuss how to check the convergence of the QCs, and show a few examples. A brief summary and outlook is given in Sec.\ref{section:conclusions}. 
The matrix elements for the derived QCs are partially given in Appendix \ref{appendix:MatrixElements}.  A supplement is supplied from which the full QCs can be retrieved~\cite{supp}.
Details of the group theory used for the derivation are provided in Appendix \ref{appendix:GroupTheory}. 

\section{Quantization Condition}\label{section:QuantizationConditions}
There are different avenues to arrive at the same QCs, up to exponentially suppressed corrections. In this work, we work with the simplest approach, namely, non-relativistic quantum mechanics. Despite its simplicity, the approach captures all the essential ideas and ingredients in the construction of QCs.
The derivation is fairly involved mathematically and requires special functions (zeta functions) and extensive use of group theory.
The introduction of spin-1/2 degrees of freedom adds even more complexity than the zero spin case and requires double-cover group theory. 
There exist previous studies on systems with half-integer spin~\cite{Bernard:2008ax,Gockeler:2012yj,Li:2012bi,Lee:2017igf} in limited scopes. A comprehensive reexamination aims not only to serve as an independent check of existing QCs in the literature but also as a foundation for new QCs at higher partial waves and in elongated boxes.
In the following, we outline the essential steps in a reasonably self-contained  derivation.

The starting point is two-particle scattering in the center of mass (CM) frame described by the Schr\"odinger equation,
\beq
\left[-\frac{\hbar^2}{2\mu}\nabla^2 + V(r)\right]\psi(\bm r) = E \psi(\bm r),
\label{eq:scheq}
\eeq
where $\mu = \frac{m_1 m_2}{m_1+m_2}$ is the reduced mass of the system. 
We omit the spin degree of freedom in the discussion until near the end when a change of basis can be used to transform the matrix elements from a spin-0 system to a spin-1/2 one. 
In infinite volume, the wavefunction in the asymptotic region has the form of an incoming plane wave plus an outgoing spherical wave,
\beq
\psi(\bm r) = e^{ikz} + f(\theta,k) \frac{e^{ikr}}{r},
\eeq
where $k$ is the relative CM momentum and is related to the CM energy by $E = \frac{\hbar^2 k^2}{2\mu}$. The scattering amplitude can be expressed in partial-wave expansion,
\beq
f(\theta,k)=\sum_{l=0}^\infty (2l+1)f_l(k)P_l(\cos\theta).
\eeq 
Phase shifts enter via the partial-wave amplitudes,
\beq
f_l(k)\equiv \frac{e^{2i\delta_l}-1}{2 i k} 
= \frac{S_l-1}{2 i k}
= \frac{T_l}{k}
= \frac{K_l}{k(1-iK_l)},
\eeq 
where alternative definitions via S matrix, T matrix and K matrix are also indicated.
Phase shift $\delta_l$ is a real valued function of the interaction energy and encodes information about the nature of the interaction
between the two particles, such as whether the force is repulsive ($\delta < 0$) or attractive ($\delta > 0$), whether a resonance is formed in the scattering process (a sharp change across $\pi/2$), and so on.

The standard procedure to determine the phase shift is by matching the asymptotic solution (where the interaction vanishes) in terms of spherical Bessel functions $\psi \propto [\alpha_lj_l(kr) + \beta_ln_l(k)]$ (or exterior solution),
and the interior solution to arrive at the expression,
\beq
e^{2i\delta_l(k)} = \frac{\alpha_l(k) + i \beta_l(k)}{\alpha_l(k) - i \beta_l(k)}.
\label{Eq:phase shift}
\eeq

In finite volume, a similar procedure can be followed. The main difference is the need to accommodate periodic boundary conditions imposed by the box. 
The details were first worked out by L\"uscher \cite{Luscher:1990ux}. 
The setup assumes that the box, of length $L$, is large enough such that the interaction range, $R$, cannot wrap around the periodic boundary of the box $R < L/2$. 
The resulting quantization condition can be expressed in a single matrix equation that includes all partial waves,
\beq
\text{det}\left[\text{diag}\{e^{2i\delta_l(k)}\} - \frac{\mathcal{M}(k,L) + i}{\mathcal{M}(k,L) - i}\right] = 0.
\label{QC}
\eeq
Here, $\text{diag}\{e^{2i\delta_l(k)}\}$ is a shorthand for a diagonal matrix of all partial waves. The $\mathcal{M}(k,L)$ is a purely mathematical, dimensionless,  Hermitian matrix function of CM momentum and box size. Its derivation for half-integer angular momentum systems is the focus of this work.
The condition establishes a connection between quantized energy levels in a periodic box and phase shifts in infinite volume. In fact, the quantization condition is fairly general: It does not matter how the energy spectrum is generated in the box, be it quantum mechanics, effective field theories, or lattice QCD. The quantization condition will remain the same up to exponentially suppressed finite-volume corrections.

To obtain an analytic expression for the $\mathcal{M}(k,L)$ matrix, one starts by identifying a solution to the wave function in the exterior where the Schr\"odinger equation has the form of a Helmholtz equation,
\beq
(\nabla^2 + k^2) \psi(\bm r) = 0.
\label{Helmholtz}
\eeq
A solution to this equation that satisfies the periodic boundary conditions is given by the Green's function,
\beq
G(\bold{r},k^2) = \frac{1}{L^3}\sum_{\bm p}\frac{e^{i\bm p\cdot \bm r}}{\bm p^2 - k^2},
\eeq
where the sum runs over the quantized box momenta. A complete basis can be formed by taking its derivatives (see Ref.\cite{Luscher:1990ux}), 
\beq
G_{lm}(\bold{r},k^2) = \mathcal{Y}_{l,m}(\nabla) G(\bold{r},k^2),
\eeq
where $ \mathcal{Y}_{lm}(\bold{r}) = r^l Y_{lm}(\theta,\phi)$ are the homogeneous harmonic polynomials. In order to match the exterior solution with the interior, one needs to expand $G_{lm}$ in terms of $n_l, j_l,$ and $Y_{lm}$,
\beqs
G_{lm}(\bm r, k^2) &= \frac{(-1)^l k^{l+1}}{4\pi} [n_l(kr) Y_{lm}(\theta,\phi)\\
&+\sum_{l'=0}^{\infty}\sum_{m'=-l'}^{l'}\mathcal{M}_{lm,l'm'}j_{l'}(kr)Y_{l'm'}(\theta,\phi)],
\eeqs
where the $\mathcal{M}$ matrix is introduced as a way to connect to phase shifts in similar fashion to that in infinite volume. Performing the interior-exterior matching procedure with the wave function expanded in this basis leads to,
\beq
\psi = \sum_{lm}d_{lm}G_{lm}(\bm r, k^2) = \sum_{lm} c_{lm}\left[a_l j_l + b_l n_l \right]Y_{lm}.
\eeq
By equating the coefficients of $n_l$ and $j_l$, the following condition holds, 
\beq
\sum_{l'm'}c_{l'm'}\left[b_{l'}\mathcal{M}_{l'm',lm} - a_{l'}\delta_{ll'}\delta_{mm'}\right]=0.
\eeq
Considering only nontrivial solutions results in the determinant equation,
\beq
    \det\left[B\mathcal{M} - A\right] = 0,
\eeq
where $A$ and $B$ are diagonal matrices from $a_l$ and $b_l$, respectively. The matrix version of Eq.(\ref{Eq:phase shift}) gives the connection to the phase shifts,
\beq
e^{2i\delta} = \frac{A + iB}{A - iB},
\eeq
and yields the QC as given in Eq.(\ref{QC}). 

The matrix $\mathcal{M}$ emerging from the matching procedure takes the explicit form,
\beq
\mathcal{M}_{l m,l' m'} = \frac{(-1)^l}{\eta \pi^{3/2}}\sum_{j=|l-l'|}^{l+l'} \sum_{s=-j}^j \frac{i^j}{q^{j+1}}Z_{j s}(q^2, \eta) C_{lm, js, l' m'},
\eeq
where $\eta$ is the z-elongation factor for non-cubic box geometry,
and the tensor coefficient are the Wigner 3j symbols,
\beqs
C_{lm,js,l'm'}&=(-1)^{m'}i^{l-j+l'}\sqrt{(2l+1)(2j+1)(2l'+1)}
\\
&\times 
\begin{pmatrix}
    l & j & l'\\
    0 & 0 & 0
\end{pmatrix}
\begin{pmatrix}
    l & j & l'\\
    m & s & -m'
\end{pmatrix}.
\eeqs
It is expressed in terms of zeta functions defined by,
\beq
Z_{lm}(q^2, \eta) = \sum_{\Tilde{\bold{n}}}\frac{\mathcal{Y}_{lm}(\Tilde{\bold{n}})}{\Tilde{\bold{n}}^2 - q^2},
\label{eq:Zeta}
\eeq
where  the summation index $\Tilde{\bold{n}}$ and $q$ are given by,
\beq
\Tilde{\bold{n}} = (n_x, n_y, n_z/\eta), \;\;\;\;\; q = \frac{kL}{2\pi}.
\eeq

The projection to half-integer angular momentum is achieved by a straightforward change of basis by coupling to spin-1/2 by $\bm J=\bm \ell + {\bm 1\over 2}$,
\beqs
& \mathcal{M}_{JlM,J'l'M'} = \\
& \sum_{m,m'}\sum_{\sigma,\sigma'}\braket{lm,\frac{1}{2}\sigma|JM}\braket{l'm',\frac{1}{2}\sigma'|J'M'}  \mathcal{M}_{lm,l'm'}.
\eeqs
where $\ket{JlM}$ is the coupled basis of total angular momentum $J$, orbital angular momentum $\ell$, and $J$ sub-states $M$.
Both the zeta function and the matrix $\mathcal{M}$ are mathematical functions dependent on dimensionless variables. One can note that the poles of the zeta function $\Tilde{\bold{n}} = q^2$ correspond to non-interacting energies in the box. Interactions cause deviations from the poles and it is these deviations that are responsible for the phase shifts in the QCs.

We employ a commonly-used shorthand notation in the presentation of the QCs, 
\beq
w_{lm}(q^2, \eta) \equiv \frac{Z_{lm}(q^2, \eta)}{\eta \pi^{3/2}q^{l+1}}.
\label{eq:wlm}
\eeq
This allows us to express $\mathcal{M}$ as a linear combination of $w_{lm}$ with purely numerical coefficients. 

\subsection{Symmetry-adapted quantization conditions}
In finite volume, the continuous rotational symmetry is reduced to the finite symmetry group of the box so the quantization condition in Eq.\eqref{QC} must take on the same symmetry properties. Multiplying both sides of Eq.\eqref{QC} by $\det[\mathcal{M}-i]$ gives,
\beq
\det[e^{2i\delta}(\mathcal{M}-i) - (\mathcal{M}+i)] = 0.
\eeq
After distributing and refactoring it becomes,
\beq
\det\left[(e^{2i\delta} - 1)\left(\mathcal{M}-i\frac{(e^{2i\delta}+1)}{(e^{2i\delta}-1)}\right)\right] = 0.
\eeq
Dropping the constant factor, the resulting matrix quantization condition can be written in standard form in terms of matrix elements,
\beq
\det\left[\mathcal{M}_{JlM,J'l'M'} - \delta_{JJ'}\delta_{ll'}\delta_{MM'}\cot \delta_{Jl}\right]=0,
\label{QC1}
\eeq
where the delta functions should not be confused with phase shifts $\delta_{Jl}$.

The QC in Eq.\eqref{QC1} can be projected to the space of irreducible representations (irreps) of the box symmetry. This projection block-diagonalizes the QC matrix. The determinant then factors into smaller, independent ones, and partial-wave mixing is confined to what the symmetry actually permits. If the box energy levels are irrep-projected in matching fashion,  it allows one to study the scattering system on an irrep by irrep basis.

The projection is achieved by a decomposition into the basis vectors of the symmetry group, expressed as,
\beq
 \ket{\Gamma \alpha J l n} = \sum_{M}C^{\Gamma \alpha n}_{JlM}\ket{JlM},
\eeq
where $\Gamma$ stands for a given irrep, $\alpha$ runs from 1 to the dimension of the irrep, $n$ runs from 1 to $n(\Gamma,J)$, the multiplicity of $J$ in irrep $\Gamma$. Using this decomposition, the matrix is projected onto, 
\begin{align}
&\braket{\Gamma\alpha J l n|\mathcal{M}|\Gamma' \alpha' J' l' n'} = \label{eq:Mdefinition} 
\\ &\sum_{MM'}(C^{\Gamma \alpha n}_{JlM})^*C^{\Gamma' \alpha' n'}_{J'l'M'}\mathcal{M}_{JlM,J'l'M'}
=\delta_{\Gamma \Gamma'}\delta_{\alpha \alpha'}\mathcal{M}^{\Gamma}_{Jln,J'l'n'}, \nonumber
\end{align}
where Shur's lemma in group representation theory is used to achieve irrep orthogonality. Since the components $\alpha$ are not observables, one can average over them for multidimensional irreps, leading to the final form,
\beq
\label{Eq:QC1}
\prod_{\Gamma}\det\left[\mathcal{M}_{Jln,J'l'n'}^\Gamma - \delta_{JJ'}\delta_{ll'}\delta_{nn'}\cot\delta_{Jl}\right] = 0.
\eeq
We see that the QC is completely factorized by the irreps of symmetry groups in the box. 

The construction of the matrix elements $\mathcal{M}^{\Gamma}_{Jln,J'l'n'}$ is technically involved, relying on standard but complicated group-theory steps. 
We have derived the QCs up to $J=11/2$. The QCs are rather lengthy, whose presentation is relegated to Appendix~\ref{appendix:MatrixElements}. Even so, only partial results up to $J=5/2$ are printed in the tables.
To facilitate the use of the QCs,
we supply a supplemental file~\cite{supp} from which all of the QCs derived in this work can be recovered in an easy-to-read and easy-to-use format that also eliminates typos if the QCs are copied. For completeness, all group theory details used to derive the QCs are placed in Appendix \ref{appendix:GroupTheory}.

In the following, we discuss a few aspects of the QCs.
\subsubsection{Kinematics}
We define $\bm P$ as the total momentum of the two-particle system in the lab frame,
\beq
\bm{P} = \bm{p_1} + \bm{p_2}.
\eeq
The momenta in the periodic box are quantized,
\beqs\label{p3}
\bm{P} &= \frac{2\pi}{L}\Tilde{\bm{d}} = \frac{2\pi}{L}(d_x,d_y,d_z/\eta),
\\
\bm{p}_1 &= \frac{2\pi}{L}\Tilde{\bm{n}}_1 = \frac{2\pi}{L}(n_{1x},n_{1y},n_{1z}/\eta),
\\
\bm{p}_2 &= \frac{2\pi}{L}\Tilde{\bm n}_2 = \frac{2\pi}{L}(n_{2x},n_{2y},n_{2z}/\eta).
\eeqs
The tilde indicates the possibility of an elongation of the box in the $z$ direction via the factor $\eta$.
Sometimes, the momenta can be referenced by the triplet of integers, with the above definition in mind.
Rest frame refers to $\bm P=0$ (the two particles have back-to-back momentum in the lab frame). Moving frames refer to non-zero $\bm P$. In the literature, they are sometimes referenced simply by the boost vector $\Tilde{\bm{d}}$, or simply $\bm d$ (suppressing the tilde).  

In the lab frame the total energy is,
\beq
E_{lab} = \frac{p_1^2}{2m_1} + \frac{p_2^2}{2m_2}.
\label{elab}
\eeq
In the CM frame, the total energy is given by
\beq
E_{cm} = \frac{k^2}{2\tilde{m}},
\eeq
where $\tilde{m}$ is the reduced mass and $\bm k$ the relative CM momentum (the two particles are back-to-back in the CM frame by definition).

For moving frames the CM energy is always lower than the lab energy,
\beq
E_{cm} = E_{lab} - \frac{P^2}{2(m_1+m_2)},
\eeq
rendering moving frames a good option to reach smaller CM energies in the QCs.

We can find a relation for $\bm k$ in terms of the lab momenta through a Galilean transformation. Assuming that particle 1 has the same sign as $\bm P$ we have,
\beqs
\bm p_1 = \bm k + \frac{m_1}{m_1 + m_2}\bm P,
\\
\bm p_2 = -\bm k + \frac{m_2}{m_1 + m_2}\bm P.
\eeqs
Solving for $\bm k$ gives,
\beqs
\bm k &= \frac{1}{2}(\bm p_1 - \bm p_2) + \frac{m_2 - m_1}{2(m_1 + m_2)}\bm P,
\\
&= \bm p_1 + \frac{1}{2}\left(\frac{m_2 - m_1}{m_1 + m_2} -1  \right)\bm P.
\eeqs
In terms of the box momenta, it reads:
\beq
\bm k = \frac{2\pi}{L}\big( \tilde{\bm n} - \frac{1}{2}A\tilde{\bm d}\big),
\eeq
where $\bm p_1$ is represented by $\tilde{\bm n} = (n_x,n_y,n_z/\eta)$, and 
\beq
A \equiv 1 + \frac{m_2 - m_1}{m_2 + m_1}.
\label{A}
\eeq
If instead particle 2 has the same sign as $\bm P$ then this amounts to a swapping of $m_1$ and $m_2$ in A. We will show below that this has no practical effect on the quantization condition.

In the moving frame, the CM wavefunction picks up a complex phase under periodic boundary condition \cite{Rummukainen:1995vs,Fu:2011xz, Leskovec:2012gb,Gockeler:2012yj},
\beq
\psi(\bm r + \bm n L) = e^{i\pi A \tilde{\bm n}\cdot\Tilde{\bm d}} \psi(\bm r ).
\eeq
Because the phase factor depends on the particle masses and the boost $\bm d$ it can be difficult to implement in practice. We will show by working in the lab frame one can avoid this condition and make connection to CM frame through kinematics.

In moving frames, the summation grid for the zeta functions in Eq. \eqref{eq:Zeta} needs to be modified to incorporate the boost,
\beq
Z_{lm}(q^2,\Tilde{\bm d}, \eta) = \sum_{\tilde{\bm n}\in P_{\Tilde{\bm d}}(\eta)}\frac{\mathcal{Y}_{lm}(\tilde{\bm n})}{\tilde{\bm n}^2 - q^2}.
\eeq
The associated shorthand retains the same form,
\beq
w_{lm}(q^2, \Tilde{\bm d}, \eta) \equiv \frac{Z_{lm}(q^2, \Tilde{\bm d},\eta)}{\eta \pi^{3/2}q^{l+1}}.
\label{wlmboost}
\eeq
The new summation grid is,
\beq
P_{\Tilde{\bm d}}(\eta) = \left\{ \tilde{\bm n} \in \mathbb{R}^3  | \tilde{\bm n} = \hat{\eta}^{-1}(\bm m - \frac{1}{2}A \bm d), \bm m \in \mathbb{Z}^3 \right\}.
\label{gridboost}
\eeq
The $z$-elongation projector $\hat{\eta}^{-1}$ is defined as  $\hat{\eta}^{-1}\bm b = (b_x,b_y,b_z/\eta)$ for any $\bm b$ vector. The evaluation of these zeta functions is non-trivial but has been discussed in Refs. 
 \cite{Leskovec:2012gb,Gockeler:2012yj,Feng:2004ua,Guo:2016zos}.

We can now see how the QC is affected by unequal masses. The mass dependence is contained in the zeta function. Should we swap the labels of $m_1$ and $m_2$ this will only change the $A$ factor in the summation grid, producing an overall minus sign in the summation index. It leads to an overall sign change in the zeta function which does not change the form of the QC.  However, the order does matter for the total energy of the system, see Eq.\eqref{elab}. 

\subsection{Group symmetry of rest and moving frames}
The symmetry group of the cubic box for rest frames is the octahedral group, $O$ with 24 group elements. Including parity (improper rotations) the group is doubled to $O_h$ with 48 elements. With spin-1/2, it is further doubled to $^2O_h$ with 96 elements. In the $z$-elongated box, the corresponding groups are $D_4$ with 8 elements, $D_{4h}$ with 16 elements, and $^2D_{4h}$ with 32 elements. If moving frames are considered, these symmetry groups are further reduced to the so-called \textit{little groups}. The little group corresponding to a moving frame is the subgroup that preserves the direction of motion. That is, if $R$ is a rotation in the group, then $R\bold{d} = \bold{d}$. 
Therefore, the little group depends on the direction of motion of the two-particle system and the box geometry. For spin-1/2 systems, the double cover little groups are named $^2C_{nv}$ where $n$ denotes that the box has an $n$-fold rotational symmetry about the direction of motion. For example, the little group for a moving frame in the $z$ direction is $^2C_{4v}$, whereas the little group for for a moving frame in the $xy$ plane is $^2C_{2v}$. 
Table \ref{tab:LittleGroups} summarizes the little groups in cubic and elongated boxes. We will consider three distinct little groups in this work: $^2C_{2v}$,$^2C_{3v}$, and $^2C_{4v}$. Note that there is no $^2C_{3v}$ but the possibility of $\bm d=(1,0,0)$ little group $^2C_{2v}$ in the elongated box.
Details of all the double cover groups considered in this work can be found in Appendix~\ref{appendix:GroupTheory}.

\begin{table}[h]
\caption{Double cover little groups for moving frames in cubic and elongated boxes. The lowest distinct patterns for the boost vector $\bm d=(n_x,n_y,n_z)$ are shown, but integer multiples of $\bm d$ belong to the same little groups. Furthermore, all momenta related via a lattice symmetry with the ones below have the same little groups (this means all permutations for cubic, and $n_x$ and $n_y$ permutations for $z$ elongated). 
}
\label{tab:LittleGroups}    
$                                
\renewcommand{\arraystretch}{1.3}
\begin{array}{c | c | c }
\toprule                         
 & \text{cubic  } (^2O_h) & \text{$z$ elongated  }  (^2D_{4h})      \\
\hline                           
 ^2C_{4v} & (0,0,1)  & (0,0,1)  \\
 \hline      
 ^2C_{3v} & (1,1,1)  &    \text{none}  \\
 \hline      
 ^2C_{2v} & (1,1,0)  &  (1,0,0), (1,1,0)     \\
\bottomrule                      
\end{array}                      
$                                   
\end{table}       

When moving from the infinite rotational symmetries of the infinite volume to the reduced symmetries of the box one ends up with a many-to-one matching from $J$ to the box-irreps. This means that each QC only couples to a subset of $J$ values in an infinite tower. This is in full display in the summary Table~\ref{tab:ChiSquareSummary}. The lowest partial wave phase shift can be predicted using the energy levels in the box if the higher partial waves can be neglected. This is known as the {\em L\"uscher formula}. As the symmetry is reduced the mixing of partial waves increases. For example, in the rest frame the gap between the lowest two $J$ in the $G_{1g}$ irrep in the cubic box is 3 ($J=\frac{1}{2}$ and $J=\frac{7}{2}$), whereas in the elongated box it is 1 ($J=\frac{1}{2}$ and $J=\frac{3}{2}$). For moving frames, the partial waves mix (for fixed J, the irrep couples to $J = l + s$ \textit{and} $J = l - s$), causing this gap to always be 1.

\subsubsection{Simplification of QCs}
There are symmetries in the $w_{lm}$ functions that can be used to simplify the matrix elements. They arise from how the spherical harmonics transform under the group operations.
\begin{enumerate}[(i)]
    \item 
    The standard property $Y_{l-m} = (-1)^mY^*_{lm}$ translates directly to $w_{l-m} = (-1)^mw^*_{lm}$.\label{enum:i}
\item
When a system is invariant under a mirror reflection about the $xy$ plane, it leads to $Y_{lm}(\theta,\phi)=Y_{lm}(\pi-\theta,\phi)=(-1)^{(l-m)} Y_{lm}(\theta,\phi)$. Which means 
$w_{lm}=0 \text{  for  } l-m = \text{odd.  In particular  } w_{l0}=0  \text{  for  } l = 1,3,5,\cdots.  $
This is valid for all systems with parity symmetry, which leads 
to a separation into sectors by parity. Note that in moving frames, parity is no longer a symmetry.\label{enum:ii}
\item 
Due to the $e^{im\phi}$ dependence in $Y_{lm}$, a system that is invariant under a $\pi/2$ rotation about the $z$ axis has the constraint that $e^{im\pi/2} = 1$. This means that $w_{lm} = 0$ for $m \neq 0,4,8,\cdots,$ regardless of $l$.
\label{enum:iii}
\item 
When the system is invariant under a mirror reflection about the $xz$ plane then
$Y_{lm}(\theta,\phi)=Y_{lm}(\theta,2\pi-\phi)= Y^*_{lm}(\theta,\phi)$.
This means all the $w_{lm}$ functions are real valued. However, the matrix elements $\mathcal{M}$ can have complex valued coefficients depending on basis vectors.\label{enum:iv}
\item 
Invariance under a swapping of $x \leftrightarrow y$ gives the constraint that $w_{lm} = e^{i\frac{\pi}{2}m}w^*_{lm}$. This constrains the relations between the real and imaginary parts of $w_{lm}$ depending on $m \mod 4$. \label{enum:v}
\item 
Similar to constraint \ref{enum:v}, invariance under a mapping of $x \rightarrow -x$ leads to the constraint that $w_{lm} = e^{i\pi m}w^*_{lm}$. This causes the $w_{lm}$ to either be real for even $m$ or purely imaginary for odd $m$. \label{enum:vi}
\end{enumerate}
These properties greatly simplify the presentation of the matrix elements. See Table \ref{tab:nonzeroZeta} for a summary of all the constraints in each frame. Moreover, the $\mathcal{M}$ matrix is Hermitian, so we only need to list the lower triangular part of the matrix.

An important consideration in the use of these matrix elements is to note the conventions and notations used. There are multiple conventions in use in the literature for basis vectors, the definition of the $w$ functions, etc. But since the QCs are invariant under a change of basis, the physical content is the same. Numerically, they all produce the same determinant.

In moving frames, condition \ref{enum:ii} no longer applies. For any moving frame other than $\bm d = (0,0,1)$, condition \ref{enum:iii} fails; and for moving frames with any momentum in the $y$ direction, condition \ref{enum:iv} fails as well. This greatly reduces the simplification of the matrix elements. However, we can find a few more simplifications through the closure relation of the zeta functions \cite{Luscher:1990ux},
\beq\label{eq:closure}
\sum_{m' = l}^l D_{m'm}^{(l)}(\alpha,\beta,\gamma) Z_{lm'} = Z_{lm},
\eeq
where $D$ is the Wigner rotation matrix for a transformation in the little group. Because this relation stems from the spherical harmonics we can apply it to all box geometries and boost directions. Using this relation for each group element gives a constraint though each constraint is not independent. A table detailing these constraints for various boost directions is given in Table \ref{tab:closurerelations}. The closure relation only provides new relations for $\bm d = (0,0,0)$ and $\bm d = (1,1,1)$ in the cubic box. All other simplifications from the closure relation are equivalent to previous simplifications.

\begin{table}[]
    \caption{Simplifications of the w$_{lm}$ in various frames for both cubic and z-elongated boxes due to symmetries of the spherical harmonics. We use the notation w$_{lm} \equiv$ wr$_{lm} + I$wi$_{lm}$. In the rightmost column we indicate which listed rule is responsible for the simplification. Rule \ref{enum:i} applies in all cases, as such, we only look at these rules in relation to $m \geq 0$.}
    \centering
    \begin{tabular}{c|c|c}
        $d$ &  w$_{lm}$ Simplifications & Rule\\
        \hline
         \{0,0,0\} &  w$_{lm} = $ wr$_{lm}$ & \ref{enum:iv}\\ 
         & w$_{lm} \neq 0$ for $l$ even and $m \equiv 0 \mod{4}$ & \ref{enum:ii} \& \ref{enum:iii}\\
         \hline
         \{0,0,1\} & w$_{lm} = $ wr$_{lm}$ & \ref{enum:iv}\\
         & w$_{lm} \neq 0$ for $m \equiv 0 \mod{4}$ & \ref{enum:iii}\\
         \hline
         \{1,1,0\} & wr$_{lm} = 0$ for $l-m$ odd & \ref{enum:ii}\\
         & wi$_{lm} = 0$  for $m \equiv 0 \mod{4}$ & \ref{enum:v}\\
         & wi$_{lm} = $ wr$_{lm}$ for $m \equiv 1 \mod{4}$ &\ref{enum:v} \\
         & wr$_{lm} = 0$  for $m \equiv 2 \mod{4}$ & \ref{enum:v}\\
         & wi$_{lm} = $ -wr$_{lm}$ for and $m \equiv 3 \mod{4}$ & \ref{enum:v}\\
         \hline
         \{1,1,1\}  & wi$_{lm} = 0$  for $m \equiv 0 \mod{4}$ & \ref{enum:v}\\
         & wi$_{lm} = $ wr$_{lm}$ for $m \equiv 1 \mod{4}$ & \ref{enum:v}\\
         & wr$_{lm} = 0$  for $m \equiv 2 \mod{4}$ & \ref{enum:v}\\
         & wi$_{lm} = $ -wr$_{lm}$ for and $m \equiv 3 \mod{4}$ & \ref{enum:v}\\
         \hline
    \end{tabular}
    \label{tab:nonzeroZeta}
\end{table}

\begin{table}
\caption{Additional relations derived from the closure condition in Eq.\eqref{eq:closure} 
for  rest frame $d= \{0,0,0\} $ and  moving frame $d= \{1,1,1\} $ in cubic box. They are used to further simply the matrix elements in the two cases.}
\label{tab:closurerelations}          
              
$                                
\renewcommand{\arraystretch}{1.4}
\begin{array}{cc}
\toprule                         
 & d= \{0,0,0\}, \text{ group } ^2O_h   \\
\hline           
 \text{} & \text{w}_{44}\to \sqrt{\frac{5}{14}} \text{w}_{40} \\
 \text{} & \text{w}_{64}\to -\sqrt{\frac{7}{2}} \text{w}_{60} \\
 \text{} & \text{w}_{84}\to \frac{1}{3} \sqrt{\frac{14}{11}} \text{w}_{80} \\
 \text{} & \text{w}_{88}\to \frac{1}{3} \sqrt{\frac{65}{22}} \text{w}_{80} \\
 \text{} & \text{w}_{104}\to -\sqrt{\frac{66}{65}} \text{w}_{100} \\
 \text{} & \text{w}_{108}\to -\sqrt{\frac{187}{130}} \text{w}_{100} \\
\hline     
     &  d= \{1,1,1\}, \text{ group } ^2C_{3v}   \\      
  \hline                  
 \text{} & \text{wr}_{11}\to -\frac{\text{w}_{10}}{\sqrt{2}} \\
 \text{} & \text{wi}_{22}\to -\text{wr}_{21} \\
 \text{} & \text{wr}_{31}\to \frac{\sqrt{3} \text{w}_{30}}{4} \\
 \text{} & \text{wr}_{33}\to -\frac{1}{4} \sqrt{5} \text{w}_{30} \\
 \text{} & \text{wi}_{42}\to 2 \sqrt{2} \text{wr}_{41} \\
 \text{} & \text{wr}_{43}\to \sqrt{7} \text{wr}_{41} \\
 \text{} & \text{w}_{44}\to \sqrt{\frac{5}{14}} \text{w}_{40} \\
 \text{} & \text{wr}_{53}\to \sqrt{\frac{5}{7}} \text{w}_{50}+3 \sqrt{\frac{3}{14}} \text{wr}_{51} \\
 \text{} & \text{w}_{54}\to -\sqrt{\frac{5}{14}} \text{w}_{50}-\frac{8 \text{wr}_{51}}{\sqrt{21}} \\
 \text{} & \text{wr}_{55}\to \sqrt{\frac{5}{42}} \text{wr}_{51}-\frac{\text{w}_{50}}{\sqrt{7}} \\
 \text{} & \text{wi}_{62}\to \frac{1}{4} \left(-\sqrt{10} \text{wr}_{61}-6 \text{wr}_{63}\right) \\
 \text{} & \text{wi}_{66}\to \frac{1}{44} \left(2 \sqrt{55} \text{wr}_{63}-9 \sqrt{22} \text{wr}_{61}\right) \\
 \text{} & \text{w}_{64}\to -\sqrt{\frac{7}{2}} \text{w}_{60} \\
 \text{} & \text{wr}_{65}\to \sqrt{\frac{6}{11}} \text{wr}_{61}+\sqrt{\frac{15}{11}} \text{wr}_{63} \\
 \text{} & \text{wi}_{76}\to \frac{363 \text{wi}_{72}-32 i \left(\sqrt{21} \text{w}_{70}-\sqrt{6} \text{wr}_{71}+9
   \sqrt{2} \text{wr}_{73}\right)}{33 \sqrt{143}} \\
 \text{} & \text{w}_{74}\to -\frac{25 \sqrt{42} \text{w}_{70}-248 \sqrt{3} \text{wr}_{71}+120 \text{wr}_{73}}{66
   \sqrt{11}} \\
 \text{} & \text{wr}_{75}\to \frac{14 \sqrt{42} \text{w}_{70}-94 \sqrt{3} \text{wr}_{71}+21 \text{wr}_{73}}{33
   \sqrt{11}} \\
 \text{} & \text{wr}_{77}\to -\frac{1}{11} \sqrt{\frac{13}{11}} \left(2 \sqrt{6} \text{w}_{70}+\sqrt{21}
   \text{wr}_{71}+2 \sqrt{7} \text{wr}_{73}\right) \\
 \text{} & \text{wi}_{82}\to \frac{1}{2} \left(\sqrt{66} \text{wr}_{83}-\sqrt{70} \text{wr}_{81}\right) \\
 \text{} & \text{wi}_{86}\to 3 \sqrt{\frac{33}{26}} \text{wr}_{81}-\sqrt{\frac{35}{26}} \text{wr}_{83} \\
 \text{} & \text{w}_{84}\to \frac{1}{3} \sqrt{\frac{14}{11}} \text{w}_{80} \\
 \text{} & \text{wr}_{85}\to 3 \sqrt{\frac{15}{13}} \text{wr}_{83}-2 \sqrt{\frac{77}{13}} \text{wr}_{81} \\
 \text{} & \text{wr}_{87}\to 2 \sqrt{\frac{21}{13}} \text{wr}_{83}-\sqrt{\frac{55}{13}} \text{wr}_{81} \\
 \text{} & \text{w}_{88}\to \frac{1}{3} \sqrt{\frac{65}{22}} \text{w}_{80} \\
\bottomrule                      
\end{array}                      
$                                                 
\end{table}           

\subsection{From non-relativistic to relativistic QCs}
Although the QCs are derived in a non-relativistic framework, they can be transformed into their relativistic versions with only a few small modifications. 

First,  relativistic kinematics must replace the non-relativistic one.
\beq
E_{lab} = \sqrt{\bm p_1^2+m_1^2} + \sqrt{\bm p_2^2+m_2^2},
\eeq
\beq
E_{cm} = \sqrt{\bm k^2+m_1^2} + \sqrt{\bm k^2+m_2^2}.
\eeq
The two are related by, 
\beq
E_{lab} = \sqrt{E_{cm}^2 + \bm P^2}.
\eeq
Boost velocity and relativistic factor are given by, 
\beq
\bm v={\bm P\over E_{lab}}, \quad \gamma={1\over \sqrt{1-v^2}}={E_{lab}\over E_{cm}}.
\eeq
The kinematical sequence in a calculation typically goes like this: for a given boost $\bm P$, first $E_{lab}$ is determined in the box (Schr\"odinger equation, lattice QCD, etc), then $\gamma$ factor, then $E_{cm}$, then CM $k$, then zeta function via dimensionless $q$.

Second, the summation grid in Eq.\eqref{gridboost} of the zeta function is modified by, 
\beq
P_{\bm d}(\eta) = \left\{ \tilde{\bm n} \in \mathbb{R}^3  | \tilde{\bm n} = \hat{\gamma}^{-1}\hat{\eta}^{-1}(\bm m - \frac{1}{2}A \bm d), \bm m \in \mathbb{Z}^3 \right\}.
\eeq
where the $\hat{\gamma}^{-1}$ projection is applied in the direction of the boost $\bm d$ and the factor in Eq.\eqref{A} is replaced by,
\beq
A=1+{m_1^2-m_2^2 \over E_{cm}^2}.
\eeq
Third, the $w_{lm}$ in Eq.\eqref{wlmboost} picks up a $\gamma$ factor (see e.g. Ref~\cite{Gockeler:2012yj}),
\beq
w_{lm}(q^2, \bm d, \eta) \equiv \frac{Z_{lm}(q^2, \bm d, \eta)}{\gamma \eta \pi^{3/2}q^{l+1}}.
\label{eq:wlmrel}
\eeq

With these changes, all the QCs in this work can be used in relativistic studies. 

\section{Two-particle energies in the box}\label{section:two-part energies}
\label{sec:scheq}
L\"uscher’s method establishes a relationship between the phase shifts in infinite volume and the energy spectrum of two-particle states in finite volume with periodic boundary conditions. Consequently, a comparison of the energy spectrum predicted by the quantization conditions with the spectrum obtained from an independent lattice Hamiltonian calculation provides a robust check of the newly derived QCs. In this section, we detail how to obtain the energy spectrum from solving the Schr\"odinger equation for a test potential. 
\subsection{Reduction of the lattice Hamiltonian} 

We want to obtain the energy spectrum in the continuum but in finite volume.
Consider the general case of a box with dimensions $L\times L\times \eta L$ where $\eta$ is the elongation factor in the z-direction.
The Schr\"{o}dinger equation $H\Psi=E\Psi$ in the box frame (lab frame) with periodic boundary conditions has the Hamiltonian,
\beq
H=-\frac{\hbar^2 }{ 2m_1} \nabla_1^2  - \frac{\hbar^2 }{ 2m_2} \nabla_2^2 + V_L(|\bm r_1 - \bm r_2|).
\label{eq:latH}
\eeq
Here $V_L$ is periodic version of the infinite-volume potential $V$,
\beq
V_L(|\bm r_1 - \bm r_2|) = \sum_{\bm n_1, \bm n_2} V(| \bm r_1 +\bm n_1 L- \bm r_2 -\bm n_2 L |).
\label{eq:pot-box}
\eeq
Visually, the continuous space gets tiled into an infinite number of $\eta L^3$ boxes in which the potential is replicated. 
We solve the problem in a single box with periodic boundary conditions.
Under this scenario, the potential is no longer rotationally symmetric. Instead, it takes on the symmetry of the box.
For a system with spin, the potential has a spin-orbit interaction, in addition to the central part,
\beq
V({\bm r,\bm p}) = V_c(r) + V_{so}(r) \bm l\cdot \bm s,
\label{eq:pot}
\eeq
where $\bm s= \bm \sigma{\hbar\over 2}$ is the spin-1/2 operator in terms of Pauli matrices. This potential is implicitly dependent on the vectors $\bm r$ and $\bm p$ via the $\bm l$ operator.
In spin space and Cartesian components it reads,
\beqs
&V({\bm r,\bm p})  = V_c(r) \left( \begin{array}{cc} 1   & 0   \\0  & 1 \\\end{array} \right) 
+ V_{so}(r)
\left( \begin{array}{cc}
  l_z   &  l_x-i l_y   \\
l_x+ i l_y  & - l_z \\
\end{array} \right) {\hbar\over 2} \\
& \text{with } l_x=yp_z-zp_y, l_y=zp_x-xp_z,  l_z=xp_y-yp_x.
\eeqs
Note that the potential is no longer local due to the presence of spin-orbit coupling. 

The orbital angular momentum relative to CM is defined by,
\beq
\bm l=\bm r_1\times \bm p_1 + \bm r_2\times \bm p_2 - \bm R\times \bm P,
\label{eq:L6}
\eeq
where $\bm R=(m_1\bm r_1 +m_2 \bm r_2)/(m_1+m_2)$ is the CM position and $\bm P=\bm p_1 + \bm p_2$ the total momentum of the system.
$l$ can be expressed in relative variables as,
\beq
\bm l=(\bm r_1 - \bm r_2)\times {m_2\bm p_1 - m_1 \bm p_2\over m_1+m_2}\equiv \bm r\times \bm p,
\label{eq:L3}
\eeq
where $\bm r$ is the relative position and 
$\bm p=\mu (\bm v_1-\bm v_2)$ the reduced momentum between the two particles. 
The momentum operator is $\bm p = -i\hbar \nabla$.

Implementing periodic boundary conditions requires the wave functions to satisfy,
\beq 
\Psi(\bm r_1 +\bm n_1 L, \bm r_2 +\bm n_2 L,m_s) =\Psi(\bm r_1, \bm r_2,m_s),
\label{eq:bc}
\eeq
where $\bm n_1=(n_{1x}, n_{1y}, n_{1z} \eta)$ and $\bm n_2=(n_{2x}, n_{2y}, n_{2z} \eta)$. Note that the $\bm n$ index appearing in periodic boundary conditions has $\eta$ multiplied in the $z$-component, while it is divided in the context of discrete momenta, such as Eq.\eqref{p3}.
The wavefunction carries an additional quantum number $m_{s}=\pm 1/2$ for the spin-1/2 particle. 

We solve Eq.\eqref{eq:latH} numerically by discretizing the box into a lattice of $N_x \times N_y \times N_z$ grid points with isotropic lattice spacing $a$. Both cubic ($N_x = N_y = N_z$) and z-elongated lattices ($N_x = N_y = N$ and $N_z = \eta N$) are considered. However, this `brute force' numerical approach presents a significant challenge: The Hilbert space for the two-particle state grows as $N^6$, making the problem computationally intractable for all but the smallest lattices. This rapid growth arises because the Hamiltonian, as formulated, includes all possible momentum sectors simultaneously (or all possible rest and moving frames). Since our interest lies in probing only a handful of specific momentum sectors, this comprehensive formulation becomes an unnecessary computational hurdle. To address this challenge and to focus on individual momentum sectors more efficiently, we switch to a basis consisting of total momentum $\bold{P}$ and relative coordinates $\bold{r}$ in the lab frame,
\beq 
\ket{\bm P,\bm r,m_s}=\sum_{\bm m} e^{i \bm P \cdot \bm m} \ket{\bm m, \bm m+ \bm r,m_s},
\label{eq:Pr}
\eeq
where the $\ket{\bm n_1, \bm n_2,m_s}$ notation is the ket in the position representation for two particles. The change of basis is realized by projecting to $\bm P$ by summing over all positions separated by $\bm r$. In this decomposition, the `6D basis' $\ket{\bm n_1, \bm n_2,m_s}$ of all momenta is reduced to a `3D basis' $\ket{\bm P,\bm r, m_s}$ of fixed momentum. The Hilbert space  grows with $N^3$ In the new basis, giving a significant improvement in scalability and making studies on larger lattices more feasible.

To discretize the lattice Hamiltonian we approximate the gradient operator by a two-point symmetric stencil,
\beq
\nabla\psi(\bold{r\mu}) = \sum^3_{k = 1} \frac{\psi(\bold{r} + a \hat{k}) - \psi(\bold{r} - a\hat{k})}{2a} + \mathcal{O}(a^2).
\eeq
Similarly, using a centered three-point stencil, the Laplacian operator is,
\beq
\nabla^2 \psi(\bm r) = \sum_{k=1}^3 {\psi(\bm r+a\hat k)-2\psi(\bm r)+\psi(\bm r-a\hat k)\over a^2} + \mathcal{O}(a^2). 
\eeq
\begin{widetext} 
The change of basis leads to the reduced problem $H\psi(\bm P,\bm r,m_s)=E\psi(\bm P,\bm r,m_s)$ where the lattice Hamiltonian is given by,
\beqs
& H\ket{\bm P, \bm r,m_s} =\\& -\frac{\hbar^2}{ 2} \sum_{k=1}^{3} \frac{1}{a^2}   
\left[ 
\left( \frac{e^{i P_k a} }{ m_1} + \frac{1}{m_2} \right) \ket{\bm P,\bm r + a\hat{k},m_s }
+\left( \frac{e^{-iP_k a}}{m_1} + \frac{1}{m_2} \right) \ket{\bm P ,\bm r -a\hat{k}, m_s}
- 2\left( \frac{1}{ m_1} + \frac{1}{ m_2} \right) \ket{\bm P,\bm r,m_s} 
\right] \\
& + V_c\,\ket{\bm P,\bm r, m_s} + V_{so}\;\frac{-i\hbar }{4a(m_1 + m_2)}\epsilon_{ijk} \bold{\sigma}_i\bold{r}_j
\left[(m_2e^{-iaP_k} + m_1)\ket{\bold{P},\bold{r}-a\hat{k}} - (m_2e^{iaP_k} + m_1)\ket{\bold{P},\bold{r}+a\hat{k}, m_s}\right].
\label{eq:latH3}
\eeqs
The discretization error of this Hamiltonian comes in at  ${\cal O}(a^2)$.
For a system at rest ($\bm P=0$), it coincides with the familiar form in the CM frame for relative motion,
\beq
H\ket{\bm 0,\bm r,m_s} =-\frac{\hbar^2}{ 2\mu} \sum_{k=1}^{3} \frac{1}{ 2 a^2}  \bigg(  \ket{\bm 0, \bm r+a\hat{k},m_s} + \ket{\bm 0, \bm r-a\hat,m_s} -2 \ket{\bm 0,\bm r,m_s} \bigg) + V(\bm r,\bm p)\ket{\bm 0,\bm r,m_s} \,,
\label{eq:latH3cm}
\eeq
where $\mu=m_1m_2/(m_1+m_2)$ is the reduced mass and $V$ is the potential  in Eq.\eqref{eq:pot}.
We emphasize that the Hamiltonian in Eq.\eqref{eq:latH3} is more general than Eq.\eqref{eq:latH3cm}: it stays in the lab frame so the standard  periodic boundary condition 
$ 
\ket{\bm P, \bm r +\bm n L,m_s}=\ket{\bm P, \bm r,m_s}
$
applies in the reduced basis, and moving frames are automatically incorporated with $\bm P$ as an input in the Hamiltonian.

The low-lying spectrum of the reduced Hamiltonian can be obtained quickly for lattices up to $24^3$ on a laptop computer.
To accelerate the convergence to the continuum we use an improved version with a centered six-point stencil for the gradient and a centered seven-point stencil for the Laplacian,
\beqs
& H\ket{\bm P, \bm r, m_s} = \\
&-\frac{\hbar^2}{ 2} \sum_{k=1}^{3} \frac{-1}{ 180 a^2}  \Big[
-2\left( \frac{e^{3iP_k a} }{ m_1} + \frac{1}{ m_2} \right) \ket{\bm P,\bm r+3a\hat{k}, m_s}
+27\left( \frac{e^{2iP_k a} }{ m_1} + \frac{1}{ m_2} \right) \ket{\bm P, \bm r+2a\hat{k}, m_s}
 \\ & 
-270\left( \frac{e^{iP_k a} }{ m_1} + \frac{1}{ m_2} \right) \ket{\bm P,\bm r+a\hat{k}, m_s}
 -2\left( \frac{e^{-3iP_k a} }{ m_1} + \frac{1}{ m_2} \right) \ket{\bm P,\bm r-3a\hat{k}, m_s}
+27\left( \frac{e^{-2iP_k a} }{ m_1} + \frac{1}{ m_2} \right) \ket{\bm P, \bm r-2a\hat{k}, m_s}
 \\ &
-270\left( \frac{e^{-iP_k a} }{ m_1} + \frac{1}{ m_2} \right) \ket{\bm P,\bm r-a\hat{k}, m_s}
+490\left( \frac{1}{ m_1} + \frac{1}{ m_2} \right) 
\ket{\bm P,\bm r, m_s}
\Big]  \\ 
& + V_c\ket{\bold{P},\bold{r}, m_s} +V_{so}\;\frac{-i\hbar }{60a}\epsilon_{ijk}\bold{\sigma}_i\bold{r}_j
\Big[(m_2e^{-3iP_ka}+m_1)\ket{\bold{P},\bold{r}-3a\hat{k}, m_s}
-9(m_2e^{-2iP_ka} + m_1)\ket{\bold{P},\bold{r}-2a\hat{k}, m_s}\\
&+45(m_2e^{-iP_ka} + m_1)\ket{\bold{P},\bold{r}-a\hat{k}, m_s}
-(m_2e^{3iP_ka}+m_1)\ket{\bold{P},\bold{r}+3a\hat{k}, m_s}
+9(m_2e^{2iP_ka} + m_1)\ket{\bold{P},\bold{r}+2a\hat{k}, m_s}\\
&-45(m_2e^{iP_ka} + m_1)\ket{\bold{P},\bold{r}+a\hat{k}, m_s}
\Big].
\label{eq:latH7}
\eeqs
The discretization error of this Hamiltonian comes in at  ${\cal O}(a^6)$.
\end{widetext} 

Having obtained the lattice Hamiltonian in the reduced $\ket{\bm P,\bm r, m_s}$ basis, we need to take the continuum limit to obtain box levels from lattice levels. This is done by increasing the number of grid points and decreasing the lattice spacing simultaneously while keeping the box size fixed, 
\beq
\lim_{\substack{a\to 0\\ N\to \infty }} N a = L.
\eeq 
Since the discretization error is known to behave as $\mathcal{O}(a^6)$, we perform a linear extrapolation 
in $a^6$ using three lattices.

We will employ two potentials in this work. One is to check our energy determination method and another to check the quantization conditions later. 
The first is
\beq\label{pot1}
 V(\bm r_1, \bm r_2) = (-V_0 + V_1|\bold{r}_1 - \bold{r}_2|^4 + c_{\ell s}\boldsymbol{\ell} \cdot \bold{s})e^{-\beta|\bold{r}_1 - \bold{r}_2|^2},
\eeq
where applies to both 6D and 3D cases if we recall $\bm r=\bm r_1-\bm r_2$.
The parameters are set to $V_0 = 4 \GeV, V_1 = {1\over16} \fm^{-4}, c_{\ell s} = 1 \GeV^{-1}\fm ^{-2}, \beta = {1\over 8} \fm^{-2}, m_1 = m_2 = 2 \GeV$. 
This potential is chosen because it is studied in detail in a previous work~\cite{Lee:2020fbo} using an independent, Fourier basis method.
The potential admits a wide variety of bound and scattering states across partial waves and energies suitable for testing purposes.
In a box with periodic boundary conditions, the relative coordinate $\bold{r}$ must be replaced by the displacement corresponding to the shortest distance between the periodic images of the two particles. The adjustment arises from the situation where one particle is closer to an image rather than the other particle. This results in the shortest path between them corresponding to going through the boundary rather than across the box. This can be accounted for with the definition,
\beq
\bold{r}_i = \begin{cases}
\bold{r}_{1i} - \bold{r}_{2i} - \text{sign}(\bold{r}_{1i} - \bold{r}_{2i})L  & \text{if}\ |\bold{r}_{1i} - \bold{r}_{2i}| \geq \frac{L}{2}
\\
\bold{r}_{1i} - \bold{r}_{2i} & \text{if}\ |\bold{r}_{1i} - \bold{r}_{2i}| < \frac{L}{2}.
\end{cases}
\eeq

We perform two checks to verify the correctness of our implementation of the $|\bm P,\bm r,m_s\rangle$ basis. The first test ensures that this reduced basis accurately reproduces the results in the original basis as defined in Eq.\eqref{eq:latH}. To achieve this, we solve a cutoff version of the 6D Hamiltonian on a $4^3$ lattice. The modification involves imposing a hard cutoff on the potential for $|\bm r_1 - \bm r_2|$ greater than two lattice spacings. This cutoff prevents interactions between the particles and their periodic images. 
With this modified 6D Hamiltonian we solve for all 8192 eigenvalues, which encompass all 64 discrete total momentum sectors of the system,
\beqs
 \bm P &=\frac{2\pi}{ a} \left( \frac{i }{ N_x}, \frac{j }{ N_y}, \frac{k }{ N_z } \right) \text{ where } \\
 & i=0,\cdots, N_x-1, \\ 
 & j=0,\cdots, N_y-1,  \\
 &k=0,\cdots, N_z-1. 
\label{eq:P}
\eeqs
Separately, we compute the 128 eigenvalues from the projected 3D Hamiltonian in Eq.\eqref{eq:latH3} on the same lattice and potential, using each of the 64 $\bm P$'s in Eq.\eqref{eq:P} as an input. 
Perfect agreement is achieved between the 6D and 3D results, for both the three-point stencil in Eq.\eqref{eq:latH3} and seven-point stencil in Eq.\eqref{eq:latH7} versions of the Hamiltonian. The same check is carried out for $z$-elongated lattices ($N_x=N_y\neq N_z$).\

The second check is to confirm that our implementation produces the correct energy spectrum in the continuum. We solve Eq.\eqref{eq:latH7} without a potential cutoff on three lattices of $32^3$ with $a = 1.125$ fm, $40^3$ with $a = 0.9$ fm, and $48^3$ with $a = 0.75$ fm. After continuum extraction we compare our results with those found in \cite{Lee:2020fbo} which uses the same potential and parameters. We find perfect agreement.

\subsection{Symmetry-adapted energy levels in the box}

In order to match the energy levels to that of the symmetry-adapted QCs in Eq.\eqref{Eq:QC1}, we project the eigenvectors of the lattice Hamiltonian to the irreps of the same symmetry group. To do this we construct projectors $P^{\Gamma,\lambda}$ according to, 
\beq
    P^{\Gamma,\lambda}\ket{\bm P, \bm r, m_s} = \frac{d_\Gamma}{g}\sum_k [D^{\Gamma}_{\lambda,\lambda}(S_k)]^* \ket{S_k \bm P, S_k \bm r, m_s},
\eeq
where $\Gamma$ is the irrep, $\lambda$ the representation row, $d_\Gamma$ the dimension of the irrep, $g$ the total number of elements in the symmetry group, $S_k$ the symmetry operation, and $D^{\Gamma}_{\lambda,\lambda}(S_k)$ its representation matrix. We say that if the norm of a projected eigenvector, $P^{\Gamma}\ket{\bm P, \bm r, m_s}$, is nonzero then the eigenvector and its corresponding eigenvalue belong to the $\lambda$ row of the irrep $\Gamma$ (we assume no accidental degeneracies).
An equivalent way to classify the energy levels is to project the Hamiltonian matrix into an irrep first, then diagonalizes it to obtain its eigenvalues~\cite{Lee_2022}.

\section{Convergence checks} \label{section:numerical checks}
\label{sec:validation}
In this section, we carry out a numerical check of the QCs derived in Sec.~\ref{section:QuantizationConditions}. 
A QC is a function of finite-volume energy and infinite-volume phase shift.
Our strategy is to obtain the energy spectrum and phase shifts independently for the same interaction potential, then feed the phase shift into the QC to predict the energy spectrum, and compare it with the independently computed spectrum. A QC is considered converged and checked if the two spectra agree with each other. This is done by including more and more partial waves in the QC.

The standard L\"uscher formula utilizes the energy levels in the box determined from a certain method, but only retains the lowest partial-wave in the QC in order to make a prediction for its phase shift. By considering higher order partial waves, one can assess how reliable the prediction is for each QC in this approach.

The energy spectrum in the box has been obtained to high precision by solving the Schr\"odinger equation in Sec.~\ref{sec:scheq}. 
The infinite volume phase shifts are computed in the following.

\subsection{Infinite volume phase shifts}
In infinite volume, the Sch\"{o}dinger equation in Eq.(\ref{eq:scheq}) in CM frame  with a central potential $V(r)$ admits solutions with spherical symmetry. 
The solution in spherical coordinates can be expanded in partial waves 
$\psi(\bm r)=\sum_l R_l(r) Y_{lm} (\theta,\phi)$ where $Y_{lm} $ is spherical harmonics.
The radial wavefunction $R_l(r)$ can be found by solving the standard radial equation,
\beq
\left[ - { \hbar^2 \over  2\mu }  {d^2\over dr^2}
+  {l(l+1) \hbar^2 \over 2 \mu r^2}  + V(r)  \right ] u_l(r)=E u_l(r),
\label{eq:radial}
\eeq
where $u_l(r)=r R_l(r)$ is the auxiliary radial wavefunction.
Depending on the potential, the solution can be bound states with quantized energies, or scattering states with continuous energies. 
For scattering states the boundary conditions are
\beqs 
\lim_{r\to0} u_l(r) &= 0, \\
\lim_{r\to\infty} u_l(r) & \propto  e^{i\delta_{l}} \sin \left[ kr -{l\pi\over 2}+\delta_{l} \right ],
\label{eq:ubc}
\eeqs
where $k=\sqrt{2\mu E/ \hbar^2 }$ is the relative back-to-back CM momentum and  $\delta_{l}$ is the phase shift for partial-wave $l$. 
The $-l\pi/2$ term is to account for the repulsive centrifugal barrier term ${l(l+1) \hbar^2 \over 2 \mu r^2} $ and is inserted to ensure that the phase shift in this definition vanishes when the potential itself vanishes.
Conventionally, to obtain the phase shift it is necessary to solve a second-order differential equation from the origin to 
the asymptotic region, then match the solution with an appropriate sine function 
(the interior-exterior match procedure discussed in Sec.~\ref{section:QuantizationConditions}).
Here we resort to the {\it variable phase method}~\cite{Calogero1967} which requires solving only a first-order differential equation,
\beqs
&{d\tilde{\delta}_l(k,r) \over dr} = \\
&- {U(r) \over k} \left [\cos\tilde{\delta}_l(k,r) \hat{j}_l(kr) - \sin\tilde{\delta}_l(k,r)\hat{n}_l(kr) \right]^2,
\label{eq:vpm}
\eeqs
where $U(r)= {2\mu \over \hbar^2 } V(r)$, and $\hat{j}_l(kr)\equiv (kr) j_l(kr)$ and $\hat{n}_l(kr)\equiv (kr) n_l(kr)$ are modified spherical Bessel functions.
The  phase function $\tilde{\delta}_l(k,r)$ is integrated from the origin where $\tilde{\delta}_l(k,0)=0$ to the asymptotic region where the potential is negligible; then the scattering phase shift is obtained directly as the asymptotic value $\delta_l(k)=\lim_{r\to\infty} \tilde{\delta}_l(k,r)$.  

For checking the QCs in this section, we consider a second, simpler, and repulsive potential in the form,
\beq
    V(\bm r) = (V_0 + c_{\ell s}\bm \ell \cdot \bm s)e^{-\beta \bm r^2}.
    \label{eq:potso}
\eeq
The wavefunction is the eigenstate of $\{{\bm J}^2,{J_z}, {\bm \ell}^2, {\bm s}^2\}$
which we label as $|JM\ell\rangle$ with $s=1/2$ suppressed for simplicity. 
In spherical coordinates, it has the form $\psi(\bm r)=R_\ell(r) {\cal Y}_{JM\ell} (\theta,\phi)$ where 
${\cal Y}_{JM\ell} $ is the two-component spinor spherical harmonics.
\begin{figure*}[!t]
\centering
    \includegraphics[scale=0.51]{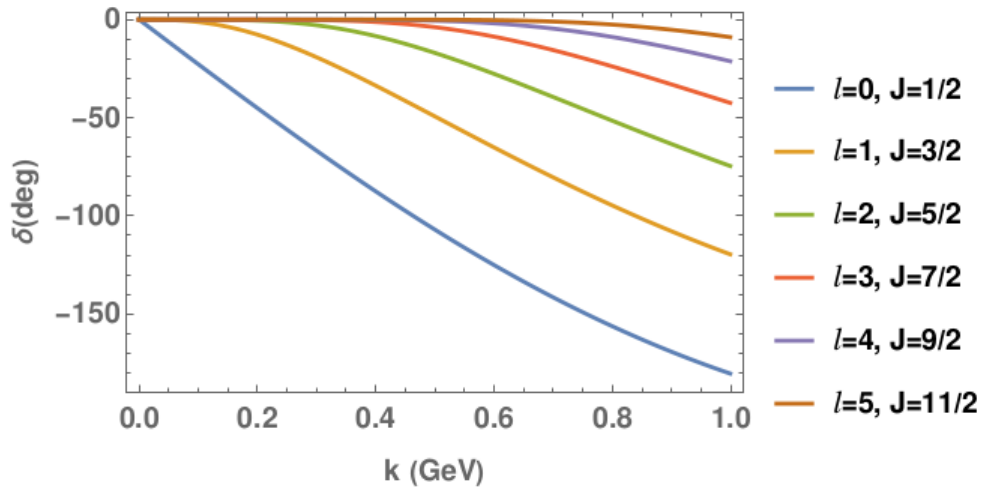}
    \includegraphics[scale=0.51]{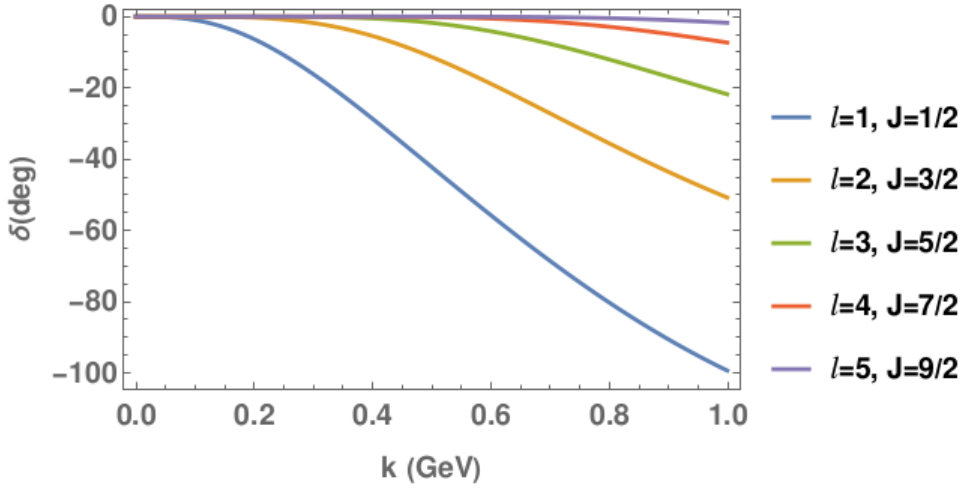}
\caption{Phase shifts as a function of CM $k$ from the test potential for partial waves up to $l = 5$ for spin-up branch in Eq.\eqref{eq:pot2} (top) and spin-down branch in Eq.\eqref{eq:pot3}  (bottom) states. The s-wave ($l = 0$) resides in the spin-up branch.}
\label{fig:phase}
\end{figure*}
The total angular momentum is $\bm J= \bm \ell + \bm s$. For a given partial wave $\ell$, there are two possible $J$ values: $J_\pm=\bm \ell \pm 1/2$. The spin-orbit coupling in this basis splits into two branches,
\beqs
\bm \ell \cdot \bm s &={1\over 2} \left [ J(J+1) -\ell (\ell+1) - {3\over 4} \right ] \\
&=
\left \{ 
\begin{array}{ll} 
\ell/ 2& \text{ for }  J=\ell+ {1\over 2} \quad (\text{all } \ell)\\
-(\ell+1) /2 & \text{ for }  J=\ell- {1\over 2}  \quad (\ell\neq 0)
\end{array}
\right.
\eeqs
The s-wave ($\ell=0$) resides in the $J=\ell+ {1\over 2}$ branch.
The potential becomes central and diagonal in this basis,
\beq
V_{J\ell}(r)=
\left( \begin{array}{cc}
V_{J=\ell+ {1\over 2}}(r)  & 0  \\
0 & V_{J=\ell- {1\over 2}}(r)  \\
\end{array} \right).
\eeq
Thus for each partial wave $l$, the two states $J=\ell\pm {1\over 2}$  can be treated separately 
in the infinite volume. In the finite volume, however, the two remain coupled as in Eq.(\ref{eq:potso}) or in the lattice Hamiltonian Eq.(\ref{eq:latH3}) or Eq.(\ref{eq:latH7}).
The two decoupled central potentials are 
\beq
V_{J=\ell+ {1\over 2}}(r)= \left (V_0 + {1\over 2}\ell \;c_{\ell s} \right ) e^{-\beta r^2},
\label{eq:pot2}
\eeq
\beq
V_{J=\ell- {1\over 2}}(r)=\left (V_0 -{1\over 2}(\ell+1) \;c_{\ell s} \right ) e^{-\beta r^2}.
\label{eq:pot3}
\eeq
We see that  the effect of the spin-orbit term is to modify the $V_0$ value. 
As $\ell$ increases, the $J=\ell+ {1\over 2}$ branch is more and more repulsive.
At the same time, the depth of the $J=\ell- {1\over 2}$ decreases with $\ell$ so the phase shift in this branch is expected to have smaller magnitudes at fixed energies.

For the parameters, we consider two particles mimicking the pion and nucleon: 
a spin-0 particle of mass $m_1 = 0.138 \GeV$ and a spin-1/2 of $m_2 = 0.94 \GeV$, and the same  $V_0 = 4 \GeV$ and $\beta = {1\over 8} \fm^{-2}$ as in the first potential in Eq.\eqref{pot1}. 
For the strength of the spin-orbit coupling, we choose a value of $c_{s\ell}=25 \GeV^{-1} \fm^{-2}$ which leads to sizable and well-separated phase shifts but still keeps
the potential in the  $J=\ell- {1\over 2}$ branch (including the centrifugal barrier) repulsive over the range of $\ell$ and $k$ values considered.
Phase shifts for partial waves up to $l = 5$ are shown in Fig.\ref{fig:phase} for both branches. The goal is to feed these phase shifts into the QCs and confirm that the energy spectrum produced is converging to the energy spectrum derived from the lattice Hamiltonian Eq.(\ref{eq:latH7}).

\subsection{Results and discussion}

\begin{figure}[!b]
\centering
    \includegraphics[scale=0.245]{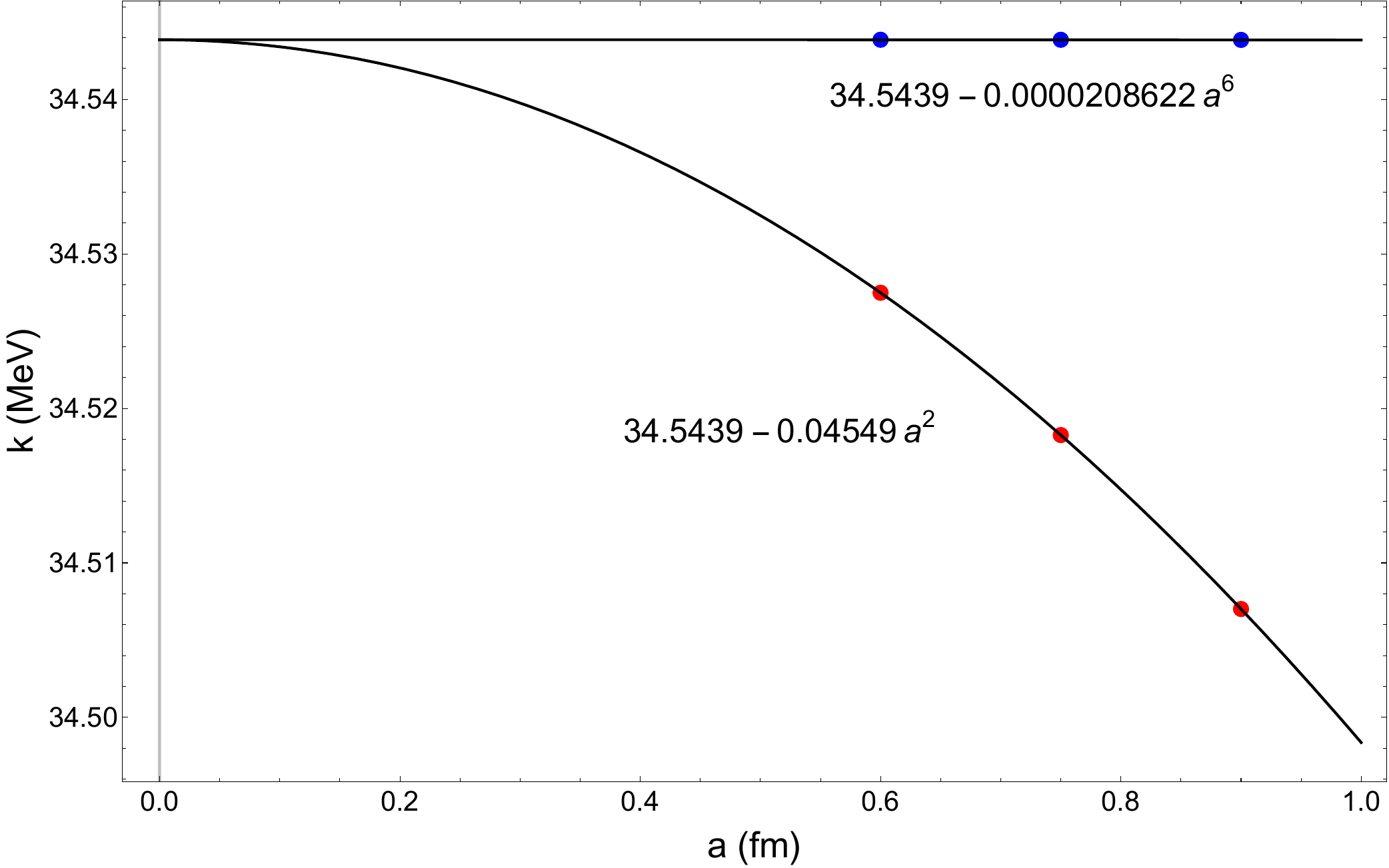}
\caption{Continuum extrapolation of the 4th lowest level in the $G_{1g}$ irrep rest frame in a cubic box of $L=36$ fm. Three lattices are used: $40^3$ with $a = 0.9 \fm$, $48^3$ with $a = 0.75 \fm$, and $60^3$ with $a = 0.6 \fm$. The red points are from the three-point stencil in Eq.\eqref{eq:latH3} and the blue points are from the seven-point stencil In Eq.\eqref{eq:latH7}. The fitted curves and forms are also displayed. }
\label{fig:3ptvs7pt}
\end{figure}

\begin{figure*}[t]
\centering
\includegraphics[width=1.0\textwidth]{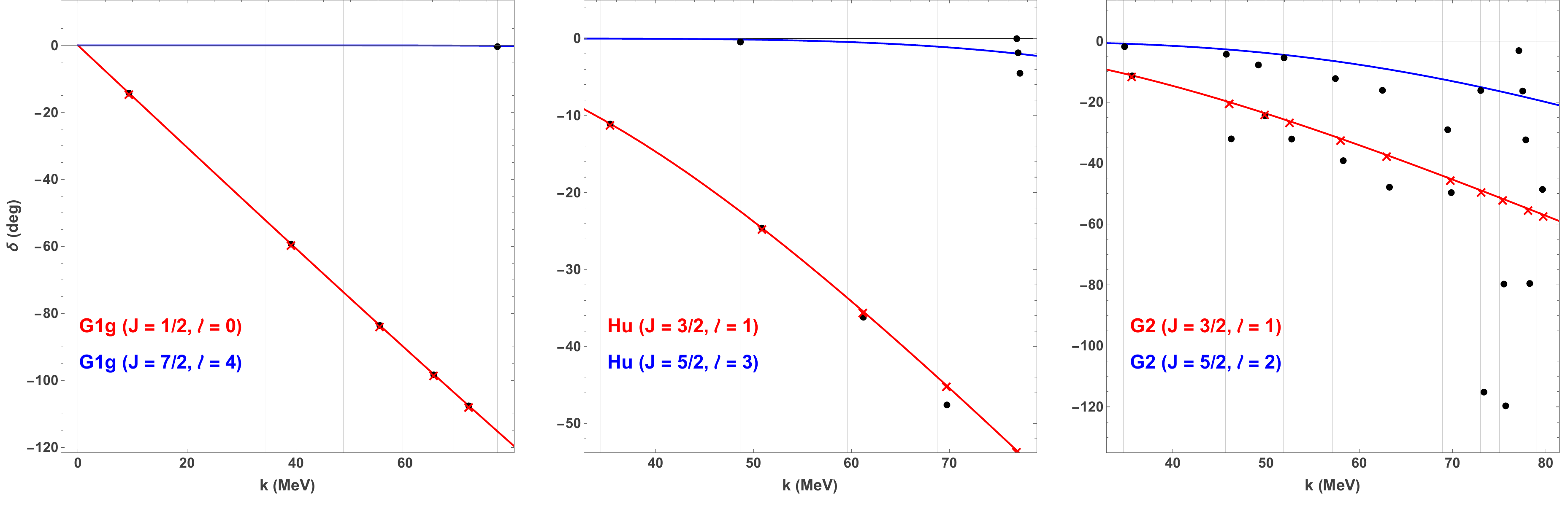}
\includegraphics[width=1.0\textwidth]{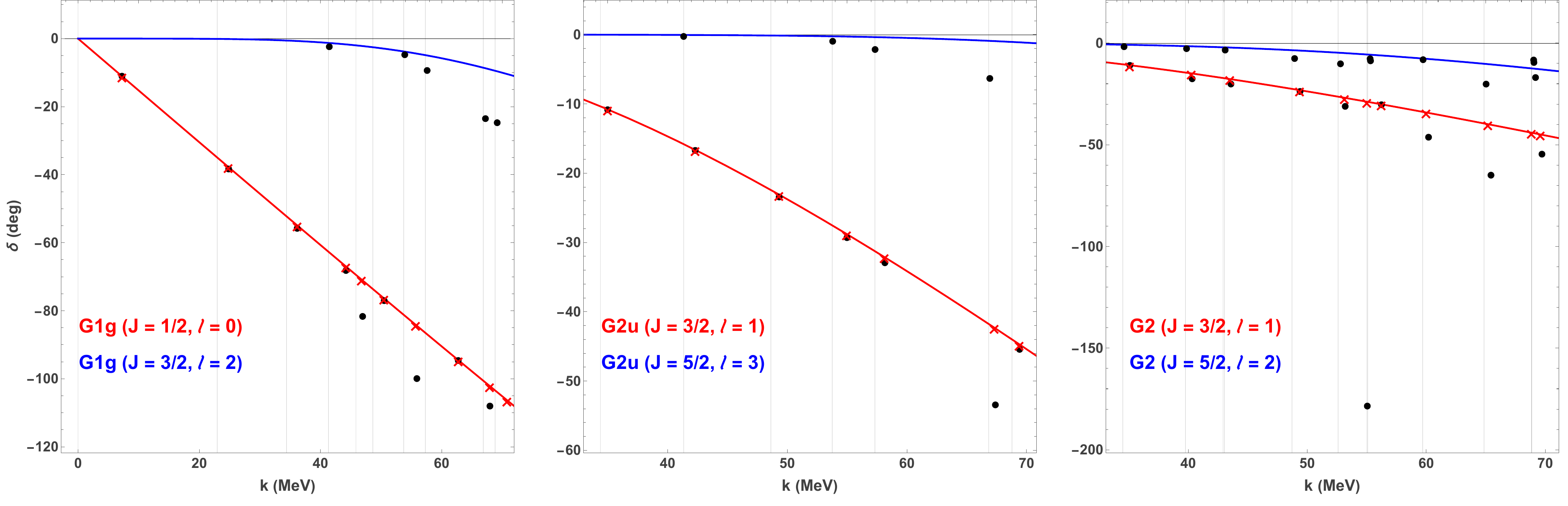}
\caption{
Top row: phase shift reconstruction from L\"uscher formula (lowest partial wave) from the box energy levels in $G_{1g}$ and $H_u$ irreps in the rest frame,  and $G_{2}$ irrep of $^2C_{4v}$ for moving frame $\bm d = (0,0,1)$, all in the cubic box. Black points are predicted phase shifts from box energy levels. Red crosses show the energy levels predicted by L\"uscher formula. The red and blue curves show the two largest phase shifts active in the irrep. Vertical gray lines indicate non-interacting levels. Bottom row: same as the top row, but 
in the elongated box.
}
\label{fig:PhasePrediction}
\end{figure*}


The range of the test potential in this section is about 12 fm so a box of size 36 fm is sufficient to make the finite-volume effects negligible. While such large volumes are not yet feasible in current lattice QCD calculations, we employ it here in order to get a precise check. We calculate the energy spectrum for three different lattices: $40^3$ with $a = 0.9$ fm, $48^3$ with $a = 0.75$ fm, and $60^3$ with $a = 0.6$ fm, then extrapolate it to the continuum limit. The extrapolation is a function of $a^6$, the error present in the Hamiltonian from the seven-point stencil approximation in Eq.\eqref{eq:latH7}. Fig.~\ref{fig:3ptvs7pt} shows an example of this continuum extrapolation alongside that  from the standard three-point stencil with $\mathcal{O}(a^2)$ error. One sees that both extrapolations reach the same result, but the seven-point function converges much faster.

For the elongated box, we use the same lattice spacings and length in the $x$ and $y$ directions, but we elongate the $z$ direction by a factor of $\eta = 1.5$. Calculation of a large number of energy states is computationally expensive due to the Hamiltonian being an $\approx 10^5 \times10^5$  matrix. For this reason, we only look at the 100 lowest eigenvalues for each QC.  Each energy level in a $d$-dimensional irrep will be $d$-fold degenerate. In some QCs, this degeneracy significantly reduces the number of unique energy values studied.

Our objective is to check the validity of the derived QCs in Eq.\eqref{Eq:QC1} to higher partial waves. In lattice QCD simulations, energy levels are used to predict phase shifts. Since each QC is a single condition that couples to an infinite tower of partial waves, generally speaking, only the lowest phase shift in a given irrep can be isolated (L\"uscher formula). One naturally wonders: what about the higher partial waves? How do they influence the prediction of the lowest partial wave?
To answer this question, we turn the typical usage of a QC in reverse: we feed the infinite volume phase shifts to the QC, then numerically find its roots, then compare the resulting levels to the box levels determined separately from the Schr\"odinger equation. We check convergence order by order according to $J$ (for fixed $J$, $\ell$ can take $J\pm 1/2$): `order 1' has only the lowest $J$ in the QC, `order 2' has the next lowest $J$ added, and so on. Usually, we check each QC up to $J = \frac{11}{2}$; if this does not reach convergence,  we then include $J = \frac{13}{2}$.

Before going to higher orders, we show the overall quality of phase shift prediction from the L\"uscher formula in Fig.~\ref{fig:PhasePrediction} (top) for selected irreps in the cubic box. L\"uscher formula is the QC truncated to include only the lowest partial wave. For this special case, we can solve the {\em inverse} problem: given an energy level, we can compute a phase shift. Note that this is not possible for any higher order truncations of QCs. The black dots in the figure indicate the solution of the inverse problem, using as inputs the box energy level computed by solving the Schr\"odinger equation. The red crosses correspond to using the L\"uscher formula to {\em compute} energy levels given the lowest partial wave.
The $G_{1g}$ irrep gives the best prediction for phase shift  $\delta_{{1\over 2}0}$ (s-wave), with the lowest 5 energy levels reproducing the phase shifts. Note also that the crosses are very close to the black points in this case. The 6th level is completely off because it occurs very close to a non-interacting energy level. In the absence of higher partial waves, this level would be exactly equal to the non-interacting energy, but it gets slightly shifted due to the presence of higher partial waves. So these shifts are not connected to the lowest partial wave, and using the L\"uscher formula produces incorrect results (in special cases, you can use these levels to extract the next partial wave~\cite{Lee_2022}).
The $H_u$ irrep gives the best prediction for phase shift  $\delta_{{3\over 2}1}$ (p-wave). Out of the 8 lowest energy levels, levels 1,3,4,5 reproduce the phase shifts, but not levels 2,6,7,8 for the same reason as in the $G_{1g}$ case (note that the second energy level has no corresponding cross). 

The $G_2$ irrep of $^2C_{4v}$ is an example of a moving frame. The story here is a bit more complex: here we have mixing between different parities and the next strongest phase-shift is quite large. In our setup, every non-interacting energy is at least doubly degenerate due to the spin, and in the presence of interactions we end up with pairs of energy levels close by, a feature that L\"uscher formula cannot capture. The spin-partners have the same $\ell$ and belong to neighboring $J$'s. In the rest frame irreps, for lower values of $J$, the neighboring $J$'s belong to different irreps. While at higher $J$, deeper in the spectrum, the irreps start to couple to neighboring $J$'s and show the spin-partners. For example, in the $G_{1g}$, the first neighbors to appear are $J = {7\over 2}$ and $J = {9\over2}$. Whereas for moving frames, the irreps couple to neighboring $J$'s from the beginning, in the $^2C_{4v}$, little group the $G_2$ irrep couples to $J = {3\over2}$ and $J = {5\over2}$. The crosses show that L\"usher formula predicts an energy level bracketed by the spin-pair. However even though the pair is very close in energy, the resultant phase shifts are very different, indicating that the inverse problem is very sensitive. The dominant phase shift is $\delta_{{3\over 2}1}$  (same as $H_u$ in the rest frame). Using L\"uscher formula here is clearly wrong. As we see later using higher order QC reproduces the correct spectrum very precisely.

\input{Tables/CubicG1gChiConvergence}

Fig.~\ref{fig:PhasePrediction} (bottom)
shows the situation for the same QCs in the elongated box.
Overall, we see that the quality is about the same as in the cubic box, even though the levels are more packed due to elongation. 
Their detailed convergence tables can be found in the supplement~\cite{supp}.

We now turn to the impact of high partial waves in the QCs. Since the effects depend on the QC under consideration to varying degrees (some are quite small), we need an objective criterion to judge the quality of convergence. To this end, we introduce a numerical $\chi^2$-measure,
\beq
\label{eq:chisq1}
\chi^2 = \frac{(k_{\text{box}}-k_{\text{QC}})^2}{(k_{\text{box}}-k_{\text{lat}})^2},
\eeq
where $k_{\text{box}}$ is the energy level from the continuum extraction of the three lattice levels, $k_{\text{QC}}$ is the level predicted from the QC by root-finding, and $k_{\text{lat}}$ is the energy level on the lattice with the finest spacing. The idea behind this criterion is to compare the difference between the predictions of the QCs and the error from the lattice. Since the difference between the box levels and levels from the finest lattice is typically $\mathcal{O}(10^{-7})$, the $\chi^2$ measure is very sensitive to small changes in the QCs. A non-convergent energy level blows up the $\chi^2$.  Note that the $\chi^2$ introduced here is not to be confused with the $\chi^2$ measure used in curve fitting. Here, convergence means $\chi^2$  approaches zero as opposed to a value close to 1.

Let us return to the $G_{1g}$ irrep in Fig.\ref{fig:PhasePrediction} as the first example to showcase how convergence is checked by the $\chi^2$ measure. Table \ref{tab:CubicG1gChiConvergence} lists the order-by-order convergence of the lowest 6 energy levels in this QC. The first 5 levels are typical of the QC convergence; as new $l$ values are included in the QC (every other order) the $\chi^2$ value drops significantly until it approaches $\approx 10^{-3}$ at which point the values in $\chi^2$ start to approach machine-precision and the precision from further orders cannot be captured here.
We see the importance of using a numerical measure to check convergence. Visually, the lowest 5 levels all fall on the red curve at order 1, giving the impression that they reproduce the infinite volume phase shifts. But Table \ref{tab:CubicG1gChiConvergence}  reveals that only level 1 has a small $\chi^2$ at order 1.  The next 4 levels only converge by order 3 which means two more partial waves are added to the lowest one in the QC.
Only even $\ell=0,4,6$ partial waves contribute because $G_{1g}$ is an even-parity irrep by $(-1)^\ell$. 
There is no multiplicity to order 4 in this QC.

Level 6 deserves special attention. It has no solution even at order 1. The reason is that this level is extremely close to a non-interacting level which 
is a singularity (pole) of the zeta functions, see Eq.(\ref{eq:Zeta}).
We say the level is `{\it pinched}' and use a red star and the non-interacting level instead to label it in the convergence table. The resolution is in the inclusion of higher partial waves which can pull the level away from the pole. These pinched points are a major indicator in the convergence analysis. A single pinched point blows up the total $\chi^2$ value for a QC.

The pinched level can give the impression that the QC is not converging. 
Or that if it is converging, it could still far from the box level. 
To check this possibility, we introduce a modified measure,
\beq
\label{eq:chisq2}
\chi_2^2 = \frac{(k_{\text{box}}-k_{\text{QC}})^2}{(k_{\text{box}} - k_{\text{free}})^2},
\eeq
where non-interacting energy level $k_{\text{free}}$ replaces $k_{lat}$ in the denominator. The idea behind this is to note that interaction does not alter the number of energy levels in the box; it merely shifts them. This $\chi_2^2$ measures the convergence of the QC against that energy shift, 
as indicated in the last line of Table \ref{tab:CubicG1gChiConvergence}.
Using this measure, we see that the pinched level is `unpinched' when higher partial waves are added, and normal convergence is observed. 
In the following, we only use the regular $\chi^2$ in the convergence tables, knowing that whenever there is a pinched level, the $\chi_2^2$ measure can properly account for its convergence.

\input{Tables/CubicHuChiConvergence}


Next, we look at the $H_u$ irrep in Fig.\ref{fig:PhasePrediction} 
which has two pinched levels (6 and 8). 
In fact, levels 6, 7, and 8 are degenerate in the non-interacting case and the interaction only separates them by a small amount,
leaving them all very close to the free particle pole. Typically, only one of these levels will be found at order 1. The inclusion of higher $J$ waves is required to achieve convergence for each level individually, as illustrated in Table \ref{tab:CubicHuChiConvergence}. 
One sees that the QC converges at order 5, as opposed to $G_{1g}$ at order 3. 
It means at order 5, phase shifts $\delta_{{3\over 2}1}$, $\delta_{{5\over 2}3}$, $\delta_{{7\over 2}3}$, $\delta_{{9\over 2}5}$, $\delta_{{11\over 2}5}$ are contributing in the QC.
This QC is an example of how higher partial waves and high precision work together to achieve convergence, especially when near degenerate levels are involved. 
Only $\ell=1,3,5$ partial waves contribute because $H_u$ is an odd-parity irrep. 
There is no multiplicity to order 5 in this QC.

The $G_2$ irrep in Fig.\ref{fig:PhasePrediction} offers a prime example for moving frames. The idea of using moving frames is to add more coverage at smaller energies than those in the rest frame. Indeed, we see about 4 times more energy levels in this QC than in $H_u$ at the same cutoff, but they do a poor job reproducing the infinite volume phase shifts. The reason is partial-wave mixing due to loss of parity in moving frames. 

In typical lattice QCD applications, one inserts the finite-volume energy levels into the quantization conditions and solves for the phase shifts. In practice, however, one can usually only isolate the lowest partial wave in a given irrep from a single QC, so the QC is commonly truncated at leading order. This truncation discards contributions from higher partial waves, i.e., it is an approximation of the full L\"uscher relation Eq.(\ref{Eq:QC1}), and it is only reliable when those higher partial waves do not contribute greatly to the energy states in question. In the rest frame this often appears to be a mild assumption, since the lowest partial waves in an irrep are typically well separated. 
In moving frames, however, the loss of parity ensures partial wave mixing; both even-$\ell$ and odd-$\ell$ appear and the distance between the lowest two partial waves in any irrep, $l_{\text{1st}} - l_{\text{2nd}}$, is guaranteed to be 1, so one should generally expect larger truncation effects in moving frames.
Even for levels that are well separated from the non-interacting poles, the reconstructed phase shift in $G_2$ is still quite poor. Within our study, this kind of degradation is typical across all moving frames. 

\input{Tables/CubicG2C4vConvergence}

Table \ref{tab:CubicG2C4vChiConvergence} shows order-by-order convergence in $G_2$ of $^2C_{4v}$ for all 22 levels displayed, 
including 4 pinched levels (16, 18, 19, 21). At order 1, $\chi^2$ is very high across the board. This is due to the fact that the levels from the finest lattice $k_{lat}$ which are used as a measuring stick in Eq.\eqref{eq:chisq1}, are already close to the box levels (to 6 significant figures). It takes 5 orders for most of the lower levels to converge; some higher levels are still not very convergent.  
At order 5, ten phase shifts $\delta_{{3\over 2}1}$, $\delta_{{3\over 2}2}$, $\delta_{{5\over 2}2}$, $\delta_{{5\over 2}3}$,  $\delta_{{7\over 2}3}$, 
$\delta_{{7\over 2}4}$,  $\delta_{{9\over 2}4}$, $\delta_{{9\over 2}5}$, $\delta_{{11\over 2}5}$, $\delta_{{11\over 2}6}$ 
are contributing in the QC (see Fig.\ref{fig:phase}).
Due to mixing, even at order 1, we have both $\delta_{{3\over 2}1}$ and $\delta_{{3\over 2}2}$.  We have to cut $\delta_{{3\over 2}2}$ to make a prediction for $\delta_{{3\over 2}1}$ Fig. \ref{fig:PhasePrediction}.
We see both even and odd $\ell$ partial waves participating due to loss of parity. 
Another symptom in moving frames is multiplicity starts early (at $J=5/2$ in this case), as opposed to no multiplicity in the rest-frame QCs $G_{1g}$ and $H_u$ below $J=11/2$. Multiplicity also leads to larger QC matrices. The matrix size is $4\times 4$ at order 4 for $G_{1g}$, and $5\times 5$ at order 5 for $H_{u}$.
For $G_2$ of $^2C_{4v}$ here, it is $20\times 20$ at order 5. The size can be found by adding up the multiplicities through the orders.
At order 2, the size is $6\times 6$, whose full form is written out as an example in Eq.\eqref{G2order2} of Appendix~\ref{appendix:MatrixElements}.
Overall, convergence is slower than in rest-frame $G_{1g}$ and $H_u$ due to partial wave mixing in  moving frames.

Lastly, we mention an interesting feature we observed in the $K_1$ and $K_2$ irreps of the $^2C_{3v}$ little group for moving frame $\bm d=(1,1,1)$ in the cubic box: they have the same convergence values (see Table~\ref{tab:ChiSquareSummary}). 
This is a consequence of the Kramer's Degeneracy Theorem \cite{Klein:Kramer}. The theorem states that for every energy eigenstate of a half-integer spin system with time reversal symmetry, one can construct another energy eigenstate of the same energy through the time reversal symmetry. As a result, every energy eigenstate must be at least doubly degenerate for half-integer spin systems with time symmetry.
This is handled naturally for most cases where the irreps are of even dimension and thus the energy levels have an even degeneracy. But in the case where a frame has one-dimensional irreps - what we have called $K$ irreps - then those irreps must form into coupled pairs with the same spectrum of energy values.
The group elements of these coupled irreps are related by $K_1(g) = K_2(g)^*$ and the basis vectors are reversed for spin-up and spin-down between them, i.e. if we write a basis vector as $v^{\Gamma}_{J,l}$ for irrep $\Gamma$ then, for fixed $J$, $v^{K_1}_{J,J\pm s}=v^{K_2}_{J,J\mp s}$.
We also observed the relations in Ref.\cite{Gockeler:2012yj}.
The property causes the convergence of these irreps to be identical when looked at $J$-by-$J$. The energy degeneracy are observed in the lattice values as well. 
Note that Kramer's theorem only applies to the 1D irreps of spin-1/2 little groups. There is no such degeneracy in the 1D irreps of spin-0 little groups (such as $A_1$ and $A_2$, or $B_1$ and $B_2$ of $C_{2v}$ ).

We have examined all 19  QCs using the same checks. Their phase shift reconstruction at order 1 and order-by-order convergence data can be found in the supplement~\cite{supp}.

\section{Summary and Outlook} \label{section:conclusions}

We derived higher-order L\"uscher quantization conditions (QCs) for the scattering of two particles with total spin $s = 1/2$. We checked the correctness of our results numerically by comparing the QC predictions with the spectrum of two-particle states in a box. This spectrum is found by solving the Schr\"odinger equation of a simple non-relativistic potential. Both the phase shifts in infinite volume and energy levels in finite volume are independently generated. 
Table~\ref{tab:ChiSquareSummary} gives an overview of all the QCs considered in this work, along with convergence checks.

\input{Tables/ChiSquareSummary}

We derived QCs for numerous scenarios, including rest frames as well as selected moving frames for cubic and elongated boxes. We worked out quantization conditions up to $J=11/2$. The QCs agree with those in the literature up to $J=7/2$~\cite{Gockeler:2012yj,Bernard:2008ax,Lee:2017igf}.
All higher orders in cubic and elongated boxes are new. 

All QCs were checked against a simple potential with phase shifts that can be readily calculated through the variable phase method. The potential and box size were chosen to satisfy two conditions: (1) the exponentially suppressed finite-volume corrections of the QCs are negligible and (2) the energy levels are sensitive to contributions from partial waves as high as $J = 11/2$. This allowed us to test with confidence high-order QCs and we find that the results agree to more than six significant figures. We performed these checks for several energy levels in each irrep. 

For the most part, we find that elongated boxes work about as well as cubic ones. This gives a cost effective way of increasing the kinematic range for a small increase in lattice volume.

Due to the smaller number of irreps that couple to systems with spin than without~\cite{Lee_2022}, there is a larger amount of partial wave mixing in the QCs. This makes the extraction of phase shifts in the meson-baryon sector more difficult than for meson-meson scattering. The loss of parity in moving frames exacerbates the problem, with most moving frames having only one or two irreps.

Looking to the future, the results and methodology developed here can be used for lattice QCD calculations, both to connect finite box energy-levels to the infinite volume amplitudes and also to lay the ground-work for such calculations. In general, energy levels computed from lattice QCD can be directly connected to phase-shifts only when QCs can be truncated to the lowest partial wave in the relevant irrep. The precision of the predicted phase-shift is influenced by the size of the next higher partial wave in the irrep. The size of this effect is not known beforehand, and it depends on the energy range and box geometry. Given a good approximation of the phase-shifts relevant for a particular channel we can determine which irreps and box geometries are most likely to have reduced mixing. In the same vein, when we have an interaction model available, assuming that we can use it to compute infinite-volume phase-shifts, we can use the QCs validated in this work  to determine which irreps and energy levels are most effective in constraining the parameters of the model. This information can be used to efficiently allocate computational resources.

\acknowledgments
This research is supported by the U.S. Department of Energy under grant  DE-FG02-95ER40907.
\newpage

\bibliography{references}

\clearpage
\appendix

\onecolumngrid
\section{Matrix elements for quantization conditions}\label{appendix:MatrixElements}

In this section, we present all the QCs derived in this work in terms of the matrix elements of $\mathcal{M}_{Jln,J'l'n'}^\Gamma$ in Eq.(\ref{Eq:QC1}). We show only partial results up to certain $J$ value for every QC in both cubic and elongated boxes here. Only unique matrix elements are displayed. Since $\mathcal{M}$ is hermitian, we only show the upper triangular elements starting at the top left, and then go down each column until we hit the diagonal, and then go to the next column. The complete matrix for any QC at any order (up to $J= \frac{11}{2}$) can be recovered on demand from a python notebook in an easy-to-read and easy-to-use format in the supplement~\cite{supp}. 

As an explicit example of how to use the following tables, we will build the 2nd order QC for the $G_2$ irrep in the $^2C_{4v}$ little group of the cubic box. If one wished to start from scratch, they would begin with the group elements in Table \ref{tab:c4vgroupelements} and follow the procedure described in Appendix \ref{appendix:basis vectors} to derive the basis vectors (Tables [\ref{tab:cubicbasisvectors} -- \ref{tab:c2vbasisvectorselongated}]). From there, the procedure detailed in Section \ref{section:QuantizationConditions} will produce the matrix elements contained in Tables [\ref{tab:CubicRestMatrixElements}--\ref{tab:C2vElongatedMatrixElements}]. To obtain the QC of the desired order, one must truncate the matrix in Eq.(\ref{Eq:QC1}). The size of the matrix depends on the multiplicity and whether partial waves mix. Referencing Table \ref{tab:ChiSquareSummary}, we find the $G_2$ irrep has $J(n)$ content $J(n) = \frac{3}{2}(1),\frac{5}{2}(2),\cdots$ and parity is not a symmetry ($l$'s mix), the 2nd order QC is then, 

\beq
 = 0.
\label{G2order2}
\eeq
\normalsize
One then looks at Table \ref{tab:C4vCubicM} for the matrix elements of the upper triangle. The lower triangle can be filled in by using the fact that these matrices are Hermitian. This would give the final form of the QC.

A reference guide with links to the tables is given here in Table~\ref{tabA}.

\begin{table*}[h]
\centering
\caption{Quick reference with links to all the QC tables in this appendix.}
\label{tabA}
\begin{adjustbox}{max width=\textwidth,center}
    \begin{tabular}{|c|c|c|c|c|c|c|}
    \toprule
& \multicolumn{3}{c|}{\textbf{Cubic Box}} & 
\multicolumn{3}{c|}{\textbf{Elongated Box}}\\
\hline
        Boost Vector & Group Elements & Basis Vectors & Matrix Elements &  Group Elements & Basis Vectors & Matrix Elements\\
        \hline
         $\bm d = (0,0,0)$ & Table \ref{tab:cubicgroupelements} & Table \ref{tab:cubicbasisvectors} & Table \ref{tab:CubicRestMatrixElements}  & Table \ref{tab:D4hgroupelements} & Table \ref{tab:elongatedbasisvectors} & Table \ref{tab:ElongatedRestMatrixElements} \\
         $\bm d = (0,0,1)$ & Table \ref{tab:c4vgroupelements} & Table \ref{tab:c4vbasisvectors} & Table \ref{tab:C4vCubicM} &  Table \ref{tab:c4vgroupelements} & Table \ref{tab:c4vbasisvectors} & Table \ref{tab:C4vElongatedM} \\
        $\bm d = (1,1,1)$ & Table \ref{tab:c3vgroupelements} & Table \ref{tab:c3vbasisvectors} & Table \ref{tab:C3vCubicM} & & & \\
        $\bm d = (1,1,0)$ & Table \ref{tab:c2vcubicgroupelements} & Table \ref{tab:c2vbasisvectorscubic} & Table \ref{tab:C2vCubicM}  & Table \ref{tab:c2vcubicgroupelements} & Table \ref{tab:c2vbasisvectorselongated} & Table \ref{tab:C2vElongatedMatrixElements}\\
        \hline
    \end{tabular}
    \end{adjustbox}
\end{table*}
 
\input{Tables/MatrixElements/CubicRestMatrixElements}

\input{Tables/MatrixElements/C4vCubicMatrixElements}

\input{Tables/MatrixElements/C3vCubicMatrixElements}

\input{Tables/MatrixElements/C2vCubicMatrixElements}

\input{Tables/MatrixElements/ElongatedRestMatrixElements}

\input{Tables/MatrixElements/C4vElongatedMatrixElements}

\input{Tables/MatrixElements/C2vElongatedMatrixElements}

\section{Group Theory Details} \label{appendix:GroupTheory}

For completeness, we give the group theory details used to derive the QCs in the previous appendix. Due to spin degree of freedom, double-cover groups must be used.

\subsection{Symmetry of the cubic box}
The symmetry group of the cubic box is the octahedral group and is denoted by $O$. This group consists of the 24 orientation preserving rotations. This group is sufficient to describe integer angular momentum in the cubic box.

To account for half-integer angular momentum in the cubic box requires the \textit{double-cover} of $O$, which we denote $^2O$. This group is constructed by viewing a $2\pi$ rotation as different from a $4\pi$ rotation. As such, the number of group elements doubles. Each element $g \in O$ now has a corresponding double cover element that we denote as $\tilde{g}$. If one describes the rotation using the rotation axis and angle, $g=R(\vec{n},\omega)$ then $\tilde{g} = R(\vec{n},(\omega+2\pi) \mod 4\pi)$. Or if described by active $z-y-z$ Euler angles $g = R(\alpha,\beta,\gamma)$ then $\tilde{g}=R(\alpha,\beta,(\gamma + 2\pi) \mod 4\pi)$. Both descriptions are provided in the tables below.

For non-moving frames, parity is another symmetry of the cubic box. This is another doubling of the group elements corresponding to $^2O_h = \,^2O \otimes\{\mathbb{E},-\mathbb{E}\}$. Where the subscript $_h$ denotes the inclusion of parity, and $\mathbb{E}$ is the identity element.

A useful tool when studying a symmetry group is to look at its irreducible representations (irreps). The number of irreps of a group is equal to the number of conjugacy classes.
One can easily show that the group $O$ has 5 conjugacy classes and thus 5 irreps. To determine the dimensionality of irreps we use the fact that the sum of the squares of the dimensions of the irreps equals the order of the group.
\beq
 |G| = \sum_i d_i^2
\eeq
where $d_i$ is the dimension of the $i$'th irrep. This gives the Diophantine equation $24 = a^2 + b^2 + c^2 + d^2 + c^2$. This has one solution with dimensions 1, 1, 2, 3, and 3. Using the convention set by Mulliken \cite{10.1063/1.1740655}, these irreps are called $A_1$, $A_2$, $E$, $T_1$, and $T_2$, respectively. $A_1$ is the trivial representation; every element is represented by 1. The $A_2$ representation can be constructed from the $A_1$ irrep using two facts. The first is that irreps are orthogonal in the sense that
\beq
\braket{\chi_i,\chi_j} = \frac{1}{|G|}\sum_{g\in G} \overline{\chi_i(g)}\chi_j(g) = \delta_{ij}.
\eeq
Where $\chi_i(g)$ is the character of the element $g$ in the $i$'th irrep. The second fact needed is that all elements in a given conjugacy class have the same character. Using these two facts, and noting that the size of the conjugacy classes are 1, 3, 8, 6, and 6 we see that the only way for the orthogonality relation to hold is for the two classes with six elements to be represented with -1.
The real meaning of the subscripts $G_{1/2}$ has to do with the symmetry properties of the irrep basis functions, but such details are not necessary to construct the irreps here.
The construction of the $E$ irrep proceeds in a similar way. One finds that the conjugacy classes must have characters 2, 2, -1, 0, and 0, respectively. One then must choose matrices with these traces that also form a homomorphism. We choose the five found in \cite{altmann1994point}. 
\scriptsize
\beqs
& e_1=\left(\begin{array}{cc}
 -1 & 0 \\
 0 & 1 \\
\end{array}\right), 
e_2=\left(\begin{array}{cc}
 -\frac{1}{2} & -\frac{\sqrt{3}}{2} \\
 \frac{\sqrt{3}}{2} & -\frac{1}{2} \\
\end{array}\right), 
 e_3=\left(\begin{array}{cc}
 -\frac{1}{2} & \frac{\sqrt{3}}{2} \\
 -\frac{\sqrt{3}}{2} & -\frac{1}{2} \\
\end{array}\right), \\
& e_4=\left(
\begin{array}{cc}
 \frac{1}{2} & -\frac{\sqrt{3}}{2} \\
 -\frac{\sqrt{3}}{2} & -\frac{1}{2} \\
\end{array}\right), 
 e_5=\left(\begin{array}{cc}
 \frac{1}{2} & \frac{\sqrt{3}}{2} \\
 \frac{\sqrt{3}}{2} & -\frac{1}{2} \\
\end{array}\right).
\eeqs
\normalsize
The $T_1$ irrep is then the standard $3\times 3$ rotation matrices and the $T_2$ follows the same pattern as the $A_1$ and $A_2$ irreps by negating the last 12 elements in the group. The $T_1$ elements are generated via
\beq
t_k = e^{-i(\bm{n}\cdot \bm{J})\omega_{k}}, \;\;\; k = 1, \cdots, 24,
\eeq
where $(J_k)_{ij} = i \epsilon_{ijk}$, and $\bm{n}$ and $\omega_k$ are the rotation axis and angle in the order given below.

The double cover adds three more irreps called $G_1,G_2$ and $H$ with dimensions 2, 2, and 4, respectively. These are the ``spinor'' irreps and are the only ones sensitive to the double-cover and antisymmetric with respect to it. That is, if $D^\Gamma(g)$ is an irrep of element $g\in \,^2O$ then $D^\Gamma(g) = -D^\Gamma(\tilde{g})$ for $\Gamma \in \{G_1, G_2, H\}$ and $D^\Gamma(g) = D^\Gamma(\tilde{g})$ for $\Gamma \in \{A_1,A_2,E,T_1,T_2\}$. The $G$ elements are generated similar to the $T$ elements but they halve the rotation angle so as to distinguish a $2\pi$ and $4\pi$ rotation.
\beq
g_k = e^{-i(\bm{n}\cdot \bm{\sigma})\omega_{k}/2}, \;\;\; k = 1, \cdots, 24.
\eeq
The four-dimensional $H$ irreps are generated by
\beq
h_k = e^{-i(\bm{n}\cdot \bm{J})\omega_{k}}, \;\;\; k = 1, \cdots, 24.
\eeq
Where $\bm{J}$ are the generators of spin-3/2
\beqs
J_x &= \begin{pmatrix}
    0 & \frac{\sqrt{3}}{2} & 0 & 0\\
    \frac{\sqrt{3}}{2} & 0 & 1 & 0\\
    0 & 1 & 0 & \frac{\sqrt{3}}{2}\\
    0 & 0 & \frac{\sqrt{3}}{2} & 0
\end{pmatrix},\\
J_y &= \begin{pmatrix}
    0 & -\frac{i\sqrt{3}}{2} & 0 & 0\\
    \frac{i\sqrt{3}}{2} & 0 & -i & 0\\
    0 & i & 0 & \frac{-i\sqrt{3}}{2}\\
    0 & 0 & \frac{i\sqrt{3}}{2} & 0
\end{pmatrix},\\
J_z &= \begin{pmatrix}
    \frac{3}{2} & 0 & 0 & 0\\
    0 & \frac{1}{2} & 0 & 0 \\
    0 & 0 & -\frac{1}{2} & 0\\
    0 & 0 & 0 & -\frac{3}{2}
\end{pmatrix}.
\eeqs

Inclusion of parity doubles the number of irreps. Forming an even and an odd version of each irrep, denoted by the subscript $_g$ and $_u$ for \textit{gerade} and \textit{ungerade} (even and odd). The even irreps do not change with parity $D^{\Gamma}(g) = D^{\Gamma}(-\mathbb{E}*g)$, whereas the odd irreps are antisymmetric with respect to the inversion $D^{\Gamma}(g) = -D^{\Gamma}(-\mathbb{E}*g)$.

We have organized the following tables in the manner described here. By first listing the elements of the standard group without double-cover or parity, and then listing the double-cover elements, and then the parity elements. We denote the `natural' representation of the group elements by $S_k$. These are the matrices that act on the standard $(x,y,z)$ coordinates. $S_k$ happens to coincide with the $T_{1u}$ irrep. A faithful representation of the $^2O_h$ group can be made with the $G_{1}$ irrep by
\beqs
\Gamma_{\text{faithful}}&=\bigl\{
D^{G_{1}}(g)\otimes I_2 \mid g\in\, ^2O
\bigr\}
\; \\ & \cup\;
\bigl\{
D^{G_{1}}(g)\otimes

\mid g\in\, ^2O
\bigr\}.
\eeqs
\input{Tables/GroupElements/2OhGroupElements}
\twocolumngrid

\subsection{Time reversal symmetry}
A natural question that arises in symmetry discussions is "What happens if we include time-reversal symmetry in our group?". After all, the addition of parity and the double-cover doubled the number of group elements and provided more irreps to project onto. Unfortunately, since the time-reversal operator is anti-unitary, it does not fit nicely into the standard representation theory. Representation theory with anti-unitary elements is known as \textit{corepresentation theory}. The theory of corepresentations was explored by Wigner \cite{WignerAntiUnitary} and Dimmock \& Wheeler \cite{DIMMOCK1962729}. In short, the inclusion of anti-unitary elements breaks certain structures of standard representation theory e.g. the number of conjugacy classes is no longer equal to the number of irreducible corepresentations (ICRs). Ref.~\cite{NewmarchGoldingCoreps} lays out the resulting structure of ICRs in terms of the irreps of the unitary part of the group. Let $\mathcal{G} = G \cup a_0 G$, with $a_0$ the anti-unitary operator. Pick an irrep $\Gamma$ of $G$.
\begin{enumerate}
    \item[a)] $\Gamma$ is real, $\Gamma = \Gamma^*$. The anti-unitary symmetry won't split the representation any further, providing no additional ICRs to project onto. Effectively, the ICR retains the same structure as the irrep in the unitary group.

    \item[b)] $\Gamma$ is "quaternionic", $\Gamma$ is equivalent to its conjugate but in a way that forces a quaternionic structure. The ICR will take the form $\Gamma \oplus \Gamma$. Simply doubling the dimension of the irrep.

    \item[c)] $\Gamma$ is complex, $\Gamma \neq \Gamma^*$. The anti-unitary element then swaps $\Gamma \leftrightarrow \Gamma^*$ and therefore the ICR will take the form $\Gamma \oplus \Gamma^*$ to encompass the extra symmetry. In the case of the $K$ irreps here this ends up merging them into one greater ICR of doubled dimension.
\end{enumerate}

In all three cases, there is no further structure to exploit from the methodology used here.

\subsection{Elongated box}

The symmetry group of the z-elongated box is $D_4$, It is the subgroup of $O$ that preserves the $(0,0,1)$ vector. The construction of the irreps follows that of the cubic box. One finds there are five irreps with dimensions $1,1,1,1,2$ and identifies them as the $A_1,A_2,B_1,B_2,$ and $E$ irreps. The double cover, $^2D_4$ doubles the group elements and adds two 2-D irreps, $G_1$ and $G_2$. Parity doubles the number of elements again and forms even and odd versions of each irrep. This brings the total number of elements to 32 with 14 irreps. We use the same notational convention as the cubic case to denote the elements in Table \ref{tab:D4hgroupelements}.

\input{Tables/GroupElements/D4hGroupElements}

\subsection{Moving frames}

When the two particle system has a nonzero total momentum, we say that it is in a moving frame. We denote the direction of total momentum as $\bm{d} = (n_x,n_y,n_z)$ with $n_i \in \mathbb{Z}$. These moving frames single out a certain direction in space, reducing the overall symmetry of the box. These subgroups are called \textit{little groups} and their elements are all elements, $k$ from $^2O_h$ (or $^2D_{4h})$ such that $S_k \bm{d} = \bm{d}$. In this work we have considered three moving frames in the cubic box and two in the elongated case: $\bm{d} = (0,0,1)$ and $\bm{d} = (1,1,0)$ in both cases and $\bm{d} = (1,1,1)$ in just the cubic box. Tables \ref{tab:cubicgroupelements} through \ref{tab:c2vcubicgroupelements} detail these little groups.

For permutations of these directions say, $\bm{d} = (1,0,0)$, then the irreps will stay the same but the group elements will change to new subgroups of the parent group. This will change the basis vectors and the overall structure of the QCs. However, physical results such as energy levels, phase shifts, and angular momentum from two similar boost directions must be the same. This change of boost direction is equivalent to a rotation of the zeta functions. 

\input{Tables/GroupElements/C4vGroupElements}

\input{Tables/GroupElements/C3vGroupElements}

\input{Tables/GroupElements/C2vCubicGroupElements}

\subsection{Basis vectors}\label{appendix:basis vectors}
The irreps of the continuum rotation group with $J = 0, \frac{1}{2},\frac{3}{2},...$, are defined on the $(2J + 1)$ dimensional space spanned by the basis vectors $\ket{JM}$. These basis vectors are the spin spherical harmonics and reduce to the standard spherical harmonics for $J \in \mathbb{Z}$. These representations, while irreducible in the continuum, become reducible into the 16 irreps of $^2O_h$. This is to say, that certain subspaces in the space spanned by $\ket{JM}$, are invariant under the symmetry transformations of the box. The basis vectors for this new subspace can be constructed from the projector
\beq\label{eq:projector}
P_{\Gamma}^{\lambda,\lambda'}\ket{JM} = \frac{d_\Gamma}{|G|}\sum_{g\in G}[D^{\Gamma}_{\lambda,\lambda'}(g)]^* \sum_{M'=-J}^J D^J_{M'M}(g)\ket{JM'}
\eeq 
where $\Gamma$ is the irrep, $\lambda$ a row from the irrep, $d_\Gamma$ the dimension of the irrep, $|G|$ is the order of the group, $D^\Gamma(g)$ the representation of group element $g$, and $D^J(g)$ the Wigner $D$ matrix evaluated at the Euler angles corresponding to group element $g$. Then, if we rotate a projected vector from row $\lambda$, then we get
\beq
S(g) P^{\lambda \lambda}_{\Gamma}\ket{JM} = D^{\Gamma}_{\lambda' \lambda}(g) P^{\lambda' \lambda}_{\Gamma}\ket{JM}.
\eeq
The recipe for calculating the basis vectors is then as follows. For a given $J$ and irrep $\Gamma$, we construct the projector for row 1. Then we perform a QR decomposition, $P^{11}_\Gamma=Q^\dagger R$, with $Q$ a unitary matrix and $R$ being upper triangular. If $R=0$ then this irrep does not couple to this $J$. If $R \neq 0$ then the $Q^\dagger$ and $R$ matrices will have the dimensions $(2J + 1) \times k$ and $k \times (2J+1)$, respectively. The rank, $k$, of these matrices is the number of linearly independent vectors for this $J$. This is what is meant by the multiplicity of the irrep in $J$.

The $k$ columns of $Q^\dagger\equiv\ket{\Gamma1Jn}$, with $n = 1,\cdots,k$ are the basis vectors corresponding to row 1. For a one-dimensional irrep, this is the end of the story. For higher dimensional irreps one can use these basis vectors to build the higher dimensions,
\beq
\ket{\Gamma \lambda' J n} = (P_{\Gamma}^{\lambda' 1} R^{-1})^T, \;\;\; \lambda' = 2,\cdots, d_\Gamma,
\eeq
where $RR^{-1} = I_{k\times k}$.

For improper rotations, one must insert a factor of $(-1)^J$ onto the Wigner $D$ function in Eq.(\ref{eq:projector}) to account for the parity of the spin-spherical harmonics. The basis vectors obtained from this procedure are orthonormal,
\beq
\braket{\Gamma'\lambda',J',n'|\Gamma \lambda J n} = \delta_{\Gamma \Gamma'}\delta_{\lambda \lambda'}\delta_{J J'}\delta_{n n'}.
\eeq

All the basis vectors up to $J = \frac{13}{2}$ are included below. Table \ref{tab:cubicbasisvectors} contains the vectors for $^2O_h$, Table \ref{tab:elongatedbasisvectors} for the $^2D_{4h}$ group in the elongated box, Table \ref{tab:c4vbasisvectors} for the little group $^2C_{4v}$ for both box geometries, Table \ref{tab:c3vbasisvectors} for the $^2C_{3v}$ little group in the cubic box, and Tables \ref{tab:c2vbasisvectorscubic} and \ref{tab:c2vbasisvectorselongated} contain the basis vectors for the $^2C_{2v}$ little group in the cubic and elongated boxes, respectively.

\input{Tables/BasisVectors/2OhBasisVectors}
\begin{widetext}
\input{Tables/BasisVectors/2D4hBasisVectors}
\input{Tables/BasisVectors/C4vBasisVectors}
\input{Tables/BasisVectors/C3vBasisVectors}
\input{Tables/BasisVectors/C2vCubicBasisVectors}
\input{Tables/BasisVectors/C2vElongatedBasisVectors}
\end{widetext}

\end{document}

%% file: Tables/CubicG1gChiConvergence.tex
\renewcommand{\arraystretch}{1.2}
\begin{table*}[!t]
\caption{Convergence of $G_{1g}$ by adding higher partial waves in the rest frame of cubic box, as measured by $\chi^2$ in Eq.\eqref{eq:chisq1}.  
The 4 orders are labeled by $Jln={1\over 2}01, {7\over 2}41, {9\over 2}41, {11\over 2}61 $ ($J$ total angular momentum, $\ell$ orbital angular momentum, $n$ multiplicity). 
All energy levels are represented by CM momentum $k$ in $\MeV$. 
The value marked with a red star (\textcolor{red}{*}) is a non-interacting level in the box replacing the `pinched' energy level not captured by the QC at this order. The last line is again the pinched 6th level but with the convergence measured by $\chi^2_2$  in Eq.\eqref{eq:chisq2}, see discussion in the text. 
}
\begin{tabular}{*{11}{c}}
\toprule
 \text{Level} & \text{Box L} & \text{Lattice} & \text{Order 1} & \text{Order 2} & \text{Order 3} & \text{Order 4} & $\chi ^2$\text{ order 1} & $\chi ^2$\text{ order 2} & $\chi ^2$\text{ order 3} & $\chi ^2$\text{ order 4} \\
 \hline
 1 & 9.35738 & 9.35737 & 9.35738 & 9.35738 & 9.35738 & 9.35738 & 0.00119665 & 0.000716157 & 0.0689808 & 0.0689826 \\
 2 & 39.0926 & 39.0926 & 39.0923 & 39.0924 & 39.0926 & 39.0926 & 13999.4 & 8759.09 & 0.00270611 & 0.00295507 \\
 3 & 55.3602 & 55.3602 & 55.36 & 55.36 & 55.3602 & 55.3602 & 4237.63 & 2659.14 & 2.7228 & 2.27047 \\
 4 & 65.2422 & 65.2422 & 65.228 & 65.2311 & 65.2422 & 65.2422 & $2.45\times 10^6$ & $1.51\times 10^6$ & 0.00339682 & 0.00415965 \\
 5 & 71.6554 & 71.6554 & 71.6529 & 71.6534 & 71.6554 & 71.6554 & 74016.4 & 45589.8 & 14.0837 & 11.9186 \\
 6 & 76.9113 & 76.9113 & $76.8827^{\textcolor{red}{*}}$ & 76.8915 & 76.9112 & 76.9112 &  & $1.72\times 10^7$ & 197.783 & 160.295 \\
 \hline
  6 & 76.9113 & 76.9113 & $76.8827^{\textcolor{red}{*}}$ & 76.8915 & 76.9112 & 76.9112 &  & 0.481 & $5.50\times 10^{-6}$ & $4.45\times 10^{-6}$\\
\end{tabular}
\label{tab:CubicG1gChiConvergence}

\end{table*}

%% file: Tables/CubicHuChiConvergence.tex
\begin{table*}[!t]
\caption{Similar to Table~\ref{tab:CubicG1gChiConvergence}, but for the $H_{u}$ irrep in the rest frame of the cubic box. 
The 5 orders are labeled by $Jln={3\over 2}11, {5\over 2}31, {7\over 2}31, {9\over 2}51, {11\over 2}51$.
All the $k_{lat}$ levels match the box levels to 6 decimal places so we omit the column to save space.}
\label{tab:CubicHuChiConvergence}
\begin{tabular}{*{12}{>{$}c<{$}}}

\toprule
 \text{Level} & \text{Box L}  & \text{Order 1} & \text{Order 2} & \text{Order 3} & \text{Order 4} & \text{Order 5} & \chi ^2\text{ order 1} & \chi ^2\text{ order 2} & \chi ^2\text{ order 3} & \chi ^2\text{ order 4} & \chi ^2\text{ order 5} \\
 1 & 35.365 & 35.3605 & 35.3627 & 35.365 & 35.365 & 35.365 & 2.21\times 10^7 & 6.12\times 10^6 & 12.25 & 10.67 & 0.49 \\
 2 & 48.6661 & 48.6249 & 48.634 & 48.6661 & 48.6661 & 48.6661 & 1.07\times 10^{11} & 6.51\times 10^{10} & 22232.7 & 11058.9 & 1.61 \\
 3 & 50.8628 & 50.8618 & 50.8623 & 50.8627 & 50.8627 & 50.8628 & 1.42\times 10^5 & 4.13 \times 10^4 & 302.29 & 152.49 & 3.18\times 10^{-4} \\
 4 & 61.2126  & 61.1715 & 61.1914 & 61.2123 & 61.2123 & 61.2126 & 1.55\times 10^8 & 4.11\times 10^7 & 7386.83 & 5711.04 & 0.0083 \\
 5 & 69.7484 & 69.6949 & 69.7211 & 69.7478 & 69.7479 & 69.7484 & 6.79\times 10^7 & 1.77\times 10^7 & 9118.62 & 7291.12 & 0.031 \\
 6 & 76.8847 & 76.8827^{\textcolor{red}{*}} & 76.8827^{\textcolor{red}{*}} & 76.8827 & 76.8833 & 76.8847 &  &  & 200496. & 96881.7 & 5.79\\
 7 & 77.0173 & 76.8827 & 77.0309 & 77.0156 & 77.0158 & 77.0173 & 3.03\times 10^8 & 3.12\times 10^6 & 47288.3 & 36145.9 & 0.31 \\
 8 & 77.2058 & 76.8827^{\textcolor{red}{*}} & 76.8827^{\textcolor{red}{*}} & 77.2056 & 77.2056 & 77.2058 &  &  & 2013.14 & 1476.96 & 0.90 \\
 \bottomrule
\end{tabular}
\end{table*}

%% file: Tables/CubicG2C4vConvergence.tex
\begin{table*}
\caption{Similar to Table~\ref{tab:CubicHuChiConvergence}, but for the $G_{2}$ irrep of the $^2C_{4v}$ little group for moving frame $\bm d=(0,0,1)$ in the cubic box.
The 5 orders are labeled in pairs by $Jln=({3\over 2}11, {3\over 2}21), ({5\over 2}22, {5\over 2}32), ({7\over 2}32, {7\over 2}42), ({9\over 2}42, {9\over 2}52), ({11\over 2}53, {1\over 2}63)$.
}
\label{tab:CubicG2C4vChiConvergence}
\begin{tabular}{*{12}{>{$}c<{$}}}
\renewcommand{\arraystretch}{3.0}
 \text{Level} & \text{Box L} & \text{Order 1} & \text{Order 2} & \text{Order 3} & \text{Order 4} & \text{Order 5} & \chi ^2\text{ order 1} & \chi ^2\text{ order 2} & \chi ^2\text{ order 3} & \chi ^2\text{ order 4} & \chi ^2\text{order 5} \\
 \hline
 1 & 34.8229 & 34.7139 & 34.8224 & 34.8228 & 34.8229 & 34.8229 & 1.44\times 10^{10} & 271490. & 2310.95 & 0.305 & 0.059 \\
 2 & 35.6299 & 35.6194 & 35.6278 & 35.6299 & 35.6299 & 35.6299 & 1.16\times 10^8 & 4.82\times 10^6 & 259.206 & 7.36 & 0.185 \\
 3 & 45.7387 & 45.6487 & 45.7258 & 45.7382 & 45.7387 & 45.7387 & 1.47\times 10^{10} & 3.00\times 10^8 & 338550. & 17.8 & 0.118 \\
 4 & 46.2681 & 46.1087 & 46.2616 & 46.268 & 46.2681 & 46.2681 & 4.52\times 10^{10} & 7.35\times 10^7 & 11384.5 & 190.2 & 0.0659 \\
 5 & 49.1788 & 48.9400 & 49.1773 & 49.1783 & 49.1788 & 49.1788 & 1.13\times 10^{11} & 5.01\times 10^6 & 498884. & 33.4 & 1.35 \\
 6 & 49.8823 & 49.852 & 49.8695 & 49.8821 & 49.8823 & 49.8823 & 5.52\times 10^8 & 9.87\times 10^7 & 31533.1 & 576.1 & 0.0240 \\
 7 & 51.9355 & 51.8563 & 51.918 & 51.9342 & 51.9355 & 51.9355 & 2.76\times 10^9 & 1.35\times 10^8 & 726572. & 425.8 & 0.016 \\
 8 & 52.7478 & 52.559 & 52.7399 & 52.7476 & 52.7478 & 52.7478 & 6.94\times 10^9 & 1.22\times 10^7 & 5631.9 & 77.1 & 0.0567 \\
 9 & 57.4367 & 57.2073 & 57.4012 & 57.4361 & 57.4367 & 57.4367 & 1.05\times 10^{11} & 2.52\times 10^9 & 644423. & 120.7 & 11.6 \\
 10 & 58.2843 & 58.0724 & 58.2793 & 58.2832 & 58.2842 & 58.2843 & 1.02\times 10^{10} & 5.68\times 10^6 & 246925. & 1087.4 & 1.56 \\
 11 & 62.4873 & 62.2651 & 62.4413 & 62.4845 & 62.4873 & 62.4873 & 2.39\times 10^{10} & 1.02\times 10^9 & 3.71\times 10^6 & 143.38 & 2.79 \\
 12 & 63.2379 & 62.9901 & 63.2149 & 63.2373 & 63.2377 & 63.2379 & 6.37\times 10^9 & 5.52\times 10^7 & 44844.8 & 3643.91 & 4.38 \\
 13 & 69.4900 & 69.1112 & 69.4810 & 69.4859 & 69.4900 & 69.4900 & 2.12\times 10^9 & 1.20\times 10^6 & 250210. & 65.4251 & 17.9 \\
 14 & 69.8728 & 69.7941 & 69.8465 & 69.872 & 69.8723 & 69.8728 & 1.46\times 10^8 & 1.64\times 10^7 & 14002.8 & 6350.0 & 0.38 \\
 15 & 73.0315 & 73.1414 & 73.0074 & 73.0255 & 73.0308 & 73.0315 & 5.88\times 10^8 & 2.84\times 10^7 & 1.79\times 10^6 & 29245.6 & 127.4 \\
 16 & 73.3718 & 72.9724^{\textcolor{red}{*}} & 73.2949 & 73.362 & 73.3714 & 73.3718 &  & 2.20\times 10^8 & 3.58\times 10^6 & 8575.0 & 8.36 \\
 17 & 75.5192 & 75.4572 & 75.4358 & 75.5181 & 75.5189 & 75.5192 & 7.20\times 10^7 & 1.30\times 10^8 & 25722.9 & 2744.0 & 89.8 \\
 18 & 75.7081 & 75.0176^{\textcolor{red}{*}} & 75.6852 & 75.703 & 75.7072 & 75.708 &  & 1.40\times 10^7 & 701237. & 21335. & 56.9 \\
 19 & 77.1195 & 77.0086^{\textcolor{red}{*}} & 77.0916 & 77.1172 & 77.1175 & 77.1195 &  & 1.69\times 10^7 & 110736. & 89199.0 & 3.24 \\
 20 & 77.5309 & 77.4664 & 77.5121 & 77.5246 & 77.5308 & 77.5308 & 5.77\times 10^7 & 4.91\times 10^6 & 552083. & 254.78 & 211.15 \\
 21 & 77.8651 & 77.0086^{\textcolor{red}{*}} & 77.8441 & 77.8517 & 77.865 & 77.865 &  & 1.29\times 10^7 & 5.22\times 10^6 & 384.6 & 84.12 \\
 22 & 78.2886 & 78.1414 & 78.2036 & 78.285 & 78.2875 & 78.2885 & 5.95\times 10^8 & 1.98\times 10^8 & 341964. & 33215.6 & 60.91 \\
\bottomrule
\end{tabular}
\end{table*}

%% file: Tables/ChiSquareSummary.tex
\renewcommand{\arraystretch}{1.7}
\begin{table*}[!t]
\caption{Summary of convergence order by order according to the average $\chi^2$ measure in Eq.\eqref{eq:chisq1} for all the QCs discussed in this work. The orders in each QC are labeled by total angular momentum and multiplicity as $J(n)$ with orbital angular momentum implicit ($\ell=J\pm 1/2$). For rest frames, $\ell$ is even for g irreps and odd for u irreps. For moving frames, both $\ell$ values are allowed. The $N$ is the number of levels inspected. Both the QC matrices and detailed convergence data can be found in the supplement~\cite{supp}.}
\label{tab:ChiSquareSummary}
\begin{tabular}{*{12}{>{$}c<{$}}}
\multicolumn{12}{C}{\textbf{Cubic Box}}\\
 \text{No.} & \text{d} & \text{Group} & \Gamma & \text{J(n)} & \text{N} & \text{Order 1} & \text{Order 2} & \text{Order 3} & \text{Order 4} & \text{Order 5} & \text{Order 6}\\
 \hline
  1 & \text{(0,0,0)} & ^2O_h & G_{1g} & \frac{1}{2},\frac{7}{2},\frac{9}{2},\frac{11}{2}\text{,...} & 6 & 4.25\times 10^5 & 3.13\times 10^6 & 35.78 & 29.09 &  &  \\
 2 &   &   & G_{1u} & \frac{1}{2},\frac{7}{2},\frac{9}{2},\frac{11}{2}\text{,...} & 5 & 1.40\times 10^7 & 26755. & 17769.9 & 0.0048 &  &  \\
 3 &   &   & G_{2g} & \frac{5}{2},\frac{7}{2},\frac{11}{2},\frac{13}{2}\text{(2),...} & 2 & 2.59\times 10^6 & 72.31 & 72.09 & 0.18 &  &  \\
 4 &   &   & G_{2u} & \frac{5}{2},\frac{7}{2},\frac{11}{2},\frac{13}{2}\text{(2),...} & 4 & 2.35\times 10^{10} & 242891. & 0.45 & 0.43 &  &  \\
 5 &   &   & H_g & \frac{3}{2},\frac{5}{2},\frac{7}{2},\frac{9}{2}\text{(2),}\frac{11}{2}\text{(2),...} & 8 & 1.58\times 10^{10} & 2.85\times 10^6 & 2.55\times 10^6 & 88.21 & 75.86 &  \\
 6 &   &   & H_u & \frac{3}{2},\frac{5}{2},\frac{7}{2},\frac{9}{2}\text{(2),}\frac{11}{2}\text{(2),...} & 8 & 1.35\times 10^{10} & 8.15\times 10^{9} & 36105. & 19840. & 1.14 &  \\
  7 & \text{(0,0,1)} & ^2C_{4v} & G_1 & \frac{1}{2},\frac{3}{2},\frac{5}{2},\frac{7}{2}\text{(2),}\frac{9}{2}\text{(3),}\frac{11}{2}\text{(3),...} & 27 & 1.56\times 10^{10} & 5.22\times 10^{8} & 3.00\times 10^7 & 1.46\times 10^6 & 4751.48. & 10.23 \\
 8 &   &   & G_2 & \frac{3}{2},\frac{5}{2}\text{(2),}\frac{7}{2}\text{(2),}\frac{9}{2}\text{(2),}\frac{11}{2}\text{(3),...} & 23 & 1.50\times 10^{10} & 2.12\times 10^8 & 8.17\times 10^5 & 8595.13 & 29.67 &  \\
 9 & \text{(1,1,1)} & ^2C_{3v} & G_1 & \frac{1}{2},\frac{3}{2},\frac{5}{2}\text{(2),}\frac{7}{2}\text{(3),}\frac{9}{2}\text{(3),}\frac{11}{2}\text{(4),...} & 34 & 7.97\times 10^{9} & 7.06\times 10^{8} & 5.03\times 10^7 & 3.47\times 10^5 & 3578.44 & 12.11 \\
 10 &   &   & K_1 & \frac{3}{2},\frac{5}{2},\frac{7}{2},\frac{9}{2}\text{(2),}\frac{11}{2}\text{(2),...} & 16 & 2.71\times 10^{9} & 2.99\times 10^7 & 5.91\times 10^5 & 1834.35 & 6.35 & \\
 11 &   &   & K_2 & \frac{3}{2},\frac{5}{2},\frac{7}{2},\frac{9}{2}\text{(2),}\frac{11}{2}\text{(2),...} & 16 & 2.71\times 10^{9} & 2.99\times 10^7 & 5.91\times 10^5 & 1834.35 & 6.35 & \\
 12 & \text{(1,1,0)} & ^2C_{2v} & G_{1} & \frac{1}{2},\frac{3}{2}\text{(2),}\frac{5}{2}\text{(3),}\frac{7}{2}\text{(4),}\frac{9}{2}\text{(5),}\frac{11}{2}\text{(6),...} & 50 & 1.29\times 10^{10} & 3.56\times 10^{9} & 5.18\times 10^7 & 6.96\times 10^5 & 5229.88 & 16.49 \\
 \hline
  \multicolumn{12}{C}{\textbf{Elongated Box}}\\
  \text{No.} & \text{d} & \text{Group} & \Gamma & \text{J(n)} & \text{N} & \text{Order 1} & \text{Order 2} & \text{Order 3} & \text{Order 4} & \text{Order 5} & \text{Order 6}\\
 \hline
 13 & \text{(0,0,0)} & ^2D_{4h} & G_{1g} & \frac{1}{2},\frac{3}{2},\frac{5}{2},\frac{7}{2}\text{(2),}\frac{9}{2}\text{(3),}\frac{11}{2}\text{(3),...} & 14 & 5.04\times 10^{9} & 8.86\times 10^{9} & 4.17\times 10^6 & 3.09\times 10^6 & 13.54 & 11.47\\
 14 &   &   & G_{1u} & \frac{1}{2},\frac{3}{2},\frac{5}{2},\frac{7}{2}\text{(2),}\frac{9}{2}\text{(3),}\frac{11}{2}\text{(3),...} & 14 & 2.81\times 10^{10} & 1.29\times 10^8 & 9.29\times 10^6 & 5579.91 & 4141.93 & 0.029 \\
 15 &   &   & G_{2g} & \frac{3}{2},\frac{5}{2}\text{(2),}\frac{7}{2}\text{(2),}\frac{9}{2}\text{(2),}\frac{11}{2}\text{(3),...} & 11 & 1.40\times 10^{10} & 1.57\times 10^7 & 9.45\times 10^6 & 41.26 & 36.13 &  \\
 16 &   &   & G_{2u} & \frac{3}{2},\frac{5}{2}\text{(2),}\frac{7}{2}\text{(2),}\frac{9}{2}\text{(2),}\frac{11}{2}\text{(3),...} & 11 & 1.90\times 10^8 & 9.36\times 10^{9} & 36881.45 & 27336.73 & 0.99 &  \\
 17 & \text{(0,0,1)} & ^2C_{4v} & G_1 & \frac{1}{2},\frac{3}{2},\frac{5}{2},\frac{7}{2}\text{(2),}\frac{9}{2}\text{(3),}\frac{11}{2}\text{(3),...} & 28 & 1.24\times 10^{10} & 1.41\times 10^{9} & 2.98\times 10^7 & 181071. & 799.89 & 1.01 \\
 18 &   &   & G_2 & \frac{3}{2},\frac{5}{2}\text{(2),}\frac{7}{2}\text{(2),}\frac{9}{2}\text{(2),}\frac{11}{2}\text{(3),...} & 22 & 1.95\times 10^{10} & 2.94\times 10^8 & 1.35\times 10^6 & 178.32 & 2.63 &  \\
 19 & \text{(1,1,0)} & ^2C_{2v} & G_1 & \frac{1}{2},\frac{3}{2},\frac{5}{2},\frac{7}{2}\text{(2),}\frac{9}{2}\text{(3),}\frac{11}{2}\text{(3),...} & 50 & 2.20\times 10^{10} & 3.08\times 10^{9} & 2.90\times 10^7 & 2.34\times 10^5 & 993.22 & 2.21 \\
 \hline
\end{tabular}
\end{table*}

%% file: Tables/MatrixElements/CubicRestMatrixElements.tex
\begin{longtable*}{CCCCCCCL}
\caption{Matrix elements for rest frame in cubic box (group $^2O_{h}$) 
in each irrep. Since the matrix elements between the even and odd parity irreps can only differ by a minus sign we have combined the tables. The top sign and left $l$ value corresponds to the even $g$ irrep while the bottom sign and the right $l$ in parenthesis gives the odd $u$ irrep. 
}
\label{tab:CubicRestMatrixElements}
\renewcommand{\arraystretch}{2.0}
\endfirsthead
\multicolumn{7}{c} {{\tablename\ \thetable{:} \it Continued}} \\
\hline \endhead
\hline \\ \endfoot
\hline \\ \endlastfoot
\toprule
\Gamma & J & l & n & J' & l' & n' & M^\Gamma_{JlnJ'l'n'} \\
\hline
  G_{1g/u} & \frac{1}{2} & \text{0(1)} & 1 & \frac{1}{2} & \text{0(1)} & 1 & \text{wr}_{00} \\
 \text{} & \frac{7}{2} & \text{4(3)} & 1 & \frac{1}{2} & \text{0(1)} & 1 & \mp \frac{4 \text{wr}_{40}}{\sqrt{21}} \\
 \text{} & \frac{7}{2} & \text{4(3)} & 1 & \frac{7}{2} & \text{4(3)} & 1 & \text{wr}_{00}+\frac{6 \text{wr}_{40}}{11}+\frac{100 \text{wr}_{60}}{33 \sqrt{13}} \\
 \text{} & \frac{9}{2} & \text{4(5)} & 1 & \frac{1}{2} & \text{0(1)} & 1 & 2 \sqrt{\frac{5}{21}} \text{wr}_{40} \\
 \text{} & \frac{9}{2} & \text{4(5)} & 1 & \frac{7}{2} & \text{4(3)} & 1 & \mp \left(\frac{12}{143} \sqrt{5} \text{wr}_{40}+\frac{56}{33} \sqrt{\frac{5}{13}} \text{wr}_{60}+\frac{224}{143}
   \sqrt{\frac{5}{17}} \text{wr}_{80}\right) \\
 \text{} & \frac{9}{2} & \text{4(5)} & 1 & \frac{9}{2} & \text{4(5)} & 1 & \text{wr}_{00}+\frac{84 \text{wr}_{40}}{143}+\frac{128 \text{wr}_{60}}{33 \sqrt{13}}+\frac{112 \text{wr}_{80}}{143
   \sqrt{17}} \\
\hline
 G_{2g/u} & \frac{5}{2} & \text{2(3)} & 1 & \frac{5}{2} & \text{2(3)} & 1 & \text{wr}_{00}-\frac{4 \text{wr}_{40}}{7} \\
\hline
\end{longtable*}

%% file: Tables/MatrixElements/C4vCubicMatrixElements.tex
\begin{widetext}
\begin{longtable*}{CCCCCCCL}
\caption{Matrix elements for moving frame $d=(0,0,1)$ in cubic box (little-group $C_{4v}$) up to $J=5/2$.  
}
\label{tab:C4vCubicM}
\renewcommand{\arraystretch}{2.0}
\endfirsthead
\multicolumn{7}{c} {{\tablename\ \thetable{:} \it Continued}} \\
\hline \endhead
\hline \\ \endfoot
\hline \\ \endlastfoot
\toprule
\Gamma & J & l & n & J' & l' & n' & M^\Gamma_{JlnJ'l'n'} \\
\hline
G_1 & \frac{1}{2} & 0 & 1 & \frac{1}{2} & 0 & 1 & \text{wr}_{00} \\
 \text{} & \frac{1}{2} & 1 & 1 & \frac{1}{2} & 0 & 1 & -\frac{i }{\sqrt{3}}\text{wr}_{10} \\
 \text{} & \frac{1}{2} & 1 & 1 & \frac{1}{2} & 1 & 1 & \text{wr}_{00} \\
 \text{} & \frac{3}{2} & 1 & 1 & \frac{1}{2} & 0 & 1 & i \sqrt{\frac{2}{3}} \text{wr}_{10} \\
 \text{} & \frac{3}{2} & 1 & 1 & \frac{1}{2} & 1 & 1 & -\sqrt{\frac{2}{5}} \text{wr}_{20} \\
 \text{} & \frac{3}{2} & 1 & 1 & \frac{3}{2} & 1 & 1 & \text{wr}_{00}+\frac{1}{\sqrt{5}}\text{wr}_{20} \\
 \text{} & \frac{3}{2} & 2 & 1 & \frac{1}{2} & 0 & 1 & \sqrt{\frac{2}{5}} \text{wr}_{20} \\
\hline
\end{longtable*}
\end{widetext}

%% file: Tables/MatrixElements/C3vCubicMatrixElements.tex
\begin{longtable*}{CCCCCCCL}
\caption{Matrix elements for moving frame $d=(1,1,1)$ in cubic box (little-group $C_{3v}$) for the three lowest partial waves in each irrep. The elements of $K_1$ and $K_2$ are related by an overall sign on certain elements. We condense these into one table where the top sign is for $K_1$ and the bottom sign is for $K_2$. Both $K_1$ and $K_2$ couple to the same $J(l)$'s.  
}
\label{tab:C3vCubicM}
\renewcommand{\arraystretch}{2.0}
\endfirsthead
\multicolumn{7}{c} {{\tablename\ \thetable{:} \it Continued}} \\
\hline \endhead
\hline \\ \endfoot
\hline \\ \endlastfoot
\toprule
\Gamma & J & l & n & J' & l' & n' & M^\Gamma_{JlnJ'l'n'} \\
\hline
 G_1 & \frac{1}{2} & 0 & 1 & \frac{1}{2} & 0 & 1 & \text{wr}_{00} \\
 \text{} & \frac{1}{2} & 1 & 1 & \frac{1}{2} & 0 & 1 & -\frac{\frac{1}{3}+i}{\sqrt{2}}\text{wr}_{10} \\
 \text{} & \frac{1}{2} & 1 & 1 & \frac{1}{2} & 1 & 1 & \text{wr}_{00} \\
 \text{} & \frac{3}{2} & 1 & 1 & \frac{1}{2} & 0 & 1 & \left(\frac{5}{6}+\frac{i}{2}\right) \text{wr}_{10} \\
 \text{} & \frac{3}{2} & 1 & 1 & \frac{1}{2} & 1 & 1 & \frac{2-i}{\sqrt{15}}\text{wr}_{21} \\
 \text{} & \frac{3}{2} & 1 & 1 & \frac{3}{2} & 1 & 1 & \text{wr}_{00}-\sqrt{\frac{2}{15}} \text{wr}_{21} \\
 \text{} & \frac{3}{2} & 2 & 1 & \frac{1}{2} & 0 & 1 & -\frac{1-2 i}{\sqrt{15}}\text{wr}_{21} \\
\end{longtable*}

%% file: Tables/MatrixElements/C2vCubicMatrixElements.tex
\begin{longtable*}{CCCCCCCL}
\caption{Matrix elements for moving frame $d=(1,1,0)$ in cubic box (little-group $^2C_{2v}$) up to $J=5/2$.  
}
\label{tab:C2vCubicM}
\renewcommand{\arraystretch}{2.0}
\endfirsthead
\multicolumn{7}{c} {{\tablename\ \thetable{:} \it Continued}} \\
\hline \endhead
\hline \\ \endfoot
\hline \\ \endlastfoot
\toprule
\Gamma & J & l & n & J' & l' & n' & M^\Gamma_{JlnJ'l'n'} \\
\hline
 G_1 & \frac{1}{2} & 0 & 1 & \frac{1}{2} & 0 & 1 & \text{wr}_{00} \\
 \text{} & \frac{1}{2} & 1 & 1 & \frac{1}{2} & 0 & 1 & (-1+i) \sqrt{\frac{2}{3}} \text{wr}_{11} \\
 \text{} & \frac{1}{2} & 1 & 1 & \frac{1}{2} & 1 & 1 & \text{wr}_{00} \\
 \text{} & \frac{3}{2} & 1 & 1 & \frac{1}{2} & 0 & 1 & (-1-i) \text{wr}_{11} \\
 \text{} & \frac{3}{2} & 1 & 1 & \frac{1}{2} & 1 & 1 & \frac{2 i}{\sqrt{5}}\text{wi}_{22} \\
 \text{} & \frac{3}{2} & 1 & 1 & \frac{3}{2} & 1 & 1 & \text{wr}_{00}-\frac{1}{\sqrt{5}}\text{wr}_{20} \\
 \text{} & \frac{3}{2} & 2 & 1 & \frac{1}{2} & 0 & 1 & -\sqrt{\frac{2}{5}} \text{wr}_{20} \\
\end{longtable*}

%% file: Tables/MatrixElements/ElongatedRestMatrixElements.tex
\begin{longtable*}{CCCCCCCL}
\caption{Matrix elements for rest frame in $z$-elongated box (group $^2D_{4h}$) up to the third partial wave in each irrep. Since the matrix elements between the even and odd parity irreps can only differ by a minus sign we have combined the tables. The top sign and left $l$ value corresponds to the even $g$ irrep while the bottom sign and the right $l$ in parenthesis gives the odd $u$ irrep. 
}
\label{tab:ElongatedRestMatrixElements}
\renewcommand{\arraystretch}{2.0}
\endfirsthead
\multicolumn{7}{c} {{\tablename\ \thetable{:} \it Continued}} \\
\hline \endhead
\hline \\ \endfoot
\hline \\ \endlastfoot
\toprule
\Gamma & J & l & n & J' & l' & n' & M^\Gamma_{JlnJ'l'n'} \\
\hline
G_{1g/u} & \frac{1}{2} & \text{0(1)} & 1 & \frac{1}{2} & \text{0(1)} & 1 & \text{wr}_{00} \\
 \text{} & \frac{3}{2} & \text{2(1)} & 1 & \frac{1}{2} & \text{0(1)} & 1 & \pm \left(\sqrt{\frac{2}{5}} \text{wr}_{20}\right) \\
 \text{} & \frac{3}{2} & \text{2(1)} & 1 & \frac{3}{2} & \text{2(1)} & 1 & \text{wr}_{00}+\frac{\text{wr}_{20}}{\sqrt{5}} \\
 \text{} & \frac{5}{2} & \text{2(3)} & 1 & \frac{1}{2} & \text{0(1)} & 1 & -\sqrt{\frac{3}{5}} \text{wr}_{20} \\
 \text{} & \frac{5}{2} & \text{2(3)} & 1 & \frac{3}{2} & \text{2(1)} & 1 & \mp \left(\frac{1}{7} \sqrt{\frac{6}{5}} \text{wr}_{20}+\frac{2}{7} \sqrt{6} \text{wr}_{40}\right) \\
 \text{} & \frac{5}{2} & \text{2(3)} & 1 & \frac{5}{2} & \text{2(3)} & 1 & \text{wr}_{00}+\frac{8 \text{wr}_{20}}{7 \sqrt{5}}+\frac{2 \text{wr}_{40}}{7} \\
 \hline
 G_{2g/u} & \frac{3}{2} & \text{2(1)} & 1 & \frac{3}{2} & \text{2(1)} & 1 & \text{wr}_{00}-\frac{\text{wr}_{20}}{\sqrt{5}} \\
\end{longtable*}

%% file: Tables/MatrixElements/C4vElongatedMatrixElements.tex
\begin{longtable*}{CCCCCCCL}
\caption{Matrix elements for moving frame $d=(0,0,1)$ in the $z$-elongated box (little-group $C_{4v}$) up to $J=7/2$.  
}
\label{tab:C4vElongatedM}
\renewcommand{\arraystretch}{2.0}
\endfirsthead
\multicolumn{7}{c} {{\tablename\ \thetable{:} \it Continued}} \\
\hline \endhead
\hline \\ \endfoot
\hline \\ \endlastfoot
\toprule
\Gamma & J & l & n & J' & l' & n' & M^\Gamma_{JlnJ'l'n'} \\
\hline
G_1 & \frac{1}{2} & 0 & 1 & \frac{1}{2} & 0 & 1 & \text{wr}_{00} \\
 \text{} & \frac{1}{2} & 1 & 1 & \frac{1}{2} & 0 & 1 & -\frac{i }{\sqrt{3}}\text{wr}_{10} \\
 \text{} & \frac{1}{2} & 1 & 1 & \frac{1}{2} & 1 & 1 & \text{wr}_{00} \\
 \text{} & \frac{3}{2} & 1 & 1 & \frac{1}{2} & 0 & 1 & i \sqrt{\frac{2}{3}} \text{wr}_{10} \\
 \text{} & \frac{3}{2} & 1 & 1 & \frac{1}{2} & 1 & 1 & -\sqrt{\frac{2}{5}} \text{wr}_{20} \\
 \text{} & \frac{3}{2} & 1 & 1 & \frac{3}{2} & 1 & 1 & \text{wr}_{00}+\frac{1}{\sqrt{5}}\text{wr}_{20} \\
 \text{} & \frac{3}{2} & 2 & 1 & \frac{1}{2} & 0 & 1 & \sqrt{\frac{2}{5}} \text{wr}_{20} \\
\end{longtable*}

%% file: Tables/MatrixElements/C2vElongatedMatrixElements.tex
\begin{longtable*}{CCCCCCCL}
\caption{Matrix elements for rest frame in the $z$-elongated box (group $^2C_{2v}$) up to $J=5/2$.
}
\label{tab:C2vElongatedMatrixElements}
\renewcommand{\arraystretch}{2.0}
\endfirsthead
\multicolumn{7}{c} {{\tablename\ \thetable{:} \it Continued}} \\
\hline \endhead
\hline \\ \endfoot
\hline \\ \endlastfoot
\toprule
\Gamma & J & l & n & J' & l' & n' & M^\Gamma_{JlnJ'l'n'} \\
\hline
  G_1 & \frac{1}{2} & 0 & 1 & \frac{1}{2} & 0 & 1 & \text{wr}_{00} \\
 \text{} & \frac{1}{2} & 1 & 1 & \frac{1}{2} & 0 & 1 & (1+i) \sqrt{\frac{2}{3}} \text{wr}_{11} \\
 \text{} & \frac{1}{2} & 1 & 1 & \frac{1}{2} & 1 & 1 & \text{wr}_{00} \\
 \text{} & \frac{3}{2} & 1 & 1 & \frac{1}{2} & 0 & 1 & -\frac{(1+i) }{\sqrt{3}}\text{wr}_{11} \\
 \text{} & \frac{3}{2} & 1 & 1 & \frac{1}{2} & 1 & 1 & \sqrt{\frac{2}{5}} \text{wr}_{20}+i \sqrt{\frac{3}{5}} \text{wr}_{21} \\
 \text{} & \frac{3}{2} & 1 & 1 & \frac{3}{2} & 1 & 1 & \text{wr}_{00}+\frac{1}{\sqrt{5}}\text{wr}_{20} \\
 \text{} & \frac{3}{2} & 2 & 1 & \frac{1}{2} & 0 & 1 & -\frac{2 i}{\sqrt{5}}\text{wi}_{22} \\
\end{longtable*}

%% file: Tables/GroupElements/2OhGroupElements.tex
\begin{longtable*}{C|CCCCCCCCCCCCC|CCCCCCCC}
\caption{Group elements of $^2O_h$}
\label{tab:cubicgroupelements}
\renewcommand{\arraystretch}{2.0}
\endfirsthead
\multicolumn{7}{c} {{\tablename\ \thetable{:} \it Continued}} \\
\hline \endhead
\hline \\ \endfoot
\hline \\ \endlastfoot
\toprule
 \text{k} & \text{}^2O_h & \text{n} & \omega  & S_k & \text{$\{\alpha $,$\beta $,$\gamma \}$} & A_{1g} & A_{2g} & E_g & T_{1g} & T_{2g} & G_{1g} & G_{2g} & H_g & A_{1u} & A_{2u} & E_u & T_{1u} & T_{2u} & G_{1u} & G_{2u} & H_u \\
 \hline
 1 & \mathbb{E} & \text{$\{$1, 0, 0$\}$} & 4 \pi  & \text{t}_1 & \{0,0,0\} & 1 & 1 & \mathbf{1} & \text{t}_1 & \text{t}_1 & \text{g}_1 & \text{g}_1 & \text{h}_1 & 1 & 1 & \mathbf{1} & \text{t}_1 & \text{t}_1 & \text{g}_1 & \text{g}_1 & \text{h}_1 \\
 2 & \textit{C}_{\text{2x}} & \text{$\{$1, 0, 0$\}$} & \pi  & \text{t}_2 & \{0,\pi ,\pi \} & 1 & 1 & \mathbf{1} & \text{t}_2 & \text{t}_2 & \text{g}_2 & \text{g}_2 & \text{h}_2 & 1 & 1 & \mathbf{1} & \text{t}_2 & \text{t}_2 & \text{g}_2 & \text{g}_2 & \text{h}_2 \\
 3 & \textit{C}_{\text{2y}} & \text{$\{$0, 1, 0$\}$} & \pi  & \text{t}_3 & \{0,\pi ,0\} & 1 & 1 & \mathbf{1} & \text{t}_3 & \text{t}_3 & \text{g}_3 & \text{g}_3 & \text{h}_3 & 1 & 1 & \mathbf{1} & \text{t}_3 & \text{t}_3 & \text{g}_3 & \text{g}_3 & \text{h}_3 \\
 4 & \textit{C}_{\text{2z}} & \text{$\{$0, 0, 1$\}$} & \pi  & \text{t}_4 & \{0,0,\pi \} & 1 & 1 & \mathbf{1} & \text{t}_4 & \text{t}_4 & \text{g}_4 & \text{g}_4 & \text{h}_4 & 1 & 1 & \mathbf{1} & \text{t}_4 & \text{t}_4 & \text{g}_4 & \text{g}_4 & \text{h}_4 \\
 5 & \textit{C}_{31}^+ & \text{$\{$1, 1, 1$\}$} & \frac{2 \pi }{3} & \text{t}_5 & \left\{0,\frac{\pi }{2},\frac{\pi }{2}\right\} & 1 & 1 & e_3 & \text{t}_5 & \text{t}_5 & \text{g}_5 & \text{g}_5 & \text{h}_5 & 1 & 1 & e_3 & \text{t}_5 & \text{t}_5 & \text{g}_5 & \text{g}_5 & \text{h}_5 \\
 6 & \textit{C}_{32}^+ & \text{$\{$-1, -1, 1$\}$} & \frac{2 \pi }{3} & \text{t}_6 & \left\{\pi ,\frac{\pi }{2},\frac{7 \pi }{2}\right\} & 1 & 1 & e_3 & \text{t}_6 & \text{t}_6 & \text{g}_6 & \text{g}_6 & \text{h}_6 & 1 & 1 & e_3 & \text{t}_6 & \text{t}_6 & \text{g}_6 & \text{g}_6 & \text{h}_6 \\
\end{longtable*}

%% file: Tables/GroupElements/D4hGroupElements.tex
\begin{table*}
\caption{Group table of  $D_{4h}$ for rest frame in elongated box. The inversion is labeled by preceding with the letter $I$ in the $k$ and $O_{h}$ columns. The $E$ representations are given in terms of Pauli matrices.}
\label{tab:D4hgroupelements}
              
$                                
\renewcommand{\arraystretch}{1.15}
\begin{array}{c | c c c c c c c c c c c c| c c c c c c c}
\toprule
 \text{k} & \text{}^2D_{4h} & \text{n} & \omega  & S_k & \text{$\{\alpha $,$\beta $,$\gamma \}$} & A_{1g} & A_{2g} & B_{1g} & B_{2g} & E_g & G_{1g} & G_{2g} & A_{1u} & A_{2u} & B_{1u} & B_{2u} & E_u & G_{1u} & G_{2u} \\
 1 & \mathbb{E} & \text{$\{$1, 0, 0$\}$} & 4 \pi  & \text{t}_1 & \{0,0,0\} & 1 & 1 & 1 & 1 & \mathbf{1} & \text{g}_1 & \text{g}_1 & 1 & 1 & 1 & 1 & \mathbf{1} & \text{g}_1 & \text{g}_1 \\
 2 & \text{C}_{\text{2x}} & \text{$\{$1, 0, 0$\}$} & \pi  & \text{t}_2 & \{0,\pi ,\pi \} & 1 & 1 & -1 & -1 & \sigma _3 & \text{g}_2 & \text{g}_2 & 1 & 1 & -1 & -1 & \sigma _3 & \text{g}_2 & \text{g}_2 \\
 3 & \text{C}_{\text{2y}} & \text{$\{$0, 1, 0$\}$} & \pi  & \text{t}_3 & \{0,\pi ,0\} & 1 & 1 & -1 & -1 & -\sigma _3 & \text{g}_3 & \text{g}_3 & 1 & 1 & -1 & -1 & -\sigma _3 & \text{g}_3 & \text{g}_3 \\
 4 & \text{C}_{\text{2z}} & \text{$\{$0, 0, 1$\}$} & \pi  & \text{t}_4 & \{0,0,\pi \} & 1 & 1 & 1 & 1 & -\mathbf{1} & \text{g}_4 & \text{g}_4 & 1 & 1 & 1 & 1 & -\mathbf{1} & \text{g}_4 & \text{g}_4 \\
 5 & \text{C}_{\text{4z}}^+ & \text{$\{$0, 0, 1$\}$} & \frac{\pi }{2} & \text{t}_{15} & \left\{0,0,\frac{\pi }{2}\right\} & 1 & -1 & -1 & 1 & -i \sigma _2 & \text{g}_{15} & \text{-g}_{15} & 1 & -1 & -1 & 1 & -i \sigma _2 & \text{g}_{15} & \text{-g}_{15} \\
 6 & \text{C}_{\text{4z}}^- & \text{$\{$0, 0, -1$\}$} & \frac{\pi }{2} & \text{t}_{18} & \left\{0,0,\frac{7 \pi }{2}\right\} & 1 & -1 & -1 & 1 & i \sigma _2 & \text{g}_{18} & \text{-g}_{18} & 1 & -1 & -1 & 1 & i \sigma _2 & \text{g}_{18} & \text{-g}_{18} \\
 7 & \text{C}_{\text{2a}} & \text{$\{$1, 1, 0$\}$} & \pi  & \text{t}_{19} & \left\{0,\pi ,\frac{\pi }{2}\right\} & 1 & -1 & 1 & -1 & \sigma _1 & \text{g}_{19} & \text{-g}_{19} & 1 & -1 & 1 & -1 & \sigma _1 & \text{g}_{19} & \text{-g}_{19} \\
 8 & \text{C}_{\text{2b}} & \text{$\{$1, -1, 0$\}$} & \pi  & \text{t}_{20} & \left\{0,\pi ,\frac{7 \pi }{2}\right\} & 1 & -1 & 1 & -1 & -\sigma _1 & \text{g}_{20} & \text{-g}_{20} & 1 & -1 & 1 & -1 & -\sigma _1 & \text{g}_{20} & \text{-g}_{20} \\
\hline
 9 & \tilde{\mathbb{E}} & \text{$\{$1, 0, 0$\}$} & 2 \pi  & \text{t}_1 & \{0,0,2 \pi \} & 1 & 1 & 1 & 1 & \mathbf{1} & \text{-g}_1 & \text{-g}_1 & 1 & 1 & 1 & 1 & \mathbf{1} & \text{-g}_1 & \text{-g}_1 \\
 10 & \tilde{\text{C}_{\text{2x}}} & \text{$\{$1, 0, 0$\}$} & 3 \pi  & \text{t}_2 & \{0,\pi ,3 \pi \} & 1 & 1 & -1 & -1 & \sigma _3 & \text{-g}_2 & \text{-g}_2 & 1 & 1 & -1 & -1 & \sigma _3 & \text{-g}_2 & \text{-g}_2 \\
 11 & \tilde{\text{C}_{\text{2y}}} & \text{$\{$0, 1, 0$\}$} & 3 \pi  & \text{t}_3 & \{0,\pi ,2 \pi \} & 1 & 1 & -1 & -1 & -\sigma _3 & \text{-g}_3 & \text{-g}_3 & 1 & 1 & -1 & -1 & -\sigma _3 & \text{-g}_3 & \text{-g}_3 \\
 12 & \tilde{\text{C}_{\text{2z}}} & \text{$\{$0, 0, 1$\}$} & 3 \pi  & \text{t}_4 & \{0,0,3 \pi \} & 1 & 1 & 1 & 1 & -\mathbf{1} & \text{-g}_4 & \text{-g}_4 & 1 & 1 & 1 & 1 & -\mathbf{1} & \text{-g}_4 & \text{-g}_4 \\
 13 & \tilde{\text{C}_{\text{4z}}^+} & \text{$\{$0, 0, 1$\}$} & \frac{5 \pi }{2} & \text{t}_{15} & \left\{0,0,\frac{5 \pi }{2}\right\} & 1 & -1 & -1 & 1 & -i \sigma _2 & \text{-g}_{15} & \text{g}_{15} & 1 & -1 & -1 & 1 & -i \sigma _2 & \text{-g}_{15} & \text{g}_{15} \\
 14 & \tilde{\text{C}_{\text{4z}}^-} & \text{$\{$0, 0, -1$\}$} & \frac{5 \pi }{2} & \text{t}_{18} & \left\{0,0,\frac{3 \pi }{2}\right\} & 1 & -1 & -1 & 1 & i \sigma _2 & \text{-g}_{18} & \text{g}_{18} & 1 & -1 & -1 & 1 & i \sigma _2 & \text{-g}_{18} & \text{g}_{18} \\
 15 & \tilde{\text{C}_{\text{2a}}} & \text{$\{$1, 1, 0$\}$} & 3 \pi  & \text{t}_{19} & \left\{0,\pi ,\frac{5 \pi }{2}\right\} & 1 & -1 & 1 & -1 & \sigma _1 & \text{-g}_{19} & \text{g}_{19} & 1 & -1 & 1 & -1 & \sigma _1 & \text{-g}_{19} & \text{g}_{19} \\
 16 & \tilde{\text{C}_{\text{2b}}} & \text{$\{$1, -1, 0$\}$} & 3 \pi  & \text{t}_{20} & \left\{0,\pi ,\frac{3 \pi }{2}\right\} & 1 & -1 & 1 & -1 & -\sigma _1 & \text{-g}_{20} & \text{g}_{20} & 1 & -1 & 1 & -1 & -\sigma _1 & \text{-g}_{20} & \text{g}_{20} \\
 \hline
 17 & \text{I$\mathbb{E}$} & \text{$\{$1, 0, 0$\}$} & 4 \pi  & \text{-t}_1 & \{0,0,0\} & 1 & 1 & -1 & -1 & \mathbf{1} & \text{g}_1 & \text{g}_1 & -1 & -1 & 1 & 1 & -\mathbf{1} & \text{-g}_1 & \text{-g}_1 \\
 18 & \text{IC}_{\text{2x}} & \text{$\{$1, 0, 0$\}$} & \pi  & \text{-t}_2 & \{0,\pi ,\pi \} & 1 & 1 & 1 & 1 & \sigma _3 & \text{g}_2 & \text{g}_2 & -1 & -1 & -1 & -1 & -\sigma _3 & \text{-g}_2 & \text{-g}_2 \\
 19 & \text{IC}_{\text{2y}} & \text{$\{$0, 1, 0$\}$} & \pi  & \text{-t}_3 & \{0,\pi ,0\} & 1 & 1 & 1 & 1 & -\sigma _3 & \text{g}_3 & \text{g}_3 & -1 & -1 & -1 & -1 & \sigma _3 & \text{-g}_3 & \text{-g}_3 \\
 20 & \text{IC}_{\text{2z}} & \text{$\{$0, 0, 1$\}$} & \pi  & \text{-t}_4 & \{0,0,\pi \} & 1 & 1 & -1 & -1 & -\mathbf{1} & \text{g}_4 & \text{g}_4 & -1 & -1 & 1 & 1 & \mathbf{1} & \text{-g}_4 & \text{-g}_4 \\
 21 & \text{IC}_{\text{4z}}^+ & \text{$\{$0, 0, 1$\}$} & \frac{\pi }{2} & \text{-t}_{15} & \left\{0,0,\frac{\pi }{2}\right\} & 1 & -1 & 1 & -1 & -i \sigma _2 & \text{g}_{15} & \text{-g}_{15} & -1 & 1 & -1 & 1 & i \sigma _2 & \text{-g}_{15} & \text{g}_{15} \\
 22 & \text{IC}_{\text{4z}}^- & \text{$\{$0, 0, -1$\}$} & \frac{\pi }{2} & \text{-t}_{18} & \left\{0,0,\frac{7 \pi }{2}\right\} & 1 & -1 & 1 & -1 & i \sigma _2 & \text{g}_{18} & \text{-g}_{18} & -1 & 1 & -1 & 1 & -i \sigma _2 & \text{-g}_{18} & \text{g}_{18} \\
 23 & \text{IC}_{\text{2a}} & \text{$\{$1, 1, 0$\}$} & \pi  & \text{-t}_{19} & \left\{0,\pi ,\frac{\pi }{2}\right\} & 1 & -1 & -1 & 1 & \sigma _1 & \text{g}_{19} & \text{-g}_{19} & -1 & 1 & 1 & -1 & -\sigma _1 & \text{-g}_{19} & \text{g}_{19} \\
 24 & \text{IC}_{\text{2b}} & \text{$\{$1, -1, 0$\}$} & \pi  & \text{-t}_{20} & \left\{0,\pi ,\frac{7 \pi }{2}\right\} & 1 & -1 & -1 & 1 & -\sigma _1 & \text{g}_{20} & \text{-g}_{20} & -1 & 1 & 1 & -1 & \sigma _1 & \text{-g}_{20} & \text{g}_{20} \\
 \hline
 25 & \tilde{\text{I$\mathbb{E}$}} & \text{$\{$1, 0, 0$\}$} & 2 \pi  & \text{-t}_1 & \{0,0,2 \pi \} & 1 & 1 & -1 & -1 & \mathbf{1} & \text{-g}_1 & \text{-g}_1 & -1 & -1 & 1 & 1 & -\mathbf{1} & \text{g}_1 & \text{g}_1 \\
 26 & \tilde{\text{IC}_{\text{2x}}} & \text{$\{$1, 0, 0$\}$} & 3 \pi  & \text{-t}_2 & \{0,\pi ,3 \pi \} & 1 & 1 & 1 & 1 & \sigma _3 & \text{-g}_2 & \text{-g}_2 & -1 & -1 & -1 & -1 & -\sigma _3 & \text{g}_2 & \text{g}_2 \\
 27 & \tilde{\text{IC}_{\text{2y}}} & \text{$\{$0, 1, 0$\}$} & 3 \pi  & \text{-t}_3 & \{0,\pi ,2 \pi \} & 1 & 1 & 1 & 1 & -\sigma _3 & \text{-g}_3 & \text{-g}_3 & -1 & -1 & -1 & -1 & \sigma _3 & \text{g}_3 & \text{g}_3 \\
 28 & \tilde{\text{IC}_{\text{2z}}} & \text{$\{$0, 0, 1$\}$} & 3 \pi  & \text{-t}_4 & \{0,0,3 \pi \} & 1 & 1 & -1 & -1 & -\mathbf{1} & \text{-g}_4 & \text{-g}_4 & -1 & -1 & 1 & 1 & \mathbf{1} & \text{g}_4 & \text{g}_4 \\
 29 & \tilde{\text{IC}_{\text{4z}}^+} & \text{$\{$0, 0, 1$\}$} & \frac{5 \pi }{2} & \text{-t}_{15} & \left\{0,0,\frac{5 \pi }{2}\right\} & 1 & -1 & 1 & -1 & -i \sigma _2 & \text{-g}_{15} & \text{g}_{15} & -1 & 1 & -1 & 1 & i \sigma _2 & \text{g}_{15} & \text{-g}_{15} \\
 30 & \tilde{\text{IC}_{\text{4z}}^-} & \text{$\{$0, 0, -1$\}$} & \frac{5 \pi }{2} & \text{-t}_{18} & \left\{0,0,\frac{3 \pi }{2}\right\} & 1 & -1 & 1 & -1 & i \sigma _2 & \text{-g}_{18} & \text{g}_{18} & -1 & 1 & -1 & 1 & -i \sigma _2 & \text{g}_{18} & \text{-g}_{18} \\
 31 & \tilde{\text{IC}_{\text{2a}}} & \text{$\{$1, 1, 0$\}$} & 3 \pi  & \text{-t}_{19} & \left\{0,\pi ,\frac{5 \pi }{2}\right\} & 1 & -1 & -1 & 1 & \sigma _1 & \text{-g}_{19} & \text{g}_{19} & -1 & 1 & 1 & -1 & -\sigma _1 & \text{g}_{19} & \text{-g}_{19} \\
 32 & \tilde{\text{IC}_{\text{2b}}} & \text{$\{$1, -1, 0$\}$} & 3 \pi  & \text{-t}_{20} & \left\{0,\pi ,\frac{3 \pi }{2}\right\} & 1 & -1 & -1 & 1 & -\sigma _1 & \text{-g}_{20} & \text{g}_{20} & -1 & 1 & 1 & -1 & \sigma _1 & \text{g}_{20} & \text{-g}_{20} \\
 \bottomrule
\end{array}
$  
\end{table*}

%% file: Tables/GroupElements/C4vGroupElements.tex
\begin{longtable*}{C|CCCCCCCCCCCC}
\caption{Group elements of $^2C_{4v}$ for both cubic and elongated boxes. The $E$ representation is given in terms of the Pauli matrices.}
\label{tab:c4vgroupelements}
\renewcommand{\arraystretch}{2.0}
\endfirsthead
\multicolumn{7}{c} {{\tablename\ \thetable{:} \it Continued}} \\
\hline \endhead
\hline \\ \endfoot
\hline \\ \endlastfoot
\toprule
  \text{k} & \text{}^2C_{4v} & \text{n} & \omega  & S_k & \text{$\{\alpha $,$\beta $,$\gamma \}$} & A_1 & A_2 & B_1 & B_2 & \text{E} & G_1 & G_2 \\
 1 & \mathbb{E} & \text{$\{$1, 0, 0$\}$} & 4 \pi  & \text{t}_1 & \{0,0,0\} & 1 & 1 & 1 & 1 & \mathbf{1} & \text{g}_1 & \text{g}_1 \\
 2 & \text{C}_{\text{2z}} & \text{$\{$0, 0, 1$\}$} & \pi  & \text{t}_4 & \{0,0,\pi \} & 1 & 1 & 1 & 1 & -\mathbf{1} & \text{g}_4 & \text{g}_4 \\
 3 & \text{C}_{\text{4z}}^+ & \text{$\{$0, 0, 1$\}$} & \frac{\pi }{2} & \text{t}_{15} & \left\{0,0,\frac{\pi }{2}\right\} & 1 & -1 & -1 & 1 & i \sigma _2 & \text{g}_{15} & \text{-g}_{15} \\
 4 & \text{C}_{\text{4z}}^- & \text{$\{$0, 0, -1$\}$} & \frac{\pi }{2} & \text{t}_{18} & \left\{0,0,\frac{7 \pi }{2}\right\} & 1 & -1 & -1 & 1 & -i \sigma _2 & \text{g}_{18} & \text{-g}_{18} \\
 5 & \tilde{\mathbb{E}} & \text{$\{$1, 0, 0$\}$} & 2 \pi  & \text{t}_1 & \{0,0,2 \pi \} & 1 & 1 & 1 & 1 & \mathbf{1} & \text{-g}_1 & \text{-g}_1 \\
 6 & \tilde{\text{C}_{\text{2z}}} & \text{$\{$0, 0, 1$\}$} & 3 \pi  & \text{t}_4 & \{0,0,3 \pi \} & 1 & 1 & 1 & 1 & -\mathbf{1} & \text{-g}_4 & \text{-g}_4 \\
 7 & \tilde{\text{C}_{\text{4z}}^+} & \text{$\{$0, 0, 1$\}$} & \frac{5 \pi }{2} & \text{t}_{15} & \left\{0,0,\frac{5 \pi }{2}\right\} & 1 & -1 & -1 & 1 & i \sigma _2 & \text{-g}_{15} & \text{g}_{15} \\
 8 & \tilde{\text{C}_{\text{4z}}^-} & \text{$\{$0, 0, -1$\}$} & \frac{5 \pi }{2} & \text{t}_{18} & \left\{0,0,\frac{3 \pi }{2}\right\} & 1 & -1 & -1 & 1 & -i \sigma _2 & \text{-g}_{18} & \text{g}_{18} \\
 9 & \text{IC}_{\text{2x}} & \text{$\{$1, 0, 0$\}$} & \pi  & \text{-t}_2 & \{0,\pi ,\pi \} & 1 & 1 & -1 & -1 & \sigma _3 & \text{g}_2 & \text{g}_2 \\
 10 & \text{IC}_{\text{2y}} & \text{$\{$0, 1, 0$\}$} & \pi  & \text{-t}_3 & \{0,\pi ,0\} & 1 & 1 & -1 & -1 & -\sigma _3 & \text{g}_3 & \text{g}_3 \\
 11 & \text{IC}_{\text{2a}} & \text{$\{$1, 1, 0$\}$} & \pi  & \text{-t}_{19} & \left\{0,\pi ,\frac{\pi }{2}\right\} & 1 & -1 & 1 & -1 & -\sigma _1 & \text{g}_{19} & \text{-g}_{19} \\
 12 & \text{IC}_{\text{2b}} & \text{$\{$1, -1, 0$\}$} & \pi  & \text{-t}_{20} & \left\{0,\pi ,\frac{7 \pi }{2}\right\} & 1 & -1 & 1 & -1 & \sigma _1 & \text{g}_{20} & \text{-g}_{20} \\
 13 & \tilde{\text{IC}_{\text{2x}}} & \text{$\{$1, 0, 0$\}$} & 3 \pi  & \text{-t}_2 & \{0,\pi ,3 \pi \} & 1 & 1 & -1 & -1 & \sigma _3 & \text{-g}_2 & \text{-g}_2 \\
 14 & \tilde{\text{IC}_{\text{2y}}} & \text{$\{$0, 1, 0$\}$} & 3 \pi  & \text{-t}_3 & \{0,\pi ,2 \pi \} & 1 & 1 & -1 & -1 & -\sigma _3 & \text{-g}_3 & \text{-g}_3 \\
 15 & \tilde{\text{IC}_{\text{2a}}} & \text{$\{$1, 1, 0$\}$} & 3 \pi  & \text{-t}_{19} & \left\{0,\pi ,\frac{5 \pi }{2}\right\} & 1 & -1 & 1 & -1 & -\sigma _1 & \text{-g}_{19} & \text{g}_{19} \\
 16 & \tilde{\text{IC}_{\text{2b}}} & \text{$\{$1, -1, 0$\}$} & 3 \pi  & \text{-t}_{20} & \left\{0,\pi ,\frac{3 \pi }{2}\right\} & 1 & -1 & 1 & -1 & \sigma _1 & \text{-g}_{20} & \text{g}_{20} \\
\end{longtable*}

%% file: Tables/GroupElements/C3vGroupElements.tex
\begin{longtable*}{C|CCCCCCCCCCC}
\caption{Group elements of $^2C_{3v}$ in the cubic box.}
\label{tab:c3vgroupelements}
\renewcommand{\arraystretch}{2.0}
\endfirsthead
\multicolumn{7}{c} {{\tablename\ \thetable{:} \it Continued}} \\
\hline \endhead
\hline \\ \endfoot
\hline \\ \endlastfoot
\toprule
 \text{k} & \text{}^2C_{3v} & \text{n} & \omega  & S_k & \text{$\{\alpha $,$\beta $,$\gamma \}$} & A_1 & A_2 & K_1 & K_2 & \text{E} & G_1 \\
 1 & \mathbb{E} & \text{$\{$1, 0, 0$\}$} & 4 \pi  & \text{t}_1 & \{0,0,0\} & 1 & 1 & 1 & 1 & \mathbf{1} & \text{g}_1 \\
 2 & \text{C}_{31}^+ & \text{$\{$1, 1, 1$\}$} & \frac{2 \pi }{3} & \text{t}_5 & \left\{0,\frac{\pi }{2},\frac{\pi }{2}\right\} & 1 & 1 & -1 & -1 & e_3 & \text{g}_5 \\
 3 & \text{C}_{31}^- & \text{$\{$-1, -1, -1$\}$} & \frac{2 \pi }{3} & \text{t}_9 & \left\{\frac{\pi }{2},\frac{\pi }{2},3 \pi \right\} & 1 & 1 & -1 & -1 & e_2 & \text{g}_9 \\
 4 & \tilde{\mathbb{E}} & \text{$\{$1, 0, 0$\}$} & 2 \pi  & \text{t}_1 & \{0,0,2 \pi \} & 1 & 1 & -1 & -1 & \mathbf{1} & \text{-g}_1 \\
 5 & \tilde{\text{C}_{31}^+} & \text{$\{$1, 1, 1$\}$} & \frac{8 \pi }{3} & \text{t}_5 & \left\{0,\frac{\pi }{2},\frac{5 \pi }{2}\right\} & 1 & 1 & 1 & 1 & e_3 & \text{-g}_5 \\
 6 & \tilde{\text{C}_{31}^-} & \text{$\{$-1, -1, -1$\}$} & \frac{8 \pi }{3} & \text{t}_9 & \left\{\frac{\pi }{2},\frac{\pi }{2},\pi \right\} & 1 & 1 & 1 & 1 & e_2 & \text{-g}_9 \\
 7 & \text{IC}_{\text{2b}} & \text{$\{$1, -1, 0$\}$} & \pi  & \text{-t}_{20} & \left\{0,\pi ,\frac{7 \pi }{2}\right\} & 1 & -1 & -i & i & e_4 & \text{g}_{20} \\
 8 & \text{IC}_{\text{2e}} & \text{$\{$1, 0, -1$\}$} & \pi  & \text{-t}_{23} & \left\{\pi ,\frac{\pi }{2},0\right\} & 1 & -1 & i & -i & e_1 & \text{g}_{23} \\
 9 & \text{IC}_{\text{2f}} & \text{$\{$0, 1, -1$\}$} & \pi  & \text{-t}_{24} & \left\{\frac{3 \pi }{2},\frac{\pi }{2},\frac{7 \pi }{2}\right\} & 1 & -1 & -i & i & e_5 & \text{g}_{24} \\
 10 & \tilde{\text{IC}_{\text{2b}}} & \text{$\{$1, -1, 0$\}$} & 3 \pi  & \text{-t}_{20} & \left\{0,\pi ,\frac{3 \pi }{2}\right\} & 1 & -1 & i & -i & e_4 & \text{-g}_{20} \\
 11 & \tilde{\text{IC}_{\text{2e}}} & \text{$\{$1, 0, -1$\}$} & 3 \pi  & \text{-t}_{23} & \left\{\pi ,\frac{\pi }{2},2 \pi \right\} & 1 & -1 & -i & i & e_1 & \text{-g}_{23} \\
 12 & \tilde{\text{IC}_{\text{2f}}} & \text{$\{$0, 1, -1$\}$} & 3 \pi  & \text{-t}_{24} & \left\{\frac{3 \pi }{2},\frac{\pi }{2},\frac{3 \pi }{2}\right\} & 1 & -1 & i & -i & e_5 & \text{-g}_{24} \\
 \end{longtable*}

%% file: Tables/GroupElements/C2vCubicGroupElements.tex
\begin{longtable*}{C|CCCCCCCCCC}
\caption{Group elements of $^2C_{2v}$ in both the cubic and elongated boxes.}
\label{tab:c2vcubicgroupelements}
\renewcommand{\arraystretch}{2.0}
\endfirsthead
\multicolumn{7}{c} {{\tablename\ \thetable{:} \it Continued}} \\
\hline \endhead
\hline \\ \endfoot
\hline \\ \endlastfoot
\toprule
 \text{k} & \text{}^2C_{2v} & \text{n} & \omega  & S_k & \text{$\{\alpha $,$\beta $,$\gamma \}$} & A_1 & A_2 & B_1 & B_2 & G_1 \\
 1 & \mathbb{E} & \text{$\{$1, 0, 0$\}$} & 4 \pi  & \text{t}_1 & \{0,0,0\} & 1 & 1 & 1 & 1 & \text{g}_1 \\
 2 & \text{C}_{\text{2a}} & \text{$\{$1, 1, 0$\}$} & \pi  & \text{t}_{19} & \left\{0,\pi ,\frac{\pi }{2}\right\} & 1 & -1 & -1 & 1 & \text{g}_{19} \\
 3 & \tilde{\mathbb{E}} & \text{$\{$1, 0, 0$\}$} & 2 \pi  & \text{t}_1 & \{0,0,2 \pi \} & 1 & 1 & 1 & 1 & \text{-g}_1 \\
 4 & \tilde{\text{C}_{\text{2a}}} & \text{$\{$1, 1, 0$\}$} & 3 \pi  & \text{t}_{19} & \left\{0,\pi ,\frac{5 \pi }{2}\right\} & 1 & -1 & -1 & 1 & \text{-g}_{19} \\
 5 & \text{IC}_{\text{2z}} & \text{$\{$0, 0, 1$\}$} & \pi  & \text{-t}_4 & \{0,0,\pi \} & 1 & 1 & -1 & -1 & \text{-g}_4 \\
 6 & \text{IC}_{\text{2b}} & \text{$\{$1, -1, 0$\}$} & \pi  & \text{-t}_{20} & \left\{0,\pi ,\frac{7 \pi }{2}\right\} & 1 & -1 & 1 & -1 & \text{-g}_{20} \\
 7 & \tilde{\text{IC}_{\text{2z}}} & \text{$\{$0, 0, 1$\}$} & 3 \pi  & \text{-t}_4 & \{0,0,3 \pi \} & 1 & 1 & -1 & -1 & \text{g}_4 \\
 8 & \tilde{\text{IC}_{\text{2b}}} & \text{$\{$1, -1, 0$\}$} & 3 \pi  & \text{-t}_{20} & \left\{0,\pi ,\frac{3 \pi }{2}\right\} & 1 & -1 & 1 & -1 & \text{g}_{20} \\
\end{longtable*}

%% file: Tables/BasisVectors/2OhBasisVectors.tex
\begin{longtable*}{CCCCC}
\caption{Basis vectors of $^2O_h$ in terms of spherical harmonics $Y(J,M)$ up to $J = \frac{13}{2}$ for the rest frame of the cubic box. Multiplicities are labeled by $n$ and multidimensional components are indicated by $\alpha$.}
\label{tab:cubicbasisvectors}
\renewcommand{\arraystretch}{2.0}
\endfirsthead
\multicolumn{5}{C} {{\tablename\ \thetable{:} \it Continued}} \\
\hline \endhead
\hline \\ \endfoot
\hline \\ \endlastfoot
\toprule
 \Gamma  & \text{J}  & \text{n} & \alpha  & \text{Basis Vector} \\
 \hline
\text{G1} & \frac{1}{2} & 1 & 1 & Y\left(\frac{1}{2},\frac{1}{2}\right) \\
 \text{} & \text{} & \text{} & 2 & Y\left(\frac{1}{2},-\frac{1}{2}\right) \\
 \text{G1} & \frac{7}{2} & 1 & 1 & \frac{1}{2} \sqrt{\frac{5}{3}} Y\left(\frac{7}{2},-\frac{7}{2}\right)+\frac{1}{2} \sqrt{\frac{7}{3}} Y\left(\frac{7}{2},\frac{1}{2}\right) \\
 \text{} & \text{} & \text{} & 2 & -\frac{1}{2} \sqrt{\frac{7}{3}} Y\left(\frac{7}{2},-\frac{1}{2}\right)-\frac{1}{2} \sqrt{\frac{5}{3}} Y\left(\frac{7}{2},\frac{7}{2}\right) \\
 \text{G1} & \frac{9}{2} & 1 & 1 & \frac{Y\left(\frac{9}{2},-\frac{7}{2}\right)}{2 \sqrt{6}}+\frac{1}{2} \sqrt{\frac{7}{3}} Y\left(\frac{9}{2},\frac{1}{2}\right)+\frac{1}{2} \sqrt{\frac{3}{2}}
   Y\left(\frac{9}{2},\frac{9}{2}\right) \\
 \text{} & \text{} & \text{} & 2 & \frac{1}{2} \sqrt{\frac{3}{2}} Y\left(\frac{9}{2},-\frac{9}{2}\right)+\frac{1}{2} \sqrt{\frac{7}{3}}
   Y\left(\frac{9}{2},-\frac{1}{2}\right)+\frac{Y\left(\frac{9}{2},\frac{7}{2}\right)}{2 \sqrt{6}} \\
 \text{G1} & \frac{11}{2} & 1 & 1 & \frac{1}{4} \sqrt{\frac{35}{3}} Y\left(\frac{11}{2},-\frac{7}{2}\right)-\frac{Y\left(\frac{11}{2},\frac{1}{2}\right)}{2 \sqrt{2}}+\frac{1}{4} \sqrt{\frac{7}{3}}
   Y\left(\frac{11}{2},\frac{9}{2}\right) \\
\end{longtable*}

%% file: Tables/BasisVectors/2D4hBasisVectors.tex
\begin{longtable}{CCCCC|CCCCC}
\caption{Basis vectors of $^2D_{4h}$ in terms of spi spherical harmonics $Y(J,M)$ up to $J = \frac{13}{2}$ for the rest frame of the elongated box. Multiplicities are labeled by $n$ and multidimensional components are indicated by $\alpha$.}
\label{tab:elongatedbasisvectors}
\renewcommand{\arraystretch}{2.0}
\endfirsthead
\hline \endhead
\hline \\ \endfoot
\hline \\ \endlastfoot
\toprule
 \Gamma  & \text{J}  & \text{n} & \alpha  & \text{Basis Vector} & \Gamma  & \text{J} & \text{n} & \alpha  & \text{Basis Vector}\\
 \text{G1} & \frac{1}{2} & 1 & 1 & Y\left(\frac{1}{2},\frac{1}{2}\right) & \text{G2} & \frac{3}{2} & 1 & 1 & Y\left(\frac{3}{2},-\frac{3}{2}\right) \\
 \text{} & \text{} & \text{} & 2 & Y\left(\frac{1}{2},-\frac{1}{2}\right) & \text{} & \text{} & \text{} & 2 & Y\left(\frac{3}{2},\frac{3}{2}\right) \\
 \text{G1} & \frac{3}{2} & 1 & 1 & Y\left(\frac{3}{2},\frac{1}{2}\right) & \text{G2} & \frac{5}{2} & 1 & 1 & Y\left(\frac{5}{2},-\frac{3}{2}\right) \\
 \text{} & \text{} & \text{} & 2 & -Y\left(\frac{3}{2},-\frac{1}{2}\right) & \text{} & \text{} & \text{} & 2 & -Y\left(\frac{5}{2},\frac{3}{2}\right) \\
 \text{G1} & \frac{5}{2} & 1 & 1 & Y\left(\frac{5}{2},\frac{1}{2}\right) & \text{G2} & \frac{5}{2} & 2 & 1 & Y\left(\frac{5}{2},\frac{5}{2}\right) \\
 \text{} & \text{} & \text{} & 2 & Y\left(\frac{5}{2},-\frac{1}{2}\right) & \text{} & \text{} & \text{} & 2 & -Y\left(\frac{5}{2},-\frac{5}{2}\right) \\
 \text{G1} & \frac{7}{2} & 1 & 1 & Y\left(\frac{7}{2},-\frac{7}{2}\right) & \text{G2} & \frac{7}{2} & 1 & 1 & Y\left(\frac{7}{2},-\frac{3}{2}\right) \\
\end{longtable}

%% file: Tables/BasisVectors/C4vBasisVectors.tex
\begin{longtable}{CCCCCC|CCCCCC}
\caption{Basis vectors of $^2C_{4v}$ in terms of spherical harmonics $Y(J,M)$ up to $J = \frac{13}{2}$ In both cubic and $z$-elongated boxes. Multiplicities are labeled by $n$ and multidimensional components are indicated by $\alpha$.}
\label{tab:c4vbasisvectors}
\renewcommand{\arraystretch}{2.0}
\endfirsthead
\hline \endhead
\hline \\ \endfoot
\hline \\ \endlastfoot
\toprule
 \Gamma  & \text{J} & \text{l} & \text{n} & \alpha  & \text{Basis Vector} & \Gamma  & \text{J} & \text{l} & \text{n} & \alpha  & \text{Basis Vector} \\
\text{G1} & \frac{1}{2} & 0 & 1 & 1 & Y\left(\frac{1}{2},\frac{1}{2}\right) & \text{G2} & \frac{3}{2} & 1 & 1 & 1 & Y\left(\frac{3}{2},-\frac{3}{2}\right) \\
 \text{} & \text{} & \text{} & \text{} & 2 & -Y\left(\frac{1}{2},-\frac{1}{2}\right) & \text{} & \text{} & \text{} & \text{} & 2 & -Y\left(\frac{3}{2},\frac{3}{2}\right) \\
 \text{G1} & \frac{1}{2} & 1 & 1 & 1 & Y\left(\frac{1}{2},\frac{1}{2}\right) & \text{G2} & \frac{3}{2} & 2 & 1 & 1 & Y\left(\frac{3}{2},-\frac{3}{2}\right) \\
 \text{} & \text{} & \text{} & \text{} & 2 & Y\left(\frac{1}{2},-\frac{1}{2}\right) & \text{} & \text{} & \text{} & \text{} & 2 & Y\left(\frac{3}{2},\frac{3}{2}\right) \\
 \text{G1} & \frac{3}{2} & 1 & 1 & 1 & Y\left(\frac{3}{2},\frac{1}{2}\right) & \text{G2} & \frac{5}{2} & 2 & 1 & 1 & Y\left(\frac{5}{2},-\frac{3}{2}\right) \\
 \text{} & \text{} & \text{} & \text{} & 2 & -Y\left(\frac{3}{2},-\frac{1}{2}\right) & \text{} & \text{} & \text{} & \text{} & 2 & -Y\left(\frac{5}{2},\frac{3}{2}\right) \\
 \text{G1} & \frac{3}{2} & 2 & 1 & 1 & Y\left(\frac{3}{2},\frac{1}{2}\right) & \text{G2} & \frac{5}{2} & 3 & 1 & 1 & Y\left(\frac{5}{2},-\frac{3}{2}\right) \\
\end{longtable}

%% file: Tables/BasisVectors/C3vBasisVectors.tex
\begin{longtable}{CCCCCC}
\caption{Basis vectors of $^2C_{3v}$ in terms of spherical harmonics $Y(J,M)$ up to $J = \frac{13}{2}$ in the cubic box. Multiplicities are labeled by $n$ and multidimensional components are indicated by $\alpha$. The $K_2$ basis vectors are the same as the $K_1$ basis vectors simply swapping the $l$ for the corresponding $J$. For example, the basis vectors for $J = 3/2, l = 1$ in $K_1$ are the same as the $J = 3/2, l = 2$ basis vectors in $K_2$.}
\label{tab:c3vbasisvectors}
\renewcommand{\arraystretch}{2.0}
\endfirsthead
\multicolumn{6}{C} {{\tablename\ \thetable{:} \it Continued}} \\
\hline \endhead
\hline \\ \endfoot
\hline \\ \endlastfoot
\toprule
 \text{G1} & \frac{1}{2} & 0 & 1 & 1 & Y\left(\frac{1}{2},\frac{1}{2}\right) \\
 \text{} & \text{} & \text{} & \text{} & 2 & Y\left(\frac{1}{2},-\frac{1}{2}\right) \\
 \text{G1} & \frac{1}{2} & 1 & 1 & 1 & \sqrt{\frac{2}{3}} Y\left(\frac{1}{2},-\frac{1}{2}\right)+\frac{(1-i) }{\sqrt{6}}Y\left(\frac{1}{2},\frac{1}{2}\right) \\
 \text{} & \text{} & \text{} & \text{} & 2 & -\frac{(1-i) }{\sqrt{6}}Y\left(\frac{1}{2},-\frac{1}{2}\right)-i \sqrt{\frac{2}{3}} Y\left(\frac{1}{2},\frac{1}{2}\right) \\
 \text{G1} & \frac{3}{2} & 1 & 1 & 1 & \frac{1}{\sqrt{6}}Y\left(\frac{3}{2},-\frac{1}{2}\right)+\frac{(1-i) }{\sqrt{6}}Y\left(\frac{3}{2},\frac{1}{2}\right)+\frac{i }{\sqrt{2}}Y\left(\frac{3}{2},\frac{3}{2}\right) \\
 \text{} & \text{} & \text{} & \text{} & 2 & \frac{1}{\sqrt{2}}Y\left(\frac{3}{2},-\frac{3}{2}\right)+\frac{(1-i) }{\sqrt{6}}Y\left(\frac{3}{2},-\frac{1}{2}\right)+\frac{i}{\sqrt{6}} Y\left(\frac{3}{2},\frac{1}{2}\right) \\
 \text{G1} & \frac{3}{2} & 2 & 1 & 1 & \frac{1}{\sqrt{3}}Y\left(\frac{3}{2},-\frac{3}{2}\right)+\left(\frac{1}{2}-\frac{i}{2}\right) Y\left(\frac{3}{2},-\frac{1}{2}\right)+\frac{\left(\frac{1}{2}+\frac{i}{2}\right) }{\sqrt{3}}Y\left(\frac{3}{2},\frac{3}{2}\right) \\
 \text{} & \text{} & \text{} & \text{} & 2 & -\frac{\left(\frac{1}{2}-\frac{i}{2}\right) }{\sqrt{3}}Y\left(\frac{3}{2},-\frac{3}{2}\right)-\left(\frac{1}{2}+\frac{i}{2}\right) Y\left(\frac{3}{2},\frac{1}{2}\right)+\frac{1}{\sqrt{3}}Y\left(\frac{3}{2},\frac{3}{2}\right) \\
\end{longtable}

%% file: Tables/BasisVectors/C2vCubicBasisVectors.tex
\begin{longtable}{CCCCCC|CCCCCC|CCCCCC}
\caption{Basis vectors of $^2C_{2v}$ in terms of spherical harmonics $Y(J,M)$ up to $J = \frac{13}{2}$ in the cubic box. Multiplicities are labeled by $n$ and multidimensional components are indicated by $\alpha$.}
\label{tab:c2vbasisvectorscubic}
\renewcommand{\arraystretch}{2.0}
\endfirsthead
\multicolumn{6}{C} {{\tablename\ \thetable{:} \it Continued}} \\
\hline \endhead
\hline \\ \endfoot
\hline \\ \endlastfoot
\toprule
\Gamma  & \text{J} & \text{l} & \text{n} & \alpha  & \text{Basis Vector} & \Gamma  & \text{J} & \text{l} & \text{n} & \alpha  & \text{Basis Vector} & \Gamma  & \text{J} & \text{l} & \text{n} & \alpha  & \text{Basis Vector}\\
\text{G1} & \frac{1}{2} & 0 & 1 & 1 & Y\left(\frac{1}{2},-\frac{1}{2}\right) & \text{G1} & \frac{7}{2} & 4 & 4 & 1 & Y\left(\frac{7}{2},\frac{7}{2}\right) & \text{G1} & \frac{11}{2} & 5 & 5 & 1 &
   Y\left(\frac{11}{2},\frac{5}{2}\right) \\
 \text{} & \text{} & \text{} & \text{} & 2 & -i Y\left(\frac{1}{2},\frac{1}{2}\right) & \text{} & \text{} & \text{} & \text{} & 2 & i Y\left(\frac{7}{2},-\frac{7}{2}\right) & \text{} & \text{} & \text{}
   & \text{} & 2 & Y\left(\frac{11}{2},-\frac{5}{2}\right) \\
 \text{G1} & \frac{1}{2} & 1 & 1 & 1 & Y\left(\frac{1}{2},\frac{1}{2}\right) & \text{G1} & \frac{9}{2} & 4 & 1 & 1 & Y\left(\frac{9}{2},-\frac{9}{2}\right) & \text{G1} & \frac{11}{2} & 6 & 5 & 1 &
   Y\left(\frac{11}{2},\frac{7}{2}\right) \\
 \text{} & \text{} & \text{} & \text{} & 2 & Y\left(\frac{1}{2},-\frac{1}{2}\right) & \text{} & \text{} & \text{} & \text{} & 2 & -i Y\left(\frac{9}{2},\frac{9}{2}\right) & \text{} & \text{} & \text{} &
   \text{} & 2 & i Y\left(\frac{11}{2},-\frac{7}{2}\right) \\
 \text{G1} & \frac{3}{2} & 1 & 1 & 1 & Y\left(\frac{3}{2},-\frac{3}{2}\right) & \text{G1} & \frac{9}{2} & 5 & 1 & 1 & Y\left(\frac{9}{2},-\frac{7}{2}\right) & \text{G1} & \frac{11}{2} & 5 & 6 & 1 &
   Y\left(\frac{11}{2},\frac{9}{2}\right) \\
 \text{} & \text{} & \text{} & \text{} & 2 & Y\left(\frac{3}{2},\frac{3}{2}\right) & \text{} & \text{} & \text{} & \text{} & 2 & Y\left(\frac{9}{2},\frac{7}{2}\right) & \text{} & \text{} & \text{} &
   \text{} & 2 & -Y\left(\frac{11}{2},-\frac{9}{2}\right) \\
 \text{G1} & \frac{3}{2} & 2 & 1 & 1 & Y\left(\frac{3}{2},-\frac{1}{2}\right) & \text{G1} & \frac{9}{2} & 4 & 2 & 1 & Y\left(\frac{9}{2},-\frac{5}{2}\right) & \text{G1} & \frac{11}{2} & 6 & 6 & 1 &
   Y\left(\frac{11}{2},\frac{11}{2}\right) \\
\end{longtable}

%% file: Tables/BasisVectors/C2vElongatedBasisVectors.tex
\begin{longtable}{CCCCCC|CCCCCC|CCCCCC}
\caption{Basis vectors of $^2C_{2v}$ in terms of spherical harmonics $Y(J,M)$ up to $J = \frac{13}{2}$ in the elongated box. Multiplicities are labeled by $n$ and multidimensional components are indicated by $\alpha$.}
\label{tab:c2vbasisvectorselongated}
\renewcommand{\arraystretch}{2.0}
\endfirsthead
\multicolumn{6}{C} {{\tablename\ \thetable{:} \it Continued}} \\
\hline \endhead
\hline \\ \endfoot
\hline \\ \endlastfoot
\toprule
 \Gamma  & \text{J} & \text{l} & \text{n} & \alpha  & \text{Basis Vector} & \Gamma  & \text{J} & \text{l} & \text{n} & \alpha  & \text{Basis Vector} & \Gamma  & \text{J} & \text{l} & \text{n} & \alpha  & \text{Basis Vector}  \\
\text{G1} & \frac{1}{2} & 0 & 1 & 1 & Y\left(\frac{1}{2},\frac{1}{2}\right) & \text{G1} & \frac{7}{2} & 4 & 4 & 1 & Y\left(\frac{7}{2},\frac{5}{2}\right) & \text{G1} & \frac{11}{2} & 5 & 5 & 1 &
   Y\left(\frac{11}{2},\frac{7}{2}\right) \\
 \text{} & \text{} & \text{} & \text{} & 2 & Y\left(\frac{1}{2},-\frac{1}{2}\right) & \text{} & \text{} & \text{} & \text{} & 2 & Y\left(\frac{7}{2},-\frac{5}{2}\right) & \text{} & \text{} & \text{} &
   \text{} & 2 & i Y\left(\frac{11}{2},-\frac{7}{2}\right) \\
 \text{G1} & \frac{1}{2} & 1 & 1 & 1 & Y\left(\frac{1}{2},-\frac{1}{2}\right) & \text{G1} & \frac{9}{2} & 4 & 1 & 1 & Y\left(\frac{9}{2},-\frac{7}{2}\right) & \text{G1} & \frac{11}{2} & 6 & 5 & 1 &
   Y\left(\frac{11}{2},\frac{5}{2}\right) \\
 \text{} & \text{} & \text{} & \text{} & 2 & -i Y\left(\frac{1}{2},\frac{1}{2}\right) & \text{} & \text{} & \text{} & \text{} & 2 & Y\left(\frac{9}{2},\frac{7}{2}\right) & \text{} & \text{} & \text{} &
   \text{} & 2 & Y\left(\frac{11}{2},-\frac{5}{2}\right) \\
 \text{G1} & \frac{3}{2} & 1 & 1 & 1 & Y\left(\frac{3}{2},-\frac{1}{2}\right) & \text{G1} & \frac{9}{2} & 5 & 1 & 1 & Y\left(\frac{9}{2},-\frac{9}{2}\right) & \text{G1} & \frac{11}{2} & 5 & 6 & 1 &
   Y\left(\frac{11}{2},\frac{11}{2}\right) \\
 \text{} & \text{} & \text{} & \text{} & 2 & i Y\left(\frac{3}{2},\frac{1}{2}\right) & \text{} & \text{} & \text{} & \text{} & 2 & -i Y\left(\frac{9}{2},\frac{9}{2}\right) & \text{} & \text{} & \text{} &
   \text{} & 2 & -i Y\left(\frac{11}{2},-\frac{11}{2}\right) \\
 \text{G1} & \frac{3}{2} & 2 & 1 & 1 & Y\left(\frac{3}{2},-\frac{3}{2}\right) & \text{G1} & \frac{9}{2} & 4 & 2 & 1 & Y\left(\frac{9}{2},-\frac{3}{2}\right) & \text{G1} & \frac{11}{2} & 6 & 6 & 1 &
   Y\left(\frac{11}{2},\frac{9}{2}\right) \\
\end{longtable}